\documentclass[lettersize,journal]{IEEEtran}
\usepackage{algorithm, algorithmic}
\usepackage{array}
\usepackage{textcomp}
\usepackage{stfloats}
\usepackage{url}
\usepackage{verbatim}
\usepackage{graphicx}
\def\BibTeX{{\rm B\kern-.05em{\sc i\kern-.025em b}\kern-.08em
    T\kern-.1667em\lower.7ex\hbox{E}\kern-.125emX}}
\usepackage{balance}

\usepackage[numbers,sort&compress]{natbib}
\usepackage{orcidlink}
\hypersetup{hidelinks,
	pdfstartview=Fit,
	breaklinks=true}

\usepackage{color}
\usepackage{amsmath,amssymb,amsfonts}
\usepackage{booktabs}
\usepackage{multirow}
\usepackage{multicol}
\usepackage{makecell}
\usepackage{rotating}
\usepackage{wrapfig}
\usepackage{subfigure}
\usepackage{colortbl}
\usepackage{ulem}
\usepackage{array}
\usepackage{threeparttable}

\makeatletter

\newcommand{\Rmnum}[1]{\expandafter\@slowromancap\romannumeral #1@}
\makeatother
 
\begin{document}
% \title{Synthesize, Estimate and Recover: Diffusion-based Approach for Network Traffic Matrix Analysis}
\title{Diffusion Models Meet Network Management: Improving Traffic Matrix Analysis with Diffusion-based Approach}
% \author{Xinyu~Yuan\IEEEauthorrefmark{2}~\orcidlink{0009-0005-6287-5712},~\IEEEmembership{Student~Member,~IEEE}, and Yan~Qiao\IEEEauthorrefmark{2}~\orcidlink{0000-0002-4407-1762},~\IEEEmembership{Member,~IEEE}\\
% \IEEEauthorblockA{\IEEEauthorrefmark{2} School of Computer Science and Information Engineering,\\
% Hefei University of Technology, Hefei, Anhui 230009 China}
% }

\author{Xinyu~Yuan\IEEEauthorrefmark{2}, Yan~Qiao\IEEEauthorrefmark{2}\IEEEauthorrefmark{4}\thanks{
    Yan~Qiao is the corresponding author.
}, Zhenchun~Wei\IEEEauthorrefmark{2}\IEEEauthorrefmark{4}, Zeyu~Zhang\IEEEauthorrefmark{3}, Minyue~Li\IEEEauthorrefmark{2}, Pei~Zhao\IEEEauthorrefmark{2}, Rongyao~Hu\IEEEauthorrefmark{2} and Wenjing~Li\IEEEauthorrefmark{4}\\
\IEEEauthorblockA{\IEEEauthorrefmark{2}School~of~Computer~Science~and~Information~Engineering, Hefei~University~of~Technology, China\\
\IEEEauthorrefmark{3}Faculty~of~Information~Technology, Macau~University~of~Science~and~Technology, China\\
\IEEEauthorrefmark{4}State~key~Laboratory~of~Networking~and~Switching~Technology, Beijing~University~of~Posts~and~Telecommunications, China}
}

% \author{Xinyu~Yuan~\orcidlink{0009-0005-6287-5712},~\IEEEmembership{Member,~IEEE},\thanks{The preliminary version of this work has been accepted by ISCC 2023 \cite{r1}} and Yan~Qiao~\orcidlink{0000-0002-4407-1762},~\IEEEmembership{Member,~IEEE}}
% \author{IEEE Publication Technology Department
% \thanks{Manuscript created October, 2020; This work was developed by the IEEE Publication Technology Department. This work is distributed under the \LaTeX \ Project Public License (LPPL) ( http://www.latex-project.org/ ) version 1.3. A copy of the LPPL, version 1.3, is included in the base \LaTeX \ documentation of all distributions of \LaTeX \ released 2003/12/01 or later. The opinions expressed here are entirely that of the author. No warranty is expressed or implied. User assumes all risk.}}

% \markboth{IEEE TRANSACTIONS ON NETWORK AND SERVICE MANAGEMENT,~Vol.~X, No.~X, October~2023}%
% {Synthesize, Estimate and Recover: Diffusion-based Approach for Traffic Matrix Analysis}

\maketitle

\begin{abstract}
Due to network operation and maintenance relying heavily on network traffic monitoring, traffic matrix analysis has been one of the most crucial issues for network management related tasks. However, it is challenging to reliably obtain the precise measurement in computer networks because of the high measurement cost, and the unavoidable transmission loss. Although some methods proposed in recent years allowed estimating network traffic from partial flow-level or link-level measurements, they often perform poorly for traffic matrix estimation nowadays. Despite strong assumptions like low-rank structure and the prior distribution, existing techniques are usually task-specific and tend to be significantly worse as modern network communication is extremely complicated and dynamic. To address the dilemma, this paper proposed a diffusion-based traffic matrix analysis framework named Diffusion-TM, which leverages problem-agnostic diffusion to notably elevate the estimation performance in both traffic distribution and accuracy. The novel framework not only takes advantage of the powerful generative ability of diffusion models to produce realistic network traffic, but also leverages the denoising process to unbiasedly estimate all end-to-end traffic in a plug-and-play manner under theoretical guarantee. Moreover, taking into account that compiling an intact traffic dataset is usually infeasible, we also propose a two-stage training scheme to make our framework be insensitive to missing values in the dataset. With extensive experiments with real-world datasets, we illustrate the effectiveness of Diffusion-TM on several tasks. Moreover, the results also demonstrate that our method can obtain promising results even with $5\%$ known values left in the datasets.
\end{abstract}


\section{Introduction}\label{I}
\IEEEPARstart{A}{}traffic matrix (TM) is applied to track the traffic volumes between all possible pairs of network nodes, which are usually mentioned as origin to destination (OD) flows~\cite{r2}. It is a critical input for many network management tasks, including capacity planning, anomaly detection, and traffic engineering~\cite{r3}. For example, the traffic measurement can help with facing collisions, congestion in network~\cite{r72}, security hazards~\cite{r73}, and inefficient utilization of network resources~\cite{r74}. {To obtain the crucial network measurement, a direct way is leveraging flow-level monitoring tools, such as Cisco’s NetFlow/TMS~\cite{r4}, and OpenTM in the emerging software-defined network (SDN)~\cite{r5}. Unfortunately, with the continuous expansion of network scale, the complete measurement of these OD flows requires extremely high administrative costs as well as computational overhead~\cite{r6}. Moreover, not all devices in legacy networks can support SDN modules~\cite{r9}. Thus the collected flow data is usually partial, and obtaining a complete TM is still an open challenge.}

There are two types of methods to {alleviate the problem}: TM Completion and Network Tomography~\cite{r10}. Based on low-rank assumptions of real-world TMs, {the first one} generally use{s} matrix or tensor completion algorithms to recover the traffic data from sparse known entries~\cite{r16,r17}. Despite extremely low efficiency as large amounts of data need to be processed, they rely heavily on sparse assumptions while foregoing the usage of some crucial information from the whole system. To be more specific, firstly, these solutions estimate the conditional mean of the observed samples and can only work when the application data follow the Gaussian distributions~\cite{r42}, but can not handle a more complicated traffic data distribution. Secondly, they did not take the low-cost link load data and its corresponding routing information that plays a generally significant role in network management into consideration, leading to these useful and easily accessible resources not being utilized.

The second way is to estimate the flow-level traffic from the link-level measurements by means of Network Tomography (NT). This method infers fine-grained OD flows by solving a group of linear equations that involve both coarse-grained link loads and flow routing matrix. However, the key problem for NT-based traffic matrix estimation (TME) is that such linear equations are usually highly rank deficient, which means there is no unique solution of OD flows corresponding to the measured link loads. In the early years, the solutions to the NT problem are provided based on the {unrealistic} assumptions about the prior information of traffic data. Then with the development of deep learning, various neural networks have been built to learn the inverse mapping from link loads to OD flows~\cite{r59,r58}. {The simple learning-based method} is able to reconstruct the dynamic properties of network traffic via link counts and routing information without additional hypotheses. However, the solution must ensure that the routing matrix used for training and inference is consistent. The condition serves as a cornerstone in enabling NT-based traffic estimation since it is almost impossible to keep the routing information static nowadays, for example, routers configured with an adaptive routing policy often choose routing paths dynamically based on current network loads. {Besides, t}heir goal is just to solve the tomography equation, and output the complete TM by inputting only the link load data. It means that even knowing more than half of the OD pairs will not have any impact on the results, whereas they would greatly reduce the solution space.

Recently, diffusion models (DMs)~\cite{r14} have emerged as a new paradigm for generative models, theoretically underpinned by non-equilibrium thermodynamics~\cite{r18} and score-matching network~\cite{r25}. They are gaining significant popularity in a wide range of synthesis domains owing to their superior sampling quality and stable training dynamics. Nevertheless, this comes at the expense of poor scalability and increased sampling times due to the long Markov chain sequences required. 
However, we noticed that samples obtained from the DMs depend on the initial state of the sample distribution and each transition. It makes diffusion models strong candidates for producing TMs that satisfy the conditions imposed by the set of measurements via a “plug-and-play” approach that combines the diffusion model and the measurement process~\cite{r28,r29,r30}. To bridge research gaps for solving both NT and TMC problems, this paper focuses on a DM-based framework, which can generate high-quality traffic matrix given partial link loads and (or) OD pairs. Formally, we propose a versatile solution for various TM-related tasks in network management, which we call Diffusion-TM. By refining OD flows at each state during the reverse diffusion sampling, our solution only requires an off-the-shelf diffusion model to yield realistic and data-consistent results, without any extra training nor needing any modifications to model structures. Also note that most deep generative models are sensitive to massive data missing, while any large set of TM measurements is bound to have a significant number of missing values in the real-world networks. {Therefore}, we further designed an efficient two-stage strategy to alleviate the effect of missing values in diffusion model training and allowed them to be performed even when as much as over 95$\%$ of the data is missing. We demonstrate the effectiveness of our method on three different tasks: TM estimation, completion, and synthesis. 
% Unfortunately, applying DMs directly to deal with the TMs is intractable owing to the long computing time for high-dimensional traffic data, and the huge magnitude differences between OD flows. To answer the problem, We design an encoder-decoder module before the DM network to improve the performance of distribution mapping. The module provides a reversible mapping between real-TM and latent representations. On one hand, it can reduce the high dimensionality of the learning space. On the other, it scales the flows in TM to a much smaller range space without losing the dynamics of the distribution. We demonstrate the effectiveness of our method on three different tasks: traffic matrix estimation, completion and synthesis. 

The key contributions of this paper can be summarized as follows:
\begin{itemize}
    \item We not only propose a \underline{\textbf{diffusion}}-based approach for IP-network \underline{\textbf{t}}raffic \underline{\textbf{m}}atrix analysis called Diffusion-TM, but also theoretically prove its efficiency on recovering traffic matrices while capturing the traffic data distribution via a novel approximation. To the best of our knowledge, this is one of the first works that leverage DMs to analyze traffic matrix. 
    \item To ensure that the result of Diffusion-TM conforms to the desired distribution across the whole range of missing values scenarios, we provide a two-stage training scheme with additional pre-processing work and missing data aware objective. The results suggest that the framework can be applied when large amounts of missing data exist in training datasets.
    \item We conduct extensive experiments with real-world traffic trace data to evaluate the effectiveness of our approach in a wide range of TM-related problems including network tomography, traffic recovery, and synthetic data generation.
\end{itemize}

The rest of the article is organized as follows. We discuss in Section~\ref{s2} the existing literature about the TM measurement problem. In Section~\ref{s3} and Section~\ref{s4}, we introduce relevant background and basic concepts, respectively. In Section~\ref{s5}, we transform the original problem into an approximation problem and explain why it takes effect. We then formally introduce our DM-based approach and the model structure in Section~\ref{s6}. We evaluate the performance of the proposed Diffusion-TM through extensive experiments in Section~\ref{s7}. Finally, We conclude the article in Section~\ref{s8}.

% \vspace{-3mm}

\section{Related Work}\label{s2}

{The TM estimation problem is a well investigated, but still open, research topic in Software Defined Networking (SDN) networks. As an example, in~\cite{r81} the authors propose an OpenFlow-based framework to assess the TM of a network, while in~\cite{r78} a mixed measurement and estimation algorithm by exploiting the availability of flow rule counters in SDN switches is presented. However, the process of measuring the TM using SDN rule counters is often memory-intensive.} Existing estimation methods can be classified into two categories. In the first category, traffic matrix completion (TMC) was proposed to recover the missing entries from a low-rank matrix with the development of sparse techniques. Zhang et al.~\cite{r33} proposed a sparsity regularized SVD method to estimate the missing values, then they improved the algorithm by proposing sparsity regularized matrix factorization (SRMF) in~\cite{r35}. However, the performance of two-dimensional matrix-based data recovery methods is relatively low due to the matrix’s limitations in information extraction. {Despite its effectiveness, Zhou et al.~\cite{r16} started to model the network traffic data as a higher dimensional array called tensor and propose algorithms based on tensor completion for more accurate missing data recovery.} The second category uses network tomography to estimate TM from the link loads by solving the linear equations. Since the linear system is generally rank deficient, the accuracies of these methods heavily rely on the underlying assumptions in the early years. For example, Vardi et al.~\cite{r37} assumed that the traffic followed the Poisson distribution, and Zhang et al.~\cite{r38} imposed a gravity model on the TM estimation. 

{Different from these solutions, the application of deep learning algorithms to discover structural characteristics of TMs has appeared as a viable approach. For TM completion (or prediction), Xie et al. developed NTC~\cite{r17} and GT-NET~\cite{r13} by exploiting neural network based tensor decomposition models to digest the data features on the estimation result. Abdelhadi et al.~\cite{r79} presented NeuTM for TM prediction based on Long Short-Term Memory Recurrent Neural Networks (LSTM RNNs), while the method proposed in~\cite{r41} combined the forward and backward Convolutional LSTM (ConvLSTM) network to correct the input TM data. To improve the accuracy of future traffic estimation, Richard et al.~\cite{r80}, in turn, investigated TM prediction based on deep ensemble learning model utilizing multiple dimensionality reduction algorithms and a genetic algorithm (GA) in SDN enabled networks. In~\cite{r39}, the authors used graph embedding to integrate the network topology with the model input, such that model could learn more specific networking knowledge. Regrading the NT problem, the authors in \cite{r59} introduced a back-propagation neural network (BPNN) to estimate TM, then MNETME~\cite{r58} refined the output data of the network using the EM algorithm. Although their performance does not rely on additional assumptions or the spatial-temporal structure of TM, these techniques still struggle with distribution alignment.} 

{To sum up the current works, their problems can be divided into three categories: (1) Unable to capture the traffic data distribution; (2) Unable to harness known traffic and routing information simultaneously, thus lack of flexibility; (3) Unsatisfactory prerequisites such as low-rank feature, prior distribution, and immutable routing. Fortunately, the development of deep generative models provides a positive answer to these fundamental questions.} Xie et al.~\cite {r42} designed a Deep Adversarial Tensor Completion (DATC) scheme based on generative adversarial network (GAN). Besides, a matrix completion and prediction algorithm based on a combination of generative autoencoders and Hidden Markov Models was proposed in~\cite{r43}. {In particular, the authors in~\cite{r22} and~\cite{r23} used a variational autoencoder (VAE) and GAN to learn the latent distribution that is “similar” to the training set of TM, respectively.} 
The approach leverages the prior (i.e. Gaussian) space of a pre-trained generative model to solve inverse problems in a zero-shot way. More concretely, they first train a generator (or decoder) network, then optimize 
TME objective function through gradients of data which can be easily computed by the chain rule. Therefore, estimations can be updated iteratively by using simple stochastic gradient descent. However, either VAE or GAN have its inherent model defects: VAE tends to produce unrealistic and blurry samples, meanwhile, the training of GAN is often unstable.

{Finally, Yuan et al.~\citep{r1} proposed a framework to solve the TM estimation problem, using diffusion models which have a spectacular ability to capture both diversity and fidelity. In this
framework, the authors search for the optimal answer by repeating the sampling process of a latent DM. Different from it, this paper studies an entirely new DM-basd approach that only needs to run along the Markov chain just once, while considering the problem of missing traffic dataset. Moreover, we theoretically prove that our strategy is the key to significantly improve the performance of traffic reconstruction.}

% \vspace{-1mm}

\begin{figure}
    \setlength{\abovecaptionskip}{-0.2cm} 
    \centerline{\includegraphics[width=1.\linewidth]{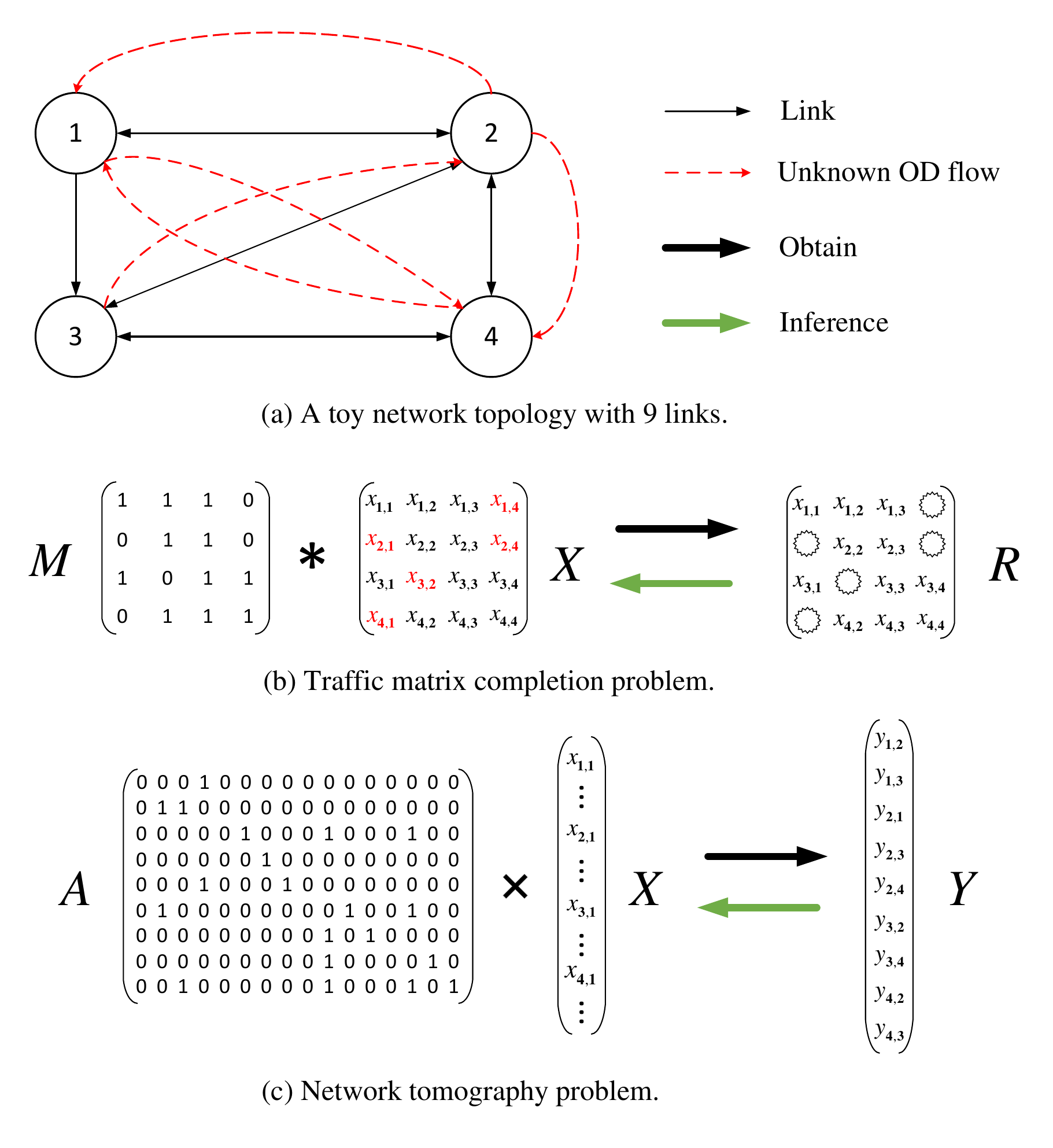}}
    \caption{\textbf{Illustration of studied problems in this paper.} We seek an estimated TM $\boldsymbol{X}$ that satisfies the conditions imposed by the set of measurements $\boldsymbol{R}$ or $\boldsymbol{Y}$. However, the considered problem is highly underdetermined .}
    \label{studied_problem}
\end{figure}

\section{System Model and Problem Formulation}\label{s3}

We start by introducing the basic notations and definitions in this section. Then, we will present the problem formulation in our article.

% \vspace{-3mm}

\subsection{System Model}

In our system model, we consider the network graph as $G=(V, E)$ where $V$ and $E$ are network nodes and links, respectively. The TM is defined as a $|V|\times |V|$ matrix where each entry represents an OD flow between a pair of nodes in the network. To facilitate calculations, we reshape TM to a vector $\{X_{1:N}\}$, where $N=|V|\times|V|$ is the total number of OD flows in TM. We denote a sequence of TMs from time point $1$ to $T$ as $\boldsymbol{X}=\left\{X_{1:N,1:T}\right\}$. 
% Throughout the paper, we use tensors, which is a higher-order generalization of a vector and a matrix, to represent these multidimensional arrays. 

% \vspace{-3mm}

\subsection{Problem Formulation}

% \subsubsection{Network Tomography}
As shown in Fig.~\ref{studied_problem}, the problem refers to the inference of unmeasured network attributes based on measurements realized at a subset
of accessible network elements. Let $\boldsymbol{Y}=\left\{Y_{1:M,1:T}\right\}$ denote the sequence of link loads, and $\boldsymbol{A}\in \mathbb{R}^{M\times N}$ denote the routing matrix, where each entry $a_{ij}$ of $\boldsymbol{A}$ has a binary value ($0$ or $1$). For deterministic routing policy, if the $j$-th flow traverses the $i$-th link, then $a_{ij}=1$; otherwise, $a_{ij}=0$. For probabilistic routing policy (such as ECMP), the value of $a_{ij}$ is within the range of $[0,1]$, representing the probability that the $j$-th flow may transverse the $i$-th link. The relationship between TM $\boldsymbol{X}$ and link load $\boldsymbol{Y}$ can be formulated as the linear equations:
\begin{equation}
\boldsymbol{A}\boldsymbol{X}=\boldsymbol{Y}.\label{NT}
\end{equation}
In most networks, the number of flows $N$ is much greater than the number of link loads $M$, leading to a highly rank-deficient system. That means Eqn.~\ref{NT} does not have a unique solution in most cases.

% \subsubsection{Traffic Matrix Completion}
As inferring $\boldsymbol{X}$ from the compressed measurements $\boldsymbol{Y}$ is a severely underdetermined task,
% additional information is needed to ensure identifiability and improve estimation accuracy. 
a more general approach is the direct flow-level measurements although only for partial traffic volumes. Let us denote an observation mask as $\boldsymbol{M}=\left\{m_{1:N,1:T}\right\}\in\left\{0,1\right\}^{N\times T}$ where $m_{n,t}=1$ if $x_{n,t}$ is observed, and $m_{n,t}=0$ if $x_{n,t}$ is unobserved. Consequently, the known-measurement matrix $\boldsymbol{R}$ which denotes the set of information that is available, is defined as 
\begin{equation}
    \boldsymbol{M}\odot\boldsymbol{X}=\boldsymbol{R},
    \label{FC}
\end{equation} 
where $\odot$ represents elementwise multiplication. 

\textbf{Problem:} Now the estimation problem for TMs can be defined as follows: given the measurement $\{\boldsymbol{Y}, \boldsymbol{R}\}$ obeying Eqn.~\ref{NT} and/or Eqn.~\ref{FC}, we aim to accurately recover the unknown traffic data, with $\{\boldsymbol{A}, \boldsymbol{M}\}$ known. And we have a set of linear constraints on the TM
\begin{equation}
\boldsymbol{\cal A}\left( \boldsymbol{X}\right) + \boldsymbol{z}=\boldsymbol{\cal Y}
\label{inverse_problem}
\end{equation}
where $\boldsymbol{\cal A}\left( \cdot \right)$ is a linear operator, the matrix $\boldsymbol{\cal Y}$ contains the available measurements, and $\boldsymbol{z}$ is the measurement noise as the Simple Network Management Protocol (SNMP) used for collecting link measurements is often noisy \cite{r44} and flow-level collection usually involves sampling at quite high rates. Specifically, we consider white Gaussian noise $\boldsymbol{z} \sim {\cal N}\left(0, \sigma_z^2\right)$ in this work. In later parts of the paper, we may also denote the linear operation as $\boldsymbol{\cal Y}=\boldsymbol{HX} + \boldsymbol{z}$, where $\boldsymbol{\cal A}\left( \boldsymbol{X} \right)$ is
replaced by a matrix operation $\boldsymbol{HX}$.

\textbf{Target:} The aim of this work is to sample points from data distribution conditioned on partially observed traffic or/and link measurements. We formulate the traffic estimation problem as a penalized least-squares problem, i.e.
\begin{equation}
    \mathop {\min }\limits_{\boldsymbol{x}} \left\| {\boldsymbol{\cal Y} - \boldsymbol{\cal A}\left( \boldsymbol{x} \right)} \right\|_2^2 - 2 \cdot {\sigma_z ^2}\log {p_0}\left( \boldsymbol{x} \right),
\end{equation}
where we model the estimation $\boldsymbol{x}$ as being drawn from prior distribution with density $p_0\left( \boldsymbol{x}\right)$. The objective can also be written as the following form with so-called posterior probability density:
\begin{equation}
    \mathop {\max }\limits_{\boldsymbol{x}} \log p\left( {\boldsymbol{x}|\boldsymbol{y}} \right), \  {\rm {s.t.}} \ \boldsymbol{y}: \boldsymbol{\cal A}\left( \boldsymbol{x}\right) + \boldsymbol{z}=\boldsymbol{\cal Y}
    \label{objective}
\end{equation}
which treats this as a maximum likelihood problem. Heuristically, we choose the solution that best fits prior distribution while satisfying the constraints.

\begin{figure}
    \setlength{\abovecaptionskip}{-0.1cm} 
    \centerline{\includegraphics[width=0.8\linewidth]{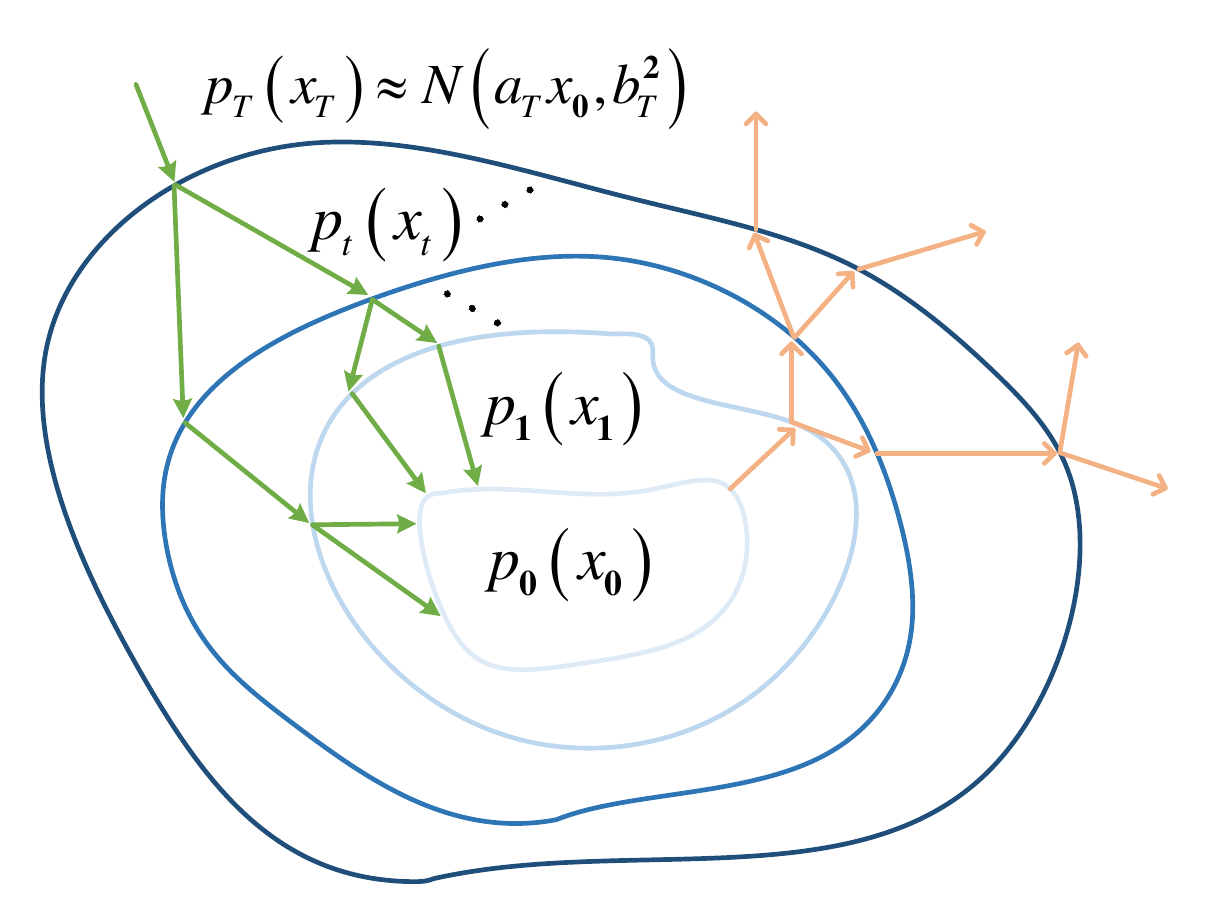}}
    \caption{\textbf{Geometrical visualization of diffusion models.} The central area represents the original data manifold which has been proved to be encircled by manifolds of noisy data $p_t\left( \boldsymbol{x_t}\right)$ \cite{r31}. The encoding (forward) process depicted by orange arrows gradually converts original data distribution $p_0\left( \boldsymbol{x_0}\right)$, into a simple isotropic Gaussian ${\cal N}\left(0,\rm{I}\right)$. While the decoding (reverse) process depicted by green arrows can be considered as transitions from $p_t\left( \boldsymbol{x_t}\right)$ to $p_{t-1}\left( \boldsymbol{x_{t-1}}\right)$ through a Markov Chain.}
    \label{diffusion_graphy}
\end{figure}

\section{Preliminaries of Diffusion Models}\label{s4}

{Suppose $p_0\left( \boldsymbol{x_0}\right)$ be the $d$-dimensional data distribution. We consider the general class of diffusion models in a continuous form that can be described with the following 
Ornstein-Uhlenbeck stochastic differential equation (SDE)}
\begin{equation}
    {d\boldsymbol{x} = \boldsymbol{f}\left( \boldsymbol{x}, t\right)dt + g\left( t\right)d\boldsymbol{w}}
\end{equation}
{where $\boldsymbol{w}$ denotes the standard Brownian motion, $\boldsymbol{f}: \mathbb{R}^d \times \mathbb{R} \to \mathbb{R}^d$, $g: \mathbb{R} \to \mathbb{R}$ is the linear drift function and scalar diffusion coefficient, respectively. The SDE results in a series of marginal distributions $\left\{p_t\left(\boldsymbol{x_t}\right)\right\}$ where $t \in \left[0,T\right]$, so that $\boldsymbol{x_T} \sim {\cal N}\left( 0,\sigma_T^2\right)$ with some constants $\sigma_T>0$.}

{It is known that the density evolution process can be reversed by another SDE with the Anderson’s theorem}
\begin{equation}
    {d\boldsymbol{x} = \left[ {\boldsymbol{f}\left( {\boldsymbol{x},t} \right) - {g^2}\left( t \right){\nabla _{\boldsymbol{x_t}}}\log {p_t}\left( \boldsymbol{x_t} \right)} \right]dt + g\left( t \right)d\boldsymbol{\bar w}}
    \label{sgm_reverse}
\end{equation}
{where $\boldsymbol{\bar w}$ is a Brownian motion running backward in time from $T$ to $0$. Then diffusion models approximate this reverse process by learning the score function ${\nabla _{\boldsymbol{x_t}}}\log {p_t}\left( \boldsymbol{x_t} \right)$. Therefore, as shown in Fig.~\ref{diffusion_graphy}, the key idea behind them is to add small amounts of noise gradually to a data point for transitions from data manifold to noisy manifolds, then train the underlying neural network to transport from the pure Gaussian to the clean area through inverting the diffusion process.} 

\begin{figure*}
    \setlength{\abovecaptionskip}{-0.2cm} 
    \centerline{\includegraphics[width=1\linewidth]{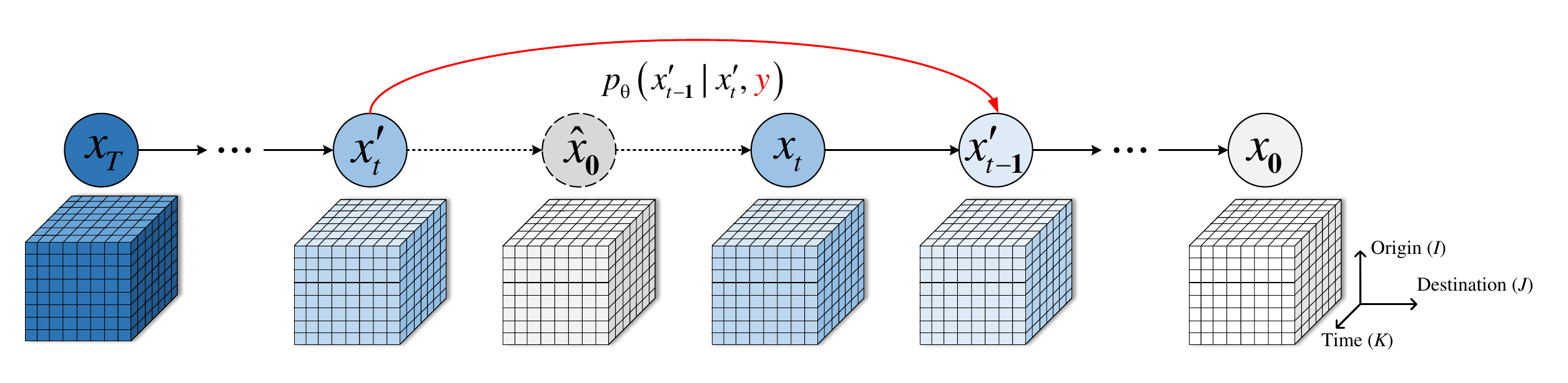}}
    \caption{\textbf{Illustration of our diffusion-based approach for solving TM estimation problems.} The reverse inference process (from right to left) iteratively denoises the target traffic matrix $\boldsymbol{x_0}$ conditioned on the measurement $\boldsymbol{y}$. Concretely, following the prediction of the estimated $\boldsymbol{\hat x_0}$ by an unconditional diffusion model, the measurement $\boldsymbol{y}$ is incorporated by solving a proximal subproblem depicted by red arrows in the VP-SDE.}
    \label{illustration}
\end{figure*}

{In this paper, we focus on Denoising Diffusion Probabilistic Models~\cite{r14} which is equivalent to the variance preserving form of the SDE (VP-SDE). To be more specific, we have the forward and reverse SDEs
as the continuous version of the diffusion process in DDPM with the choice of $\boldsymbol{f}\left( {\boldsymbol{x},t} \right) = -\beta\left(t\right) \boldsymbol{x}/2$ and $g\left( t \right) = \sqrt{\beta\left(t\right)}$. In particular, the forward process of DDPMs gradually corrupts original data $x_0 \in \mathbb{R}^d$ via a fixed Markov chain $x_0, \dots, x_T$ with each variable in $\mathbb{R}^d$ as follows:}
\begin{equation}
   {\left\{\begin{aligned}
       &\boldsymbol{x_t}|\boldsymbol{x_{t-1}} \sim {\cal N}\left(\boldsymbol{x_t};\sqrt{{\alpha _t}}{x_{t-1}},(1-{\alpha_t}){\rm I}\right), \\
       &\boldsymbol{x_t}|\boldsymbol{x_0} \sim {\cal N}\left(\boldsymbol{x_t};\sqrt{{\bar\alpha _t}}{x_0},(1-{\bar\alpha_t}){\rm I}\right), \\
       & \boldsymbol{x_t}=\sqrt{{{\bar\alpha}_t}}\boldsymbol{x_0}+\sqrt{1-{{\bar\alpha}_t}}\epsilon
   \end{aligned}\right.}
\label{forward}
\end{equation}
{with $\epsilon \sim{\cal N}(0,\rm{I})$, $\alpha_t:=1-\beta_t$ where $\beta_t \in \left(0, 1\right)$ is a variance at diffusion step $t$, scaling factor $\bar\alpha_t:=\prod\limits_{s=1}^t \alpha_s$.}

{Starting from the Gaussian noise $\boldsymbol{x_T}$, we can run the reverse process parametrized by the model ${p_\theta }\left(\boldsymbol{x_{t - 1}}|{x_t}\right): = {\cal N}(\boldsymbol{x_{t - 1}};{\mu}\left(\boldsymbol{x_t},t,\theta\right),\boldsymbol{\Sigma} \left(\boldsymbol{x_t},t,\theta\right) )$ to get $\boldsymbol{x_0}$. There are many different ways to parameterize the posterior mean ${\mu}\left(\boldsymbol{x_t},t,\theta\right)$, and the most obvious option is to predict ${\mu}\left(\boldsymbol{x_t},t,\theta\right)$ directly:}
\begin{equation}
    {{\boldsymbol{\cal L}_{o}}: = {\mathbb{E}_{t,\boldsymbol{x_0}}}\left[\frac{1}{{2{{\boldsymbol{\Sigma}^2 }\left(\boldsymbol{x_t},t,\theta\right)}}}{\left\| {\hat \mu (\boldsymbol{x_t},\boldsymbol{x_0}) - {\mu}\left(\boldsymbol{x_t},t,\theta\right)} \right\|^2}\right],}
\end{equation}
{where ${{\boldsymbol{\Sigma} }\left(\boldsymbol{x_t},t,\theta\right)}$ is often set to pre-defined time dependent constants, and $\hat \mu (\boldsymbol{x_t},\boldsymbol{x_0})$ is the mean of the posterior $p(\boldsymbol{x_t}|\boldsymbol{x_{t - 1},x_0})$ which are defined as follows:}
\begin{equation}
    {\hat \mu (\boldsymbol{x_t},\boldsymbol{x_0})=\frac{{\sqrt {{{\bar \alpha }_{t - 1}}} {\beta _t}}}{{1 - {{\bar \alpha }_t}}}\boldsymbol{x_0} + \frac{{\sqrt {{\alpha _t}} (1 - {{\bar \alpha }_{t - 1}})}}{{1 - {{\bar \alpha }_t}}}\boldsymbol{x_t}.
    \label{posterior_mean}}
\end{equation}
{Alternatively, the network could also predict $\boldsymbol{x_0}$ or $\boldsymbol{\epsilon}$ using \ref{forward} and \ref{posterior_mean}. In this work, we train our model to reconstruct input $\boldsymbol{x_0}$ itself combined with a reweighted loss function:}
\begin{equation}
{{\boldsymbol{\cal L}_{simple}}={\mathbb{E}_{t,\epsilon,\boldsymbol{x_0}}}\left[{{{\left\|{\boldsymbol{x_0}-{\boldsymbol{\hat x_0}}(\boldsymbol{x_t},t,\theta)}\right\|}^2}}\right].}
\end{equation}

The connection between the score function and the prediction in DDPMs can be formulated approximately as: $s_\theta\left( \boldsymbol{x_t}, t\right) \approx \left({\boldsymbol{x_t} - \sqrt {{{\bar \alpha }_t}} \boldsymbol{{\hat x}_0}\left( {\boldsymbol{x_t},t,\theta } \right)}\right)/\left({{1 - {{\bar \alpha }_t}}}\right)$ \cite{r32}. Data generation through denoising depends on the score function and can be seen as noise conditional score-based generation. Then the reverse process $p_\theta(\boldsymbol{x_{t-1}}|\boldsymbol{x_t})$ in DDPMs can be written as follows:
% {\setlength\abovedisplayskip{0.1cm}
% \setlength\belowdisplayskip{0.1cm}
\begin{equation}
    \begin{aligned}
        \boldsymbol{x_{t - 1}} =& \frac{{\sqrt {{\alpha _t}} (1 - {{\bar \alpha }_{t - 1}})}}{{1 - {{\bar \alpha }_t}}}\boldsymbol{x_t} + \frac{{\sqrt {{{\bar \alpha }_{t - 1}}} {\beta _t}}}{{1 - {{\bar \alpha }_t}}}\boldsymbol{\hat x_0}(\boldsymbol{x_t},t,\theta )  \\
        &+ \frac{{1 - {{\bar \alpha }_{t - 1}}}}{{1 - {{\bar \alpha }_t}}}{\beta _t}\boldsymbol{z_t}.
    \end{aligned}
    \label{ddim}
\end{equation}

In order to sample with diffusion models more quickly, \cite{r47} proposed Denoising Diffusion Implicit Models (DDIMs) that can be rewritten as:
\begin{equation}
    \begin{aligned}
        \boldsymbol{x_{t-1}} =& \sqrt {1 - \frac{{\sigma _{{\psi _t}}^2}}{{1 - {{\bar \alpha }_t}}}} \left( {\boldsymbol{x_t} - \sqrt {{{\bar \alpha }_t}} {\boldsymbol{\hat x}_0}\left( {\boldsymbol{x_t},t,\theta } \right)} \right) \\
        &+ \sqrt {{{\bar \alpha }_{t - 1}}} \boldsymbol{\hat x_0}\left( {\boldsymbol{x_t},t,\theta } \right) + {\sigma _{{\psi _t}}}\boldsymbol{z_t}
    \end{aligned}
\end{equation}
where ${\sigma _{{\psi _t}}}$ controls the stochastic degree of the diffusion process. Compared to DDPM, DDIM extended from Markovian to non-Markovian is able to generate higher-quality samples using a much fewer number of sampling steps.

\begin{table*}[htb]
\normalsize
\resizebox{\textwidth}{!}{
\begin{tabular}{l}
\toprule
\textbf{Algorithm 1} Diffusion-TM Sampling for Network Tomography\\
\midrule
\begin{minipage}{0.48\linewidth}\label{alg}
    \centering
    \begin{algorithmic}[1]
    \REQUIRE scale coefficients $\left\{\rho_t\right\}_{t=1}^T$, link loads $\boldsymbol{Y}$, and\\
    ~~~~~~~routing matrix $\boldsymbol{A}$
    \ENSURE estimated TM $\boldsymbol{\hat x_0}$
    \STATE $\boldsymbol{x_T} \sim {\cal N}(0, \rm{I})$;
    \FOR{all $t$ from $T$ to 1}
         \STATE $z \sim {\cal N}\left( 0, \rm{I}\right)$;
         \STATE $\hat s_\theta \leftarrow {\textbf{Score}}\left( {\boldsymbol{{\hat x}_0}\left( {\boldsymbol{x_t},t,\theta } \right),t} \right)$;
         \STATE $\boldsymbol{x'_{t - 1}} = \frac{1}{{\sqrt {{\alpha _t}} }}\left( {\boldsymbol{x_t} - (1 - {\alpha _t}){s_\theta }\left( {\boldsymbol{x_t},t} \right)} \right) + {\sigma _t}z$;
         \STATE $\boldsymbol{\hat x_0} \leftarrow \frac{1}{{\sqrt {{{\bar \alpha }_t}} }}\left( {\boldsymbol{x_t} + \left( {1 - {{\bar \alpha }_t}} \right)\hat s_\theta} \right)$;
         \STATE $\boldsymbol{x_{t - 1}} = \boldsymbol{x'_{t - 1}} + {\rho_t}{\nabla _{\boldsymbol{x_t}}}\left\| {\boldsymbol{Y} - \boldsymbol{A} {\boldsymbol{{\hat x}_0}}} \right\|_2^2$;
    \ENDFOR
    \STATE $\boldsymbol{\hat x_0} \leftarrow {\textbf{EM\_Optimization}}\left( \boldsymbol{\hat x_0}, \boldsymbol{Y}\right)$;
    \RETURN $\boldsymbol{\hat x_0}$;
    \end{algorithmic}
\end{minipage} \\
\bottomrule
\end{tabular}
\begin{tabular}{l}
\toprule
\textbf{Algorithm 2} Diffusion-TM Sampling for Traffic Matrix Completion\\
\midrule
\begin{minipage}{0.48\linewidth}
    \centering
    \begin{algorithmic}[1]
    \REQUIRE scale coefficients $\left\{\rho_t\right\}_{t=1}^T$, observed TM $\boldsymbol{X}^o$, and\\
    ~~~~~~~observation matrix $\boldsymbol{M}$
    \ENSURE estimated TM $\boldsymbol{\hat x_0}$
    \STATE $\boldsymbol{x_T} \sim {\cal N}(0, \rm{I})$;
    \FOR{all $t$ from $T$ to 1}
         \STATE $z \sim {\cal N}\left( 0, \rm{I}\right)$;
         \STATE $\hat s_\theta \leftarrow {\textbf{Score}}\left( {\boldsymbol{{\hat x}_0}\left( {\boldsymbol{x_t},t,\theta } \right),t} \right)$;
         \STATE $\boldsymbol{x'_{t - 1}} = \frac{1}{{\sqrt {{\alpha _t}} }}\left( {\boldsymbol{x_t} - (1 - {\alpha _t}){s_\theta }\left( {\boldsymbol{x_t},t} \right)} \right) + {\sigma _t}z$;
         \STATE $\boldsymbol{\hat x_0} \leftarrow \frac{1}{{\sqrt {{{\bar \alpha }_t}} }}\left( {\boldsymbol{x_t} + \left( {1 - {{\bar \alpha }_t}} \right)\hat s_\theta} \right)$;
         \STATE $\boldsymbol{x_{t - 1}} = \boldsymbol{x'_{t - 1}} + {\rho_t}{\nabla _{\boldsymbol{x_t}}}\left\| {\boldsymbol{M} \odot \boldsymbol{X}^o - \boldsymbol{M} \odot {\boldsymbol{{\hat x}_0}}} \right\|_2^2$;
         \STATE $\boldsymbol{\hat x_0} \leftarrow \textbf{Replace}\left( \boldsymbol{\hat x_0},\boldsymbol{X},\boldsymbol{M},t\right)$;
    \ENDFOR
    \RETURN $\boldsymbol{\hat x_0}$;
    \end{algorithmic}
\end{minipage} \\
\bottomrule
\end{tabular}}
% \label{alg}
\end{table*}

\section{Problem Variation and Solution}\label{s5}
This section explores improving the analysis of traffic matrices with diffusion models which has demonstrated very appealing performance in general distribution modeling. 
% Since the problem is massively ill-posed, it is natural to choose the solution that best fits the learnt distribution, among all traffic matrices agreeing with particular constraints. 
% But rather than generating an estimated TM by repeatedly adjusting the outputs like previous generative-model-based methods, 
We propose thinking an approach for refining the reverse process of an unconditional diffusion model for TM-related tasks. Given the learned TM distribution $p\left(\boldsymbol{x}\right)$ of DMs, the main challenge of the problem is how to conduct the mapping from $\boldsymbol{y}$ to $\boldsymbol{x}$ without explicit information on the conditional probability $p\left( \boldsymbol{x}|\boldsymbol{y}\right)$. Below we first decompose our target step by step, then we re-formulate and solve the problem by leveraging the Tweedie’s method~\cite{r77}. And finally, we theoretically show that one can find a solution to both constraint and original data consistency, so the result becomes more
accurate and stable.

\subsection{Main Idea}
Shown as Eqn.~\ref{objective}, both traffic matrix completion and traffic tomography problem could be divided into constrained generation tasks{. T}he goal is to produce TMs from the posterior distribution $p\left( \boldsymbol{x_0}|\boldsymbol{y}\right)$ given the condition $\boldsymbol{y}$ which could be observed entities or/and link loads. Therefore, we consider rewriting Eqn.~\ref{sgm_reverse} as follows for conditional transition
\begin{equation}
    d\boldsymbol{x} = \left[ {\boldsymbol{f}\left( {\boldsymbol{x},t} \right) - {g^2}\left( t \right){\nabla _{\boldsymbol{x_t}}}\log {p_t}\left( \boldsymbol{x_t}|\boldsymbol{y} \right)} \right]dt + g\left( t \right)d\boldsymbol{\bar w}.
    \label{Conditional_sde}
\end{equation}
Leveraging the diffusion model as the prior, the question here is how to compute the conditional score ${\nabla _{\boldsymbol{x_t}}}\log {p_t}\left( {\boldsymbol{x_t}|\boldsymbol{y}} \right)$.

We start with the problem-specific score which can be decomposed via Bayes’ rule as below:
\begin{equation}
    {\nabla _{\boldsymbol{x_t}}}\log {p_t}\left( {\boldsymbol{x_t}|\boldsymbol{y}} \right) = {\nabla _{\boldsymbol{x_t}}}\log {p_t}\left( {\boldsymbol{x_t}} \right) + {\nabla _{\boldsymbol{x_t}}}\log {p_t}\left( {\boldsymbol{y}|\boldsymbol{x_t}} \right),
    \label{start}
\end{equation}
where the first term can be approximated by a pre-trained score function $s_{\theta} \left(\boldsymbol{x_t}, t \right)$, and the second is a guidance term which is intractable to compute because there is no explicit dependence between $\boldsymbol{x_t}$ and $\boldsymbol{y}$. 

Thus, we have to resort to approximate ${\nabla _{\boldsymbol{x_t}}}\log {p_t}\left( {\boldsymbol{y}|\boldsymbol{x_t}} \right)$ to circumvent using the likelihood term directly. Note that the forward diffusion is able to be represented by Eqn.~\ref{forward}, we can consider an independent graphical model: $\boldsymbol{x_0} \to \boldsymbol{y},\ \boldsymbol{x_0} \to \boldsymbol{x_t}$. Then, by factorizing $p\left( \boldsymbol{y}|\boldsymbol{x_t}\right)$ as follows:
\begin{equation}
    \begin{aligned}
       p\left( {\boldsymbol{y}|{\boldsymbol{x_t}}} \right) &= \int {p\left( {\boldsymbol{y}|\boldsymbol{x_0},\boldsymbol{x_t}} \right)p\left( {\boldsymbol{x_0}|\boldsymbol{x_t}} \right)d\boldsymbol{x_0}} \\
       &= \int {p\left( {\boldsymbol{y}|\boldsymbol{x_0}} \right)p\left( {\boldsymbol{x_0}|\boldsymbol{x_t}} \right)d\boldsymbol{x_0}},
    \end{aligned}
\end{equation}
we can now transform the problem into approximating another intractable $p\left( {\boldsymbol{x_0}|\boldsymbol{x_t}} \right)$ as the likelihood of $p\left( \boldsymbol{y}|\boldsymbol{x_0}\right)$ is tractable in general. 

\subsection{Approximation Problem}
Here, our solution to above issue is to use the specialized representation of the posterior mean to obtain reasonable approximations to the true $p\left( {\boldsymbol{x_0}|\boldsymbol{x_t}} \right)$ through a generalization of \textit{Tweedie’s Formula}.

\textbf{Lemma 1} (\textbf{Tweedie’s Formula}): \textit{Given $\boldsymbol{\eta} \sim g\left( \cdot \right)$, suppose $p\left( {\boldsymbol{x}|\boldsymbol{\eta}} \right)$ belong to the exponential family distribution}
\begin{equation}
    p\left( {\boldsymbol{x}|\boldsymbol{\eta}} \right) = {p_0}\left( \boldsymbol{x} \right)\exp \left( {{\boldsymbol{\eta} ^T}F\left( \boldsymbol{x} \right) - \psi \left( \boldsymbol{\eta} \right)} \right),
\end{equation}
\textit{the unique posterior mean $\boldsymbol{\hat \eta}$ of $p\left( \boldsymbol{\eta}|\boldsymbol{x} \right)$ will satisfy}
\begin{equation}
    {\left( {{\nabla _{\boldsymbol{x}}}F\left( \boldsymbol{x} \right)} \right)^T} \boldsymbol{\hat \eta} = {\nabla _{\boldsymbol{x}}}\log p\left( \boldsymbol{x} \right) - {\nabla _{\boldsymbol{x}}}\log {p_0}\left( \boldsymbol{x} \right),
\end{equation}
\textit{where $F\left( \boldsymbol{x} \right)$ is the function of $\boldsymbol{x}$, $\boldsymbol{\eta}$ the natural or canonical parameter of the family, $\psi \left( \boldsymbol{\eta} \right)$ the cumulant generating function (which makes the density $p\left( {\boldsymbol{x}|\boldsymbol{\eta}} \right)$ integrate to 1), and ${p_0}\left( \boldsymbol{x} \right)$ the density when $\boldsymbol{\eta}=0$.}

% \textit{Proof}. 

\textbf{Conclusion}: \textit{If we suppose that $\boldsymbol{\eta}$ has been sampled from a prior distribution $g\left( \boldsymbol{\eta}\right)$, and $\boldsymbol{x}|\boldsymbol{\eta} \sim {\cal N}\left(  \boldsymbol{\eta}, \sigma^2\right)$ has been observed, we can write:}
\begin{equation}
     \mathbb{E}\left( \boldsymbol{\eta}|\boldsymbol{x} \right)=\boldsymbol{x} + \sigma^2 {\nabla _{\boldsymbol{x}}}\log p\left( \boldsymbol{x} \right).
     \label{TF_conclusion}
\end{equation}
\textit{that comes from rewriting Tweedie’s formula where}
\begin{equation}
    F\left(\boldsymbol{x} \right) = \frac{\boldsymbol{x}}{{{\sigma ^2}}}, \ \ \psi \left( \boldsymbol{\eta} \right) = \frac{\boldsymbol{\eta} }{{2{\sigma ^2}}}, \ {\rm{and}} \ \ {p_0}\left( \boldsymbol{x} \right) \sim {\cal N}\left( {0,{\sigma ^2}} \right).
\end{equation}
The proof of Lemma 1 can be found in Appendix. And the conclusion tells us that one can achieve the denoised result by computing the posterior expectation, then associated with diffusion models, we have a classic result of the formula.

\textbf{Proposition 1}: \textit{Suppose $s_{\theta}\left(\boldsymbol{x_t}, \boldsymbol{t}\right)$ minimizes the score matching loss $\mathbb{E}_{t,\boldsymbol{x_0},\boldsymbol{x_t}|\boldsymbol{x_0}} \left[\lambda_t {\left\| {{s_\theta }\left( {\boldsymbol{x_t},t} \right) - {\nabla _{\boldsymbol{x_t}}}\log {p_t}\left( {\boldsymbol{x_t}} \right)} \right\|_2^2}\right]$. For the case of reverse diffusion sampling, $p\left(\boldsymbol{x_0}|\boldsymbol{x_t} \right)$ has the unique posterior mean}
\begin{equation}
    \boldsymbol{\hat x_0} = \frac{1}{{\sqrt {{{\bar \alpha }_t}} }}\left( {\boldsymbol{x_t} + \left( {1 - {{\bar \alpha }_t}} \right)s_{\theta}\left(\boldsymbol{x_t}, \boldsymbol{t}\right)} \right)
\end{equation}

% \textit{Proof}. 

The proof of Proposition 1 can be found in Appendix. We are now able to approximate $p\left( \boldsymbol{x_0}|\boldsymbol{x_t} \right)$ with the mapping function ${\cal D}_t: \boldsymbol{x_t} \to \boldsymbol{x_0} := \boldsymbol{\hat x_0}$. Hence, the likelihood function
$p\left( {\boldsymbol{y}|\boldsymbol{x_t}} \right)$ can be replaced with $p\left( {\boldsymbol{y}|\boldsymbol{{\hat x}_0}} \right)$. Formally, we have the following approximation
\begin{equation}
    {\nabla _{\boldsymbol{x_t}}}p\left( {\boldsymbol{y}|\boldsymbol{x_t}} \right) \simeq {\nabla _{\boldsymbol{x_t}}}p\left( {\boldsymbol{y}|{\boldsymbol{\hat x}_0}} \right), \ \rm{where} \ \boldsymbol{\hat x_0} := {\mathbb E}\left[\boldsymbol{x_0}|\boldsymbol{x_t}\right]
    \label{approximation}
\end{equation}

As this paper considers cases with Gaussian noise, the likelihood function $p\left( \boldsymbol{y}|\boldsymbol{x_0}\right)$ should satisfy ${\cal N}\left({\cal A}\left( \boldsymbol{x_0}\right), \sigma_z^2 \right)$. Then using Eqn.~\ref{approximation}, we get
\begin{equation}
    {\nabla _{\boldsymbol{x_t}}}\log p\left( {\boldsymbol{y}|\boldsymbol{x_t}} \right) \simeq  - \frac{1}{{{\sigma_z^2}}}{\nabla _{\boldsymbol{x_t}}}\left\| {\boldsymbol{y} - {\cal A}\left( {\boldsymbol{{\hat x}_0}} \right)} \right\|_2^2.
    \label{up_approximation}
\end{equation}
Now putting Eqn.~\ref{start} and Eqn.~\ref{up_approximation} together, the discrete reverse diffusion under the additional guidance can be represented by
\begin{equation}
   \begin{aligned}
       \boldsymbol{x'_{t - 1}} &= \frac{1}{{\sqrt {{\alpha _t}} }}\left( {\boldsymbol{x_t} - (1 - {\alpha _t}){s_\theta }\left( {\boldsymbol{x_t},t} \right)} \right) + {\sigma _t}z, \\
       \boldsymbol{x_{t - 1}} &= \boldsymbol{x'_{t - 1}} + {\rho_t}{\nabla _{\boldsymbol{x_t}}}\left\| {\boldsymbol{y} - {\cal A}\left( {\boldsymbol{{\hat x}_0}} \right)} \right\|_2^2
   \end{aligned}
\end{equation}
where $\rho_t \buildrel \Delta \over = \left(1 - {\alpha _t}\right)/ \left({\sqrt {{\alpha _t}} }\sigma_z^2\right)$ is set as the strength of the guidance.

\begin{figure}
    \setlength{\abovecaptionskip}{-0.2cm} 
    \centerline{\includegraphics[width=1\linewidth]{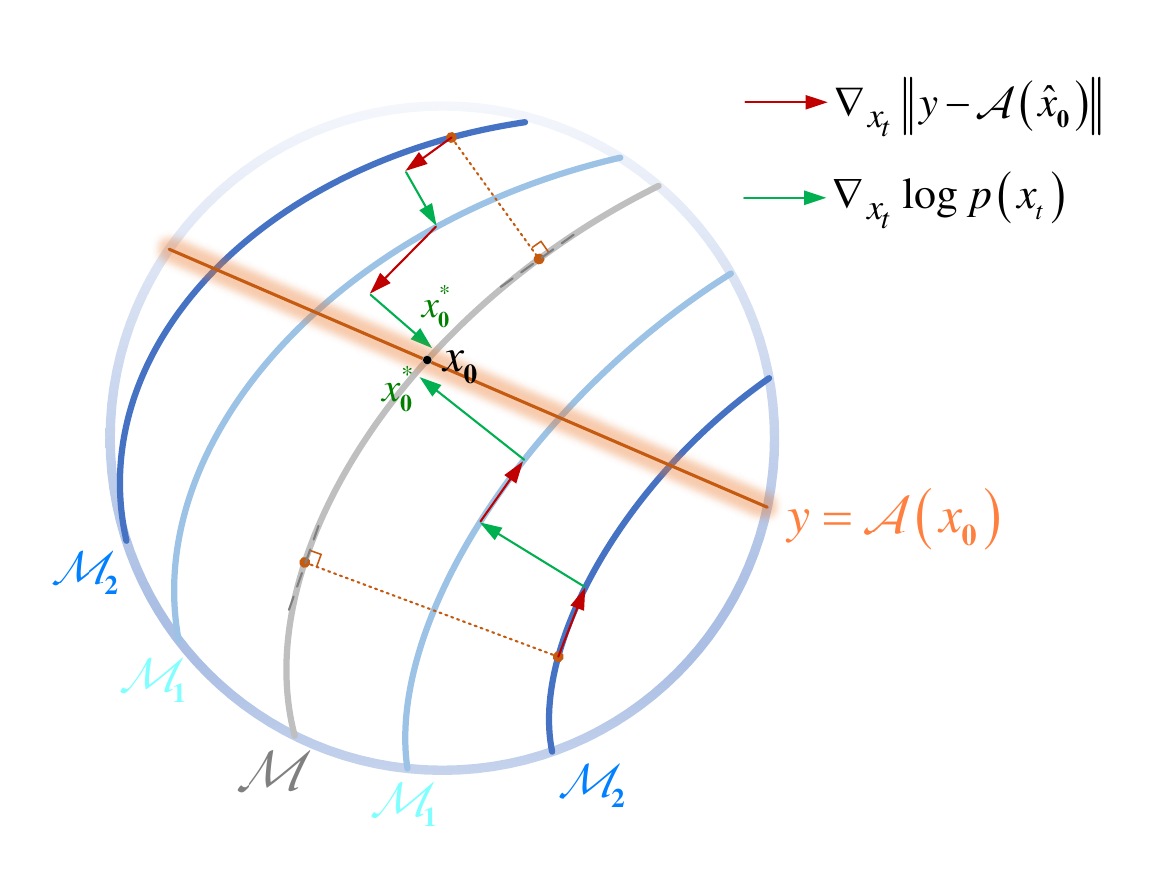}}
    \caption{\textbf{Guiding generation process toward target solutions.} Each curve represents a manifold ${\cal M}_i$ of (noisy) TM data. The proposed correction step (red arrow) alleviates reverse diffusion step (green arrow) leaving the solution space of inverse problems.}
    \label{Gradient_guided_geometry}
\end{figure}

\subsection{Theoretical Guarantee}

\textbf{Theorem 1}: \textit{The correction imposed by the gradient-based guidance at each step will not leave the data manifold ${\cal M} \subset \mathbb{R}^n$ which is the set of all traffic data points $\boldsymbol{x_0}$.}

% \textit{Proof.} 

The proof of Theorem 1 can be found in Appendix which is mainly concluded from~\cite{r31}. It indicates that reasonable estimations of all flows can be obtained by combining the score function and our gradient term, given the measurement model. 
% And 
% according to ${\rho _t}{\nabla _{\boldsymbol{x_t}}}\left\| {\boldsymbol{y} - {\cal A}\left( {\boldsymbol{{\hat x}_0}} \right)} \right\|_2^2 \in {T_s}\left( {{\cal M}, {\cal D}_t\left( \boldsymbol{x_t}\right)} \right)$, 
Concretely, the illustration of our sampling method is demonstrated in Fig.~\ref{illustration}, and we also present our scheme visually in Fig.~\ref{Gradient_guided_geometry}. The term can be considered as forcing the diffusion model to search for the optimal solution along the data manifold at every noisy state space.

\section{Diffusion-TM: Diffusion-based Approach for General Traffic Matrix Analysis}\label{s6}

% \subsection{Overview}

Based on the discussion in Section~\ref{s5}, we summarize the detailed sampling algorithm of our method called Diffusion-TM, and list {it} for traffic tomography and TM completion in Alg.~\hyperref[alg]{1} and Alg.~\hyperref[alg]{2}. Additionally, we further improve the algorithm by adding the expectation maximization iteration~\cite{r63} and additional replace-based guidance~\cite{r45} in Alg.~\hyperref[alg]{1} and Alg.~\hyperref[alg]{2} respectively, as there is still room for imposing some optimizations. The detailed descriptions will be shown later in this section.

Since these algorithms only require a pre-trained unconditional diffusion model, our proposed approach is a plug-and-play framework which does not depend on specific applications. This makes the model amenable to three different TM-related tasks at the same time: synthetic TM generation, TM recovery and tomography. Therefore, 
% we mainly present the detailed solutions of 
the choice of model architecture and training strategy for the key DM is an important issue.

First, note that there is no shortage of studies proving the network monitoring data usually have hidden spatio-temporal redundancies \cite{r35,r51} which propels us to designing models that take TM series as input for learning. And we choose a Transformer \cite{r50} that enhances the models’ ability to capture global correlation and patterns of TM sequences. 
% This way, information of the entire noisy sequence is encoded before decoding. 
As aforementioned, obtaining the complete training set of the OD traffic is usually difficult or even impossible, let alone building a set of continental sequences. To alleviate the effect of missing values in the training set, we designed a pre-processing workflow based on an autoencoder network. The module provides Diffusion-TM a coarse-grained estimation of missing OD flows to make the underlying network as insensitive to these missing values as possible. And for both two networks, only the estimation loss on observed TM is used in back propagation to update their weights.

\subsection{Expectation Maximization (EM) Algorithm for Tomography}
The Expectation Maximization (EM) algorithm can be regarded as a solution to an optimization problem with latent variables. It adopts an iterative procedure to calculate the Maximum Likelihood (ML) estimation \cite{r63,r64,r65}. Specifically, we follow the canonical form of the EM iteration for solving the NT problem proposed by \citep{r63} as follows:
\begin{equation}
    {x_j} \leftarrow \frac{{{x_j}}}{{\sum\nolimits_{i = 1}^M {{a_{ij}}} }}\sum\nolimits_{i = 1}^M {\frac{{{a_{ij}} * {y_i}}}{{\sum\nolimits_{k = 1}^N {{a_{ik}} * {x_k}} }}}
\end{equation}
where $a_{i,j}$ represents the value located in the $i$-th row and $j$-th column of the routing matrix. And $x_j$ is the $j$-th element of OD flows $X$, $y_i$ is the $i$-th element of link loads $Y$. The EM algorithm can approximate the solution of Eqn.~\ref{NT} as the iteration running on~\cite{r58}. Thus as shown in Alg.~\hyperref[alg]{1} line 9, we select the iterative procedure to further optimize the estimation generated by our diffusion model.

\subsection{Additional Replace-based Guidance for Completion}
From the above part, the EM algorithm is designed to help constrain the possible values of estimation through tomography equations. But \textit{how can additional optimization be applied to completion tasks?}

We first define the known and unknown OD-pairs of $\boldsymbol{x_t}$ as $\Omega\left(\boldsymbol{x_t}\right)$ and $\bar{\Omega}\left(\boldsymbol{x_t}\right)$ respectively. For the traffic matrix completion task, our goal is to sample from $p\left(\bar{\Omega}\left(\boldsymbol{x_0}\right)|\Omega\left(\boldsymbol{x_0}\right) = \boldsymbol{y}\right)$. A notable property of such task is that we can run the reverse process only to known dimensions since the element-wise forward noise is applied to the dimensions independently. Now back to Eqn,~\ref{Conditional_sde}, we again focus on the likelihood ${p_t}\left( \boldsymbol{x_t}|\boldsymbol{y} \right)$ which is then equal to ${p_t}\left( {\bar \Omega \left( {\boldsymbol{x_t}} \right)|\Omega \left( {\boldsymbol{x_t}} \right) = \boldsymbol{y}} \right)$. Formally, we have
\begin{equation}
    \begin{aligned}
        &{p_t}\left( {\bar \Omega \left( {\boldsymbol{x_t}} \right)|\Omega \left( {\boldsymbol{x_0}} \right) = \boldsymbol{y}} \right) :=  {p_t}\left( {\bar \Omega \left( {\boldsymbol{x_t}} \right)|Y} \right) \\
        &= \int {{p_t}\left( {\bar \Omega \left( {\boldsymbol{x_t}} \right)|\Omega \left( {\boldsymbol{x_t}} \right),Y} \right){p_t}\left( {\Omega \left( {\boldsymbol{x_t}} \right)|Y} \right)d} \Omega \left( {\boldsymbol{x_t}} \right) \\
        &= {\mathbb{E}_{\Omega \left( {\boldsymbol{x_t}} \right)|Y}}\left[ {{p_t}\left( {\bar \Omega \left( {\boldsymbol{x_t}} \right)|\Omega \left( {\boldsymbol{x_t}} \right),Y} \right)} \right] \\
        &= {\mathbb{E}_{\Omega \left( {\boldsymbol{x_t}} \right)|Y}}\left[ {{p_t}\left( {\bar \Omega \left( {\boldsymbol{x_t}} \right)|\Omega \left( {\boldsymbol{x_t}} \right)} \right)} \right] \\
        &\approx {p_t}\left( {\bar \Omega \left( {\boldsymbol{x_t}} \right)|\hat \Omega \left( {\boldsymbol{x_t}} \right)} \right) = {p_t} \left(\left[\bar \Omega \left( {\boldsymbol{x_t}} \right);\Omega \left( {\boldsymbol{x_t}} \right)\right]\right),
    \end{aligned}
\end{equation}
where $\hat \Omega \left( {\boldsymbol{x_t}} \right)$ denotes samples from ${p_t}\left( \Omega \left( {\boldsymbol{x_t}} \right)|\Omega \left( \boldsymbol{x_0} \right) = \boldsymbol{y} \right) $, and $\left[\bar \Omega \left( {\boldsymbol{x_t}} \right);\Omega \left( {\boldsymbol{x_t}} \right)\right]$ represents the concatenation of two sets of dimensions. 

That means there is space for conducting additional constraints while still leveraging the unconditional score function, and \cite{r45} proposed this general method for imputation from the jointly trained diffusion model.

To flesh this out, the samples for $\Omega \left( {\boldsymbol{x_t}} \right)$ are replaced by exact samples from the forward process $q(\Omega \left( {\boldsymbol{x_t}} \right)| \Omega \left( {\boldsymbol{x_0}} \right))$ in Eqn.~\ref{forward}, at each iteration, while the sampling procedure for updating $\bar \Omega \left( {\boldsymbol{x_t}} \right)$ is still sampling from $p_\theta(\bar \Omega \left( {\boldsymbol{x_t}} \right)| \bar \Omega \left( {\boldsymbol{x_{t+1}}} \right))$. The samples $ \Omega \left( {\boldsymbol{x_t}} \right)$ then have the correct marginal distribution, and $\bar \Omega \left( {\boldsymbol{x_t}} \right)$ will conform with $\Omega \left( {\boldsymbol{x_t}} \right)$ through the denoising process. Using this strategy, we can generate an intact sample that follows the correct conditional distribution in addition to the correct marginal. We refer to the approach as the replacement method for extra guidance during the reverse process, and run it at the end of each sampling step in Alg.~\hyperref[alg]{2}.

\subsection{Transformer-based Underlying Network}

We use a Transformer with Sigmoid Non-linear processing the final output to estimate $\boldsymbol{\hat x_0}\left(\boldsymbol{x_t},t,\theta\right)$. As shown in Fig.~\ref{Model} Model 2, the underlying network is divided into two modules, i.e., a transformer encoder and a transformer decoder. Both the encoder and decoder network consist of multiple transformer blocks. Each transformer block in encoder (decoder) contains a full attention (a full attention, a cross attention to combine encoding information) and a feed forward layer. For the diffusion embedding, we follow previous works \cite{r14} with transformer sinusoidal positional embedding to encode the diffusion step. After getting the diffusion step embedding, we sum them up and add them to each block. Specifically, the diffusion step $t$ is injected into the network using the Adaptive Layer Normalization operator, which can be written as $a_t {\rm Laynorm}(w)+b_t$ where $w$ is the intermediate activations, $a_t$ and $b_t$ are obtained from a linear projection of the diffusion embedding. 

% \subsection{Pre-processing Module}

\subsection{Training Procedure under Traffic-deficient Setting}
\begin{figure}
    \setlength{\abovecaptionskip}{-0.1cm} 
    \centerline{\includegraphics[width=1\linewidth]{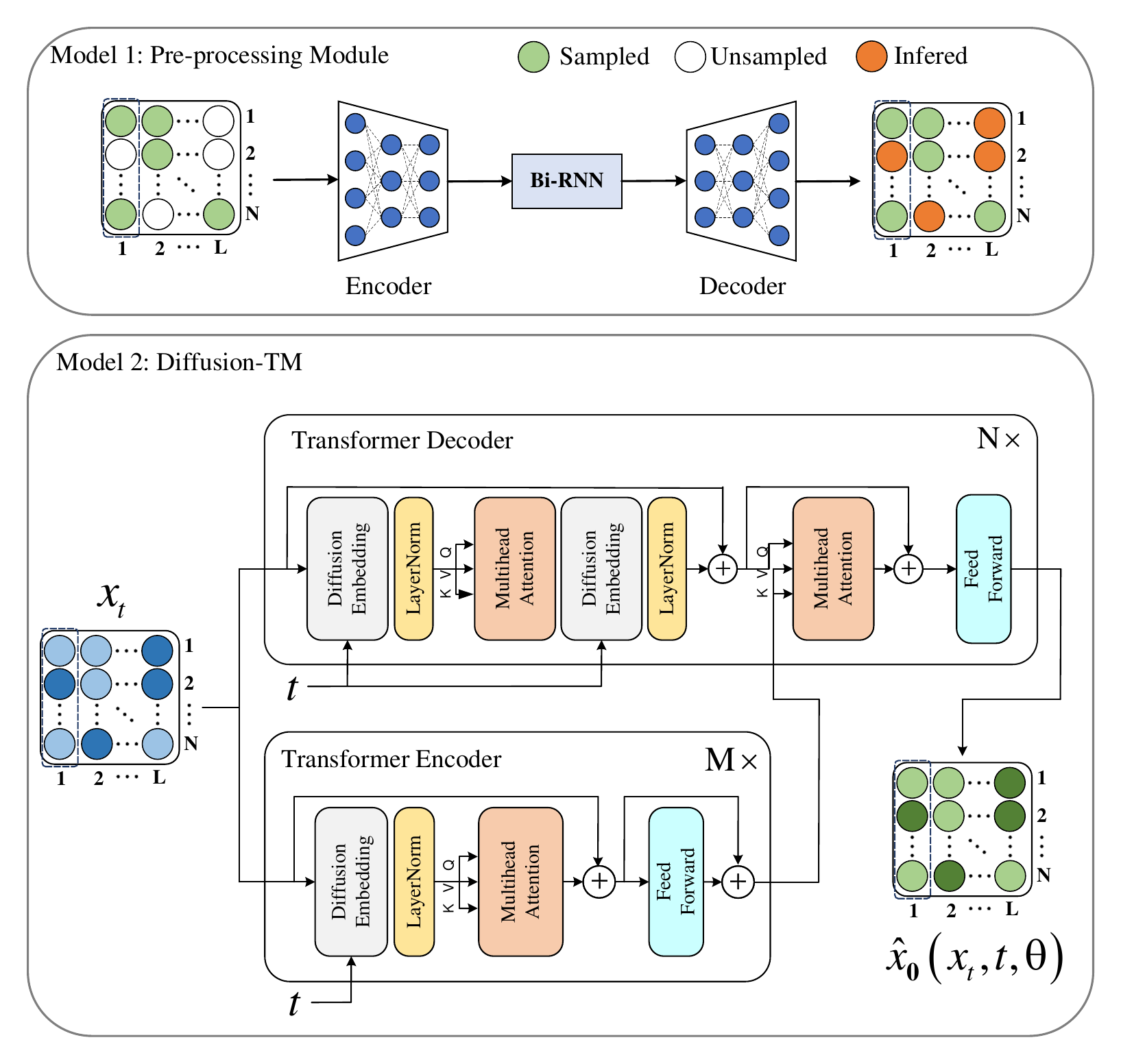}}
    \caption{\textbf{(Top) Architecture of Pre-processing Module.} The training of diffusion models starts with a AutoEncoder-based pre-processing module, which generates coarse-grained estimations of missing values in training set. \textbf{(Bottom) Overall model structure of Diffusion-TM.} The underlying Transformer is fed a TM sequence $x_t$ in a diffusion step $t$, as well as $t$ itself, then the diffusion model predicts the clean sample ${\hat x}_0$.}
    \label{Model}
\end{figure}

Our training goal is to obtain a converging diffusion model for stably generating high-quality TM samples, which in this paper is tantamount to easing the effect of missing values in the measured data. 
% We first make a rough prediction of the missing values in the training set through an autoencoder equipped with the Bi-RNN,
According to our solution overview above, the distribution of traffic data can be hard to fit for deep generative models if only very few of the OD flows were observed, while the TM is generally incomplete with lots of entries in the matrix unobserved under the measurements. Thus to minimize the influence, a pre-processing module for missing data in the training set is needed. Since traffic matrices with unobserved volumes are often close to complete data after dimensionality reduction \cite{r42}, we choose an autoencoder (AE) to obtain coarse-grained estimations of unknown flows before formal training of DMs. As illustrated in Fig.~\ref{Model} Model 1, the module contains one encoder which is composed of two fully connected layers using ReLU as Non-linear function, and one decoder which is responsible for mapping the outputs of representation space onto the original space through Sigmoid regularization. Between the two of them, a bi-directional Recurrent Neural Network (Bi-RNN), which can be implemented with either LSTM or GRU units, is established to extract per-point features. The input TM sequences after downscaling are fed to the Bi-RNN, and then their temporal information about the observed time points will be stored and passed through our decoder to produce the first-stage output.

Specifically, given a TM series represented by a sequence of points $\boldsymbol{s} = \left\{ {s_1}, {s_2}, \dots, {s_K}\right\}$, the output of the AE can be computed by
\begin{equation}
    \begin{aligned}
       {h^1_{i}} &= {\rm{ReLU}}\left(W_1 {s_i} + B_1\right),\\
       {h^2_{i}} &= {\rm{ReLU}}\left(W_2 {h^1_{i}} + B_2\right),\\
       \left[{h^3_{i}},{c_{i}}\right] &= \textbf{Bi-RNN}\left(h^2_{i-1}, \left[{h^3_{i-1}},{c_{i-1}}\right]\right),\\
       f_i &= {\rm{Sigmoid}}\left(W_3 {h^3_{i}} + B_3\right)
    \end{aligned}
\end{equation}
where $h$ and $c$ are the hidden states and the optional cell states, $W$ and $B$ are the weights and biases
of a fully connected layer. Respectively,  ${\rm{ReLU}}\left(\cdot\right)$ is the activation function of ReLU, and ${\rm{Sigmoid}}\left(\cdot\right)$ is the activation function of Sigmoid. The loss function in the pre-processing work can be written as follows:
\begin{equation}
    {\cal L}_p = \left\| {\boldsymbol{X}^o \odot \boldsymbol{M} - D\left( {\boldsymbol{X}^o} \right) \odot \boldsymbol{M}} \right\|_2^2
    \label{AE_objective}
\end{equation}
where $\boldsymbol{X}^o$ and $\boldsymbol{M}$ is a partially observed traffic data and sampling indication matrix, respectively. During the pre-training process, the loss is back propagated to deep network $D$ to update its parameters. The empirical loss guarantees that only the estimation loss on observed samples is used in back propagation, while the training errors for missing data are discarded. After pre-processing module training, we leveraged it to update the measured traffic set by replacing all the missing OD pairs with values reconstructed through our autoencoder. Then, the "complete" dataset would be used to boost the distribution learning of Diffusion-TM.

% For the training of Diffusion-TM, we did not make many changes to the original paper \cite{r14}.
In the diffusion probabilistic model training procedure, we first sample the diffusion step $t$ from a uniform distribution, and then compute its corresponding $\boldsymbol{x_t}$ through Eqn.~\ref{forward} with a random Gaussian noise $\epsilon$. But it should be noted that a direct reconstruction of $\boldsymbol{x_0}$ also takes missing values into account although they have been recovered by the pre-processing work to some extent. Thus again, we further adjust the training objective like Eqn.~\ref{AE_objective} as
\begin{equation}
    {\cal L}_{VLB} = \left\| {\boldsymbol{x_0} \odot \boldsymbol{M} - \boldsymbol{\hat x_0}\left( \boldsymbol{x_t},t,\theta \right) \odot \boldsymbol{M}} \right\|_2^2.
\end{equation}
The detailed training and sampling algorithm is shown in Alg.~\hyperref[alg_train]{3} and Alg.~\hyperref[alg_new]{4}, respectively. Through applying DDIM algorithm to accelerate the sampling, we can use a different number of iterations $S \le T$ during inference, making it possible to explicitly trade off between inference computation and output quality.

\begin{table}[htb]
\normalsize
\begin{tabular}{l}
\toprule
\textbf{Algorithm 3} Training Algorithm of Diffusion-TM\\
\midrule
\begin{minipage}{1.\linewidth}
    \label{alg_train}
    \centering
    \begin{algorithmic}[1]
    \REQUIRE training epoch of pre-processing model ($D$) $\boldsymbol{I}'_{max}$,\\
    ~~~~~~~training epoch of Diffusion-TM $\boldsymbol{I}''_{max}$,\\
    ~~~~~~~training set with missing values $\left\{\boldsymbol{X}^o_k\right\}_{k=1}^K$, and\\
    ~~~~~~~the corresponding observation matrices $\left\{\boldsymbol{M}_k\right\}_{k=1}^K$
    \ENSURE trained denoising network $\theta$
    \FOR{all $i$ from 1 to $\boldsymbol{I}'_{max}$}
        \STATE $\boldsymbol{X}^o,\boldsymbol{M} \leftarrow \textbf{Get\_Batch}\left( \left\{\boldsymbol{X}^o_k\right\}_{k=1}^K, \left\{\boldsymbol{M}_k\right\}_{k=1}^K \right)$;
        \STATE Take gradient descent step on \\
               ~~~~~${\nabla _D}\left\| {\boldsymbol{X}^o \odot \boldsymbol{M} - D\left( {\boldsymbol{X}^o} \right) \odot \boldsymbol{M}} \right\|_2^2$;
    \ENDFOR
    \STATE $\left\{\boldsymbol{X}'_k\right\}_{k=1}^K \leftarrow D\left( \left\{\boldsymbol{X}^o_k\right\}_{k=1}^K \right)$;
    \STATE $\left\{\boldsymbol{X}_k\right\}_{k=1}^K \leftarrow \textbf{Update}\left( \left\{\boldsymbol{X}'_k\right\}_{k=1}^K, \left\{\boldsymbol{X}^o_k\right\}_{k=1}^K, \left\{\boldsymbol{M}_k\right\}_{k=1}^K \right)$;
    \FOR{all $i$ from 1 to $\boldsymbol{I}''_{max}$}
        \STATE $\boldsymbol{X},\boldsymbol{M} \leftarrow \textbf{Get\_Batch}\left( \left\{\boldsymbol{X}_k\right\}_{k=1}^K, \left\{\boldsymbol{M}_k\right\}_{k=1}^K \right)$;
        \STATE $t \leftarrow {\rm{Uniform}}\left(\left\{1,\dots,T \right\}\right)$;
        \STATE $\epsilon \leftarrow {\cal K}\left(0, {\rm{I}}\right)$;
        \STATE $\boldsymbol{x_t} \leftarrow \sqrt{{{\bar\alpha}_t}}\boldsymbol{X}+\sqrt{1-{{\bar\alpha}_t}}\epsilon$
        \STATE Take gradient descent step on \\
               ~~~~~${\nabla _\theta}\left\| {\boldsymbol{X} \odot \boldsymbol{M} - \boldsymbol{\hat x_0}\left( \boldsymbol{x_t},t,\theta \right) \odot \boldsymbol{M}} \right\|_2^2$;
    \ENDFOR
    \RETURN $\theta$;
    \end{algorithmic}
\end{minipage} \\
\bottomrule
\end{tabular}
\begin{tabular}{l}
\\
\toprule
\textbf{Algorithm 4} Fast Sampling Algorithm of Diffusion-TM\\
\midrule
\begin{minipage}{1.\linewidth}
    \label{alg_new}
    \centering
    \begin{algorithmic}[1]
    \REQUIRE trained denoising network $\theta$, fast inference time \\ ~~~~~~~stride $\bigtriangleup_t$
    \ENSURE synthetic traffic matrix $\boldsymbol{x_0}$
    \STATE $\boldsymbol{x_T} \sim {\cal N}\left( 0, \rm{I}\right)$;
    \WHILE{$t > 0$}
        \STATE $\boldsymbol{z} \sim {\cal N}\left( 0, \rm{I}\right)$ if $t > \bigtriangleup_t$, else $\boldsymbol{z}=0$ and $\bigtriangleup_t = t$;
        \STATE $\boldsymbol{x_{t - \bigtriangleup_t}} = \frac{{\sqrt {{\alpha _t}} (1 - {{\bar \alpha }_{t - \bigtriangleup_t}})}}{{1 - {{\bar \alpha }_t}}}\boldsymbol{x_t} + \frac{{\sqrt {{{\bar \alpha }_{t - \bigtriangleup_t}}} {\beta _t}}}{{1 - {{\bar \alpha }_t}}}\boldsymbol{\hat x_0}(\boldsymbol{x_t},t,\theta )$  \\
        ~~~~~~~~~~~~$+ \frac{{1 - {{\bar \alpha }_{t - \bigtriangleup_t}}}}{{1 - {{\bar \alpha }_t}}}{\beta _t}\boldsymbol{z}$;
        \STATE $t \leftarrow t - \bigtriangleup_t$;
    \ENDWHILE
    \RETURN $\boldsymbol{x_0}$;
    \end{algorithmic}
\end{minipage} \\
\bottomrule
\end{tabular}
\end{table}

\vspace{-3mm}

\section{Experiments}\label{s7}
In this section, we conduct experiments to evaluate the performances of the proposed Diffusion-TM. Three tasks are considered in the experiments: synthetic traffic data generation, network tomography, and traffic matrix completion. Our source code is available at \href{https://github.com/Y-debug-sys/DTM}{\textcolor{brown}{https://github.com/Y-debug-sys/DTM}}.

\vspace{-3mm}

\subsection{Experimental Setup}

\subsubsection{Datasets} To evaluate the performance of our proposed method, we use two real-world traffic datasets.
\begin{itemize}
    \item \textbf{Abilene} \cite{r48}. Abilene is derived from the U. S. Internet2 Network. The Abilene dataset contains 12 routers, 30 directed inner links, and 24 outside links. The dataset collected the volumes of all OD flows in the network every 5 minutes from March to September 2004. We use the first 3000 samples as training data and use 672 samples in the next week for testing.
    \item \textbf{G\'{E}ANT} \cite{r49}. GÉANT is derived from the pan-European research backbone network. The GÉANT network contains 23 routers and 120 directed links. All flows in this dataset were collected in 15-minute intervals from January to April 2003. We also train models with the first 3000 time slots, and then 672 samples are used to report the results of inference.
\end{itemize}

\subsubsection{Baselines} We compare Diffusion-TM with two types of methods including 4 network tomography algorithms and 5 traffic matrix completion algorithms.
\begin{itemize}
    \item \textbf{Network Tomography algorithms}. We implement 4 algorithms with deep learning technology, among which, two methods VAE-TME~\cite{r22} and WGAN-TME~\cite{r23} are based on generative models, the other two algorithms BPTME~\cite{r59}, MNETME~\cite{r58} used a neural network to learn the inverse mapping directly. 
    % We also compare our proposed method with Tomo-gravity~\cite{r60} which is a machine learning based TME framework. 
    \item \textbf{TM completion algorithms}. The first four algorithms (NTC~\cite{r17}, NTM~\cite{r55}, NTF~\cite{r56}, CoSTCo~\cite{r57}) are based on neural tensor factorization. NTF and CoSTCo adopt multi-layered perceptron and convolution neural networks as interaction functions, respectively. NTC and NTM design interaction functions based on outer-product. The last one is DATC~\cite{r42}, a recent work that exploited autoencoder and GANs to complete the traffic data.
\end{itemize}

Almost all learning-based NT algorithms assume that the training data has zero missing data. Thus, we infill these missing values before running the baseline. Following the work of~\cite{r35}, we construct the interpolation matrix $X_{base}$ by computing row and column means of the observed traffic samples. Let $X\left(i,j\right)$ denote the value of the $i$-th OD flow pair at the $j$-th time point. Then formally, the pre-processing result is given by
\begin{equation}
    X_{base}\left(i, j\right) = {\bar X} + X_{flow}\left(i\right) + X_{time}\left(j\right),
    \label{average_pre_processing}
\end{equation}
where $\bar X$ is the mean value of traces $X$ over all observed elements, and
\begin{equation}
    \begin{aligned}
        X_{flow}\left(i\right) &= \frac{1}{m}{\sum\limits_{j = 1}^m \left({X\left( {i,j} \right)} - {\bar X} \right)}, \\
        X_{time}\left(j\right) &= \frac{1}{n}{\sum\limits_{i = 1}^n \left({X\left( {i,j} \right)} - {\bar X} \right)}
    \end{aligned}
\end{equation}

For these baselines, we reuse their released source codes in their official repositories\footnote{The codes of experiments with the tensor factorization methods are available at \href{https://github.com/MerrillLi/LightNestle}{\textcolor{teal}{https://github.com/MerrillLi/LightNestle}}. And VAE-TME is available at \href{https://github.com/MikeKalnt/VAE-TME}{\textcolor{teal}{https://github.com/MikeKalnt/VAE-TME}}.} and rely on their designed training and model selection procedures in the original paper. Regarding algorithms that do not provide any code, we follow their supplementary description of the implementation to the best of our ability.

\subsubsection{Implementation Details} We apply a grid search to find default hyper-parameters in the underlying transformer that perform well across datasets. The range considered for each hyper-parameter is the batch size tuned in [32, 64, 128], the number of attention heads in [4, 8, 16], the number of basic dimension searched in [64, 96, 128], the diffusion steps in [50, 100, 300, 500, 1000] and the guidance strength in [1e-0, 1e-1, 5e-2, 1e-2, 1e-3]. The activation function of the output layer we consider is \textit{Sigmoid}. A single Nvidia 3090 GPU is used for model training. In all of our experiments, we use $cosine$ noise scheduling and optimize our network using Adam with $(\beta_1, \beta_2) = (0.9, 0.96)$. And a linearly decay learning rate starts at 0.0008 after 500 iterations of warmup. For a fair comparison, all the deep learning based algorithms use ${\rm L1Norm}$ loss function. 
We set the sliding window size $w = 12$ for Abilene and G\'{E}ANT in all experiments. 

Finally, we replicate the experimental setup in~\cite{r70}, in which they clipped outliers larger than the $99\%$ percentile to the $99\%$ percentile, then normalize all training samples were normalized by dividing the maximum values.

\vspace{-3mm}

\subsection{Evaluation Metrics}
We implement two methods to evaluate the learnt \textbf{distribution}: 
2-dimensional visualization via t-SNE analysis and Maximum Mean Discrepancy (MMD), a kernel-based method for computing the difference of the statistics of the two sets of samples. Given samples $X:=\left\{x_1,\dots,x_n\right\}$ and $Y:=\left\{y_1,\dots,y_m\right\}$ drawn independently and identically distributed (i.i.d.) from two distributions
\begin{equation}
    \begin{aligned}
       &{\rm{MMD}}_k^2 \left( {X,Y} \right) \\
        &:= \frac{1}{{n\left( {n - 1} \right)}}\sum\limits_{i,j = 1}^n {k\left( {{x_i},{x_j}} \right)} - \frac{2}{{mn}} \sum\limits_{i = 1}^n {\sum\limits_{j = 1}^m {k\left( {{x_i},{y_j}} \right)} }\\
        &\ \ \ + \frac{1}{{m\left( {m - 1} \right)}}\sum\limits_{i,j = 1}^m k\left( {{y_i},{y_j}} \right)
    \end{aligned}
\end{equation}
Our experiments use the universal Gaussian kernel, defined as $k\left(x_i, x_j\right)={\rm exp}\left(\frac{1}{2\sigma^2}\left| {x_i-x_j} \right|^2\right)$, where $\sigma$ is the bandwidth parameter. For characteristic kernel functions, it can be proven that ${\rm{MMD}}_k \left( {p,q} \right)$ if and only if $p = q$, leading to consistent results. 

% In addition, we qualify the estimation \textbf{accuracy} using four mainstreaming metrics: Normalized Mean Absolute Error (NMAE), Normalized Root Mean Square Error (NRMSE), Temporal Related Mean Absolute Error (TRE), and Spatial Related Mean Absolute Error (SRE). Specifically, we calculate
In addition, we qualify the estimation \textbf{accuracy} using three mainstreaming metrics: Normalized Mean Absolute Error (NMAE), Normalized Root Mean Square Error (NRMSE), and Temporal Related Mean Absolute Error (TRE). Specifically, we calculate
\begin{equation}
    \begin{aligned}
       & {\rm NMAE} = \frac{{\sum\nolimits_{i,j:M\left( {i,j} \right) = 0} {\left| {X\left( {i,j} \right) - \hat X\left( {i,j} \right)} \right|} }}{{\sum\nolimits_{i,j:M\left( {i,j} \right) = 0} {\left| {X\left( {i,j} \right)} \right|} }},\\
       & {\rm NRMSE} = \frac{{\sqrt {\sum\nolimits_{i,j:M\left( {i,j} \right) = 0} {{{\left( {X\left( {i,j} \right) - \hat X\left( {i,j} \right)} \right)}^2}} } }}{{\sqrt {\sum\nolimits_{i,j:M\left( {i,j} \right) = 0} {{{\left( {X\left( {i,j} \right)} \right)}^2}} } }},\\
       & {\rm TRE}\left(j\right) = \frac{{\sum\nolimits_i {\left| {X\left( {i,j} \right) - \hat X\left( {i,j} \right)} \right|} }}{{\sum\nolimits_i {X\left( {i,j} \right)} }}\\
       % & {\rm SRE}\left(i\right) = \frac{{\sum\nolimits_j {\left| {X\left( {i,j} \right) - \hat X\left( {i,j} \right)} \right|} }}{{\sum\nolimits_j {X\left( {i,j} \right)} }}\\
    \end{aligned}
\end{equation}
where $\hat X$ is the estimated traffic matrix. For traffic matrix completion, we first drop some data from existing measurements and then only measure errors on the pseudo-missing values. 

\subsection{Learnt Distribution Visualization}
\begin{figure}[htbp]
	\centering
	\subfigure{
        \rotatebox{90}{\scriptsize{~~~Diffusion-TM}}
		\begin{minipage}[t]{0.27\linewidth}
			\centering
			\includegraphics[width=1.12\linewidth]{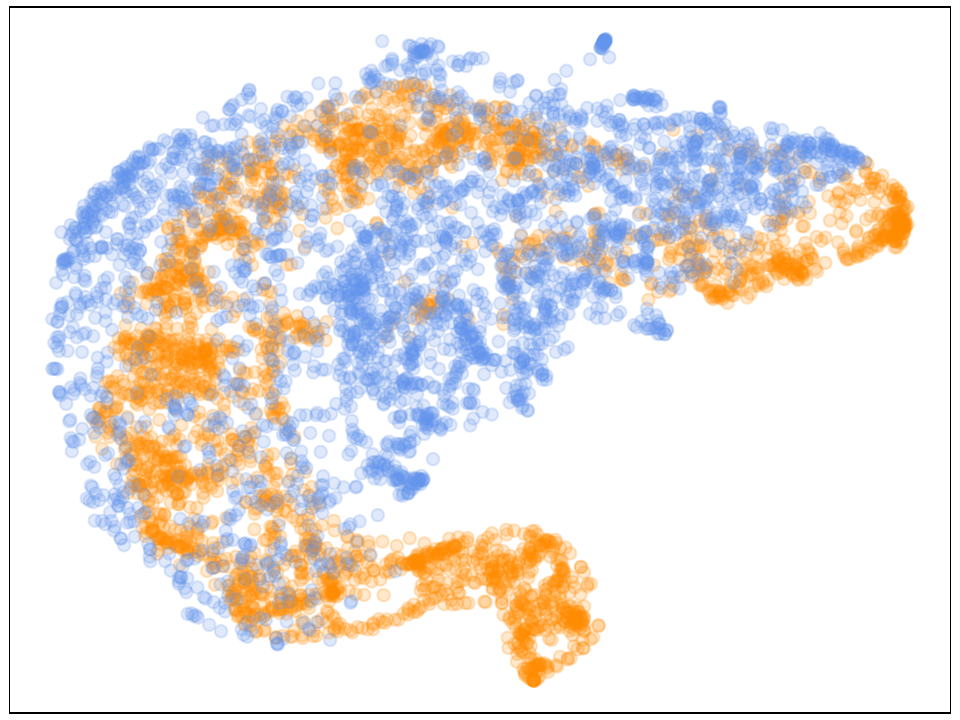}
		\end{minipage}
	}
	\subfigure{
		\begin{minipage}[t]{0.27\linewidth}
			\centering
			\includegraphics[width=1.12\linewidth]{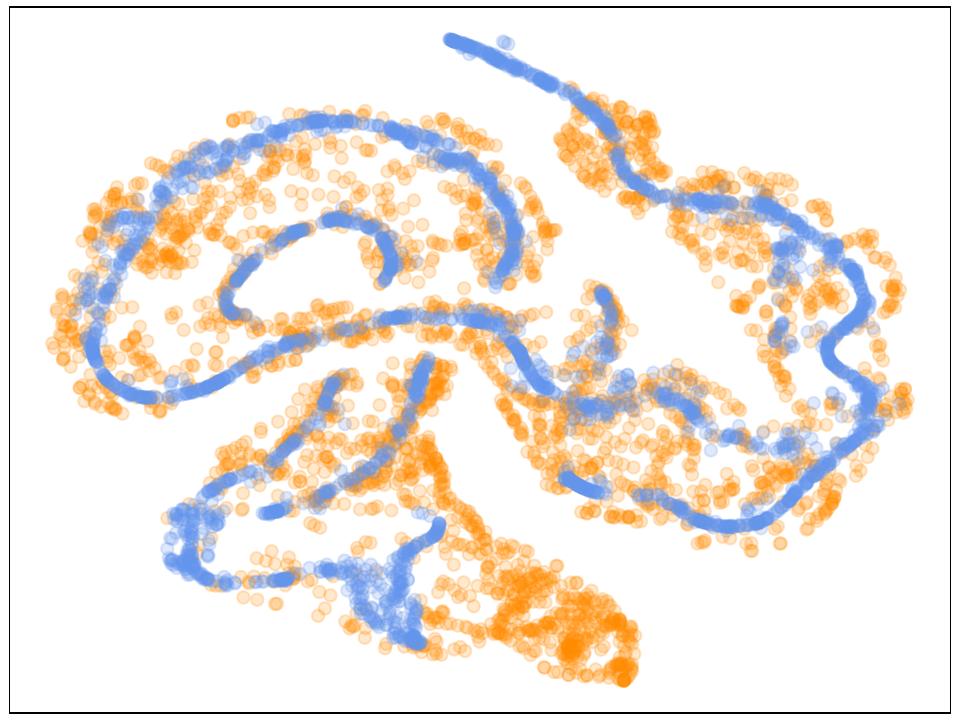}
		\end{minipage}
	}
	\subfigure{
		\begin{minipage}[t]{0.27\linewidth}
			\centering
			\includegraphics[width=1.12\linewidth]{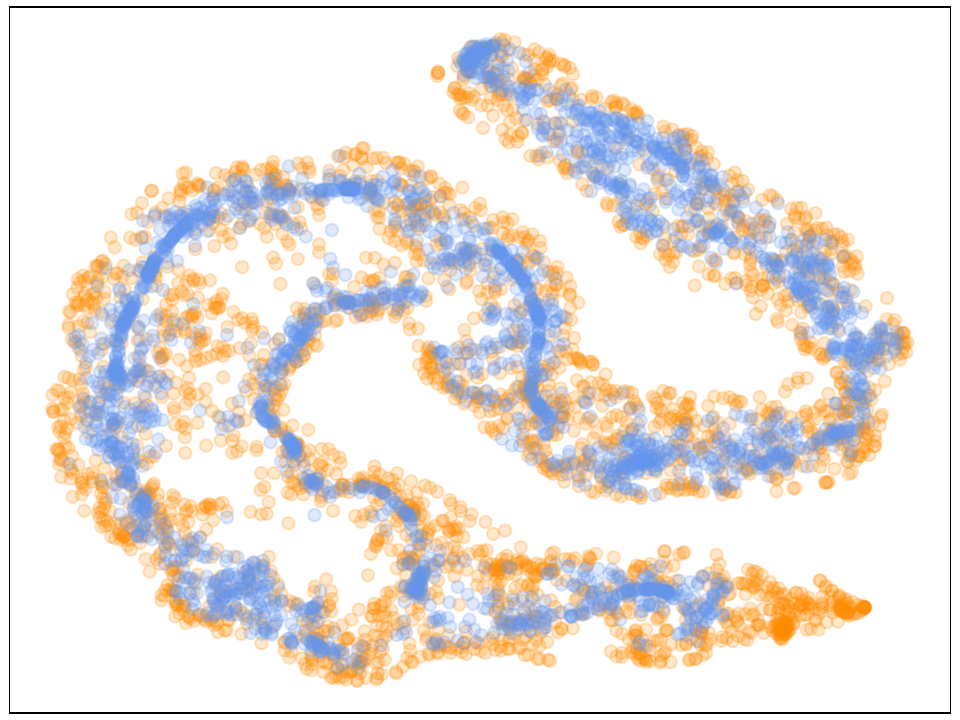}
		\end{minipage}
	}

	\vspace{-3mm}
	\setcounter{subfigure}{0}

    \subfigure{
        \rotatebox{90}{\scriptsize{~~~~~~~WGAN}}
		\begin{minipage}[t]{0.27\linewidth}
			\centering
			\includegraphics[width=1.12\linewidth]{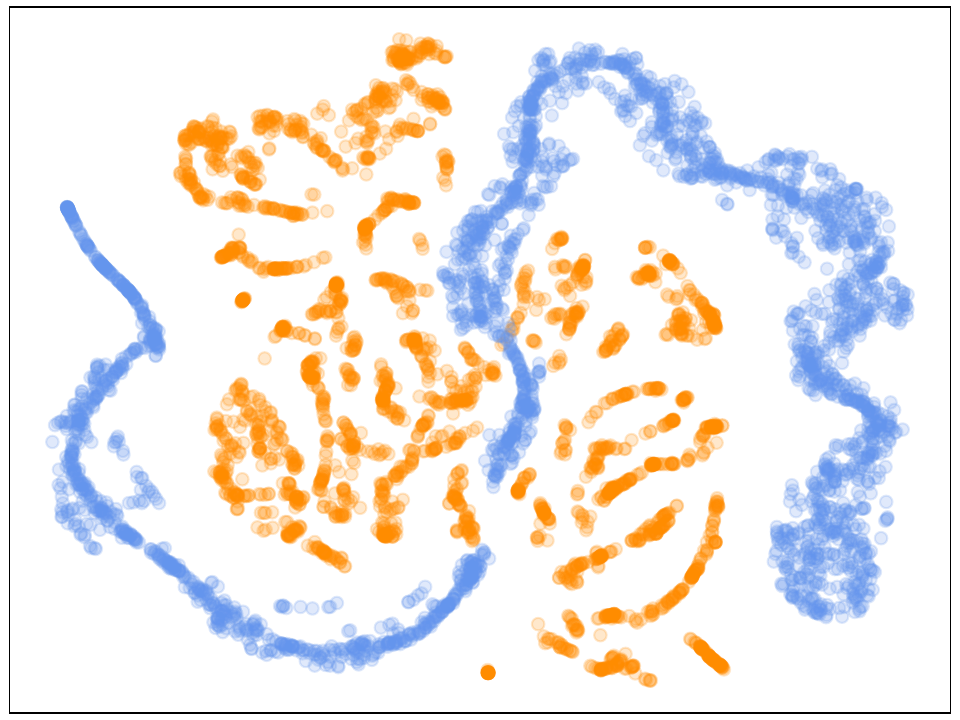}
		\end{minipage}
	}
	\subfigure{
		\begin{minipage}[t]{0.27\linewidth}
			\centering
			\includegraphics[width=1.12\linewidth]{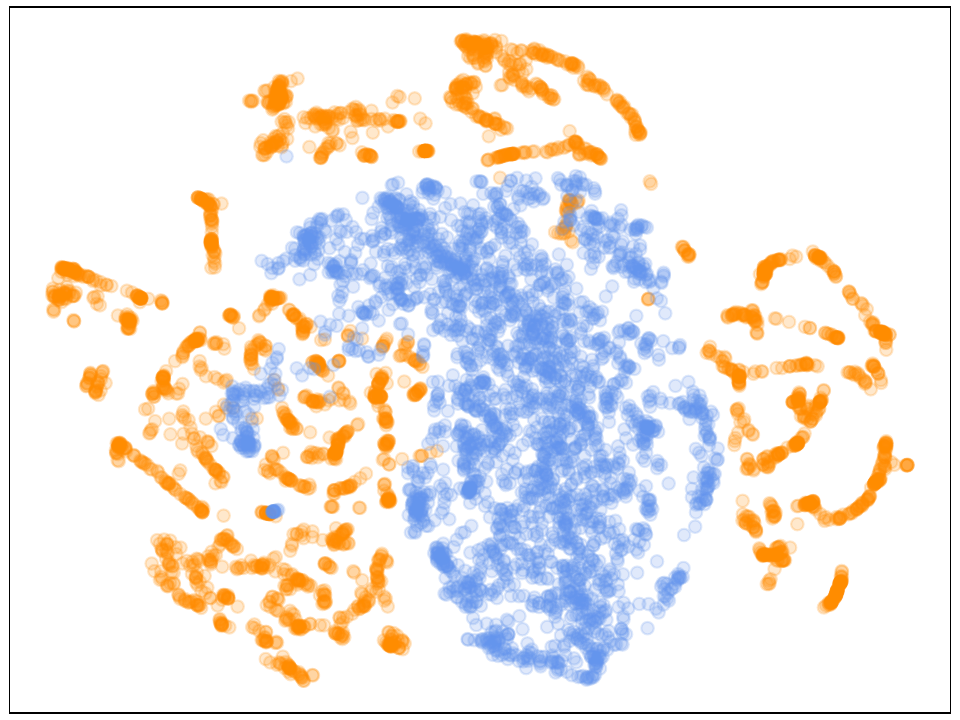}
		\end{minipage}
	}
	\subfigure{
		\begin{minipage}[t]{0.27\linewidth}
			\centering
			\includegraphics[width=1.12\linewidth]{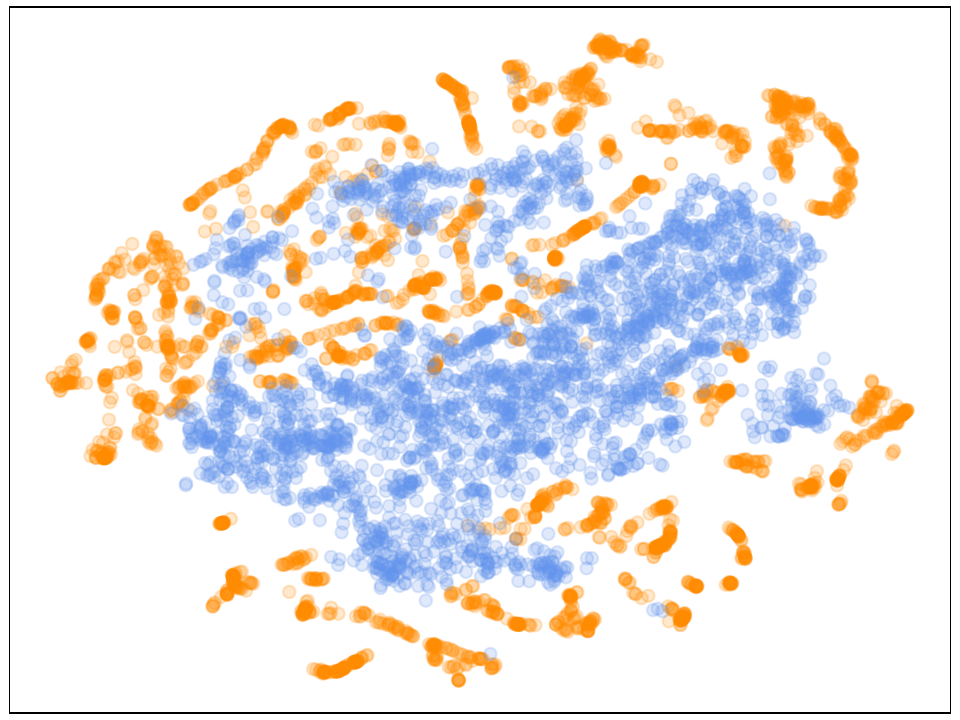}
		\end{minipage}
	}
 
        \vspace{-3mm}
	\setcounter{subfigure}{0}

 \subfigure[$2\%$]{
        \rotatebox{90}{\scriptsize{~~~~~~~~~VAE}}
		\begin{minipage}[t]{0.27\linewidth}
			\centering
			\includegraphics[width=1.12\linewidth]{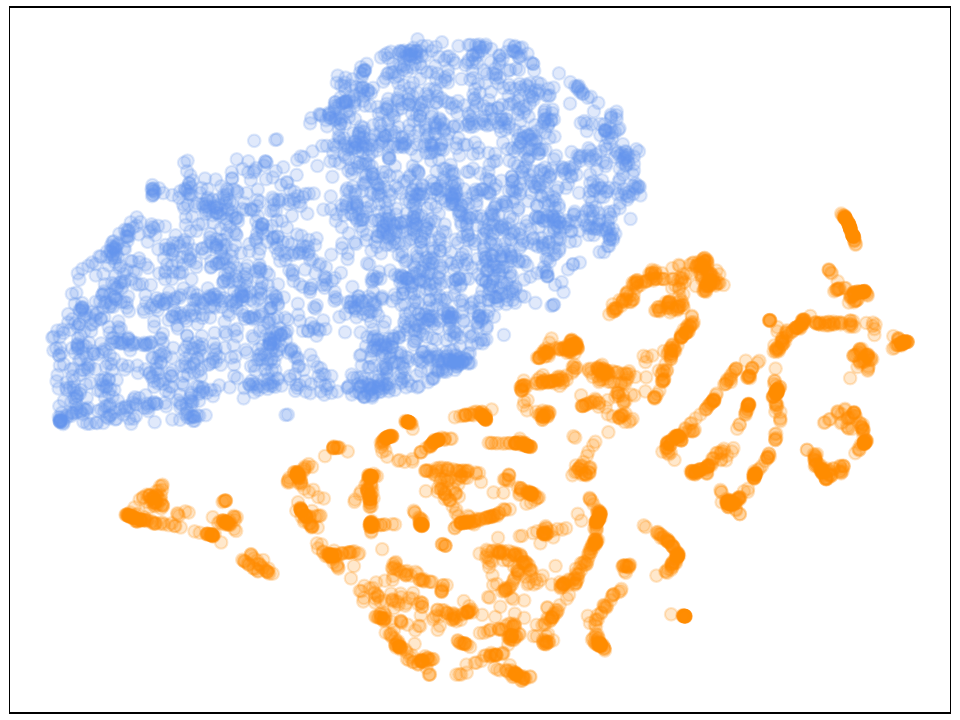}
		\end{minipage}
	}
	\subfigure[$50\%$]{
		\begin{minipage}[t]{0.27\linewidth}
			\centering
			\includegraphics[width=1.12\linewidth]{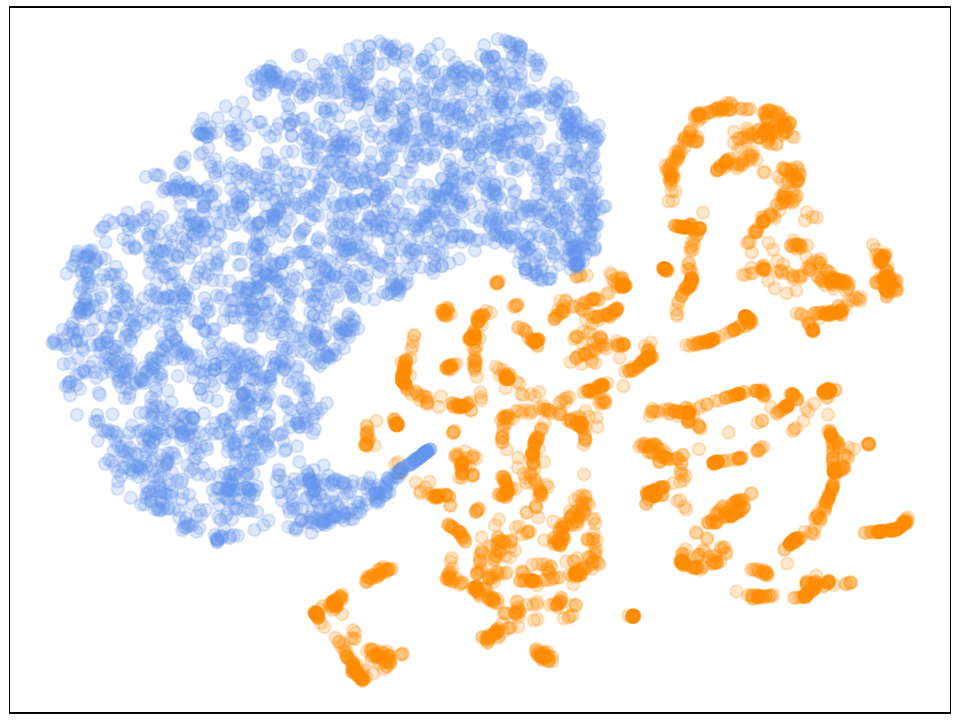}
		\end{minipage}
	}
	\subfigure[$100\%$]{
		\begin{minipage}[t]{0.27\linewidth}
			\centering
			\includegraphics[width=1.12\linewidth]{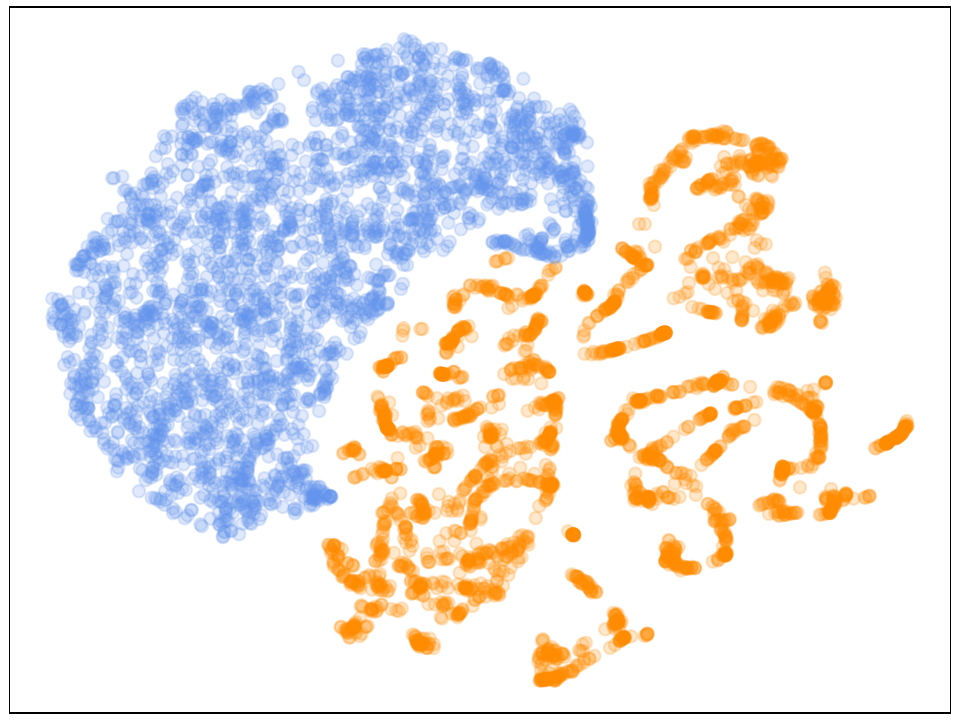}
		\end{minipage}
	}
\caption{t-SNE plots for Diffusion-TM (1st row), WGAN (2nd row), VAE (3rd row) in Abilene dataset with different sampling rates ($2\%, 50\%, 100\%$) of training traffic traces.} 
\label{tsne_abilene}
\end{figure}

\begin{figure}[htbp]
	\centering
	\subfigure{
        \rotatebox{90}{\scriptsize{~~~Diffusion-TM}}
		\begin{minipage}[t]{0.27\linewidth}
			\centering
			\includegraphics[width=1.12\linewidth]{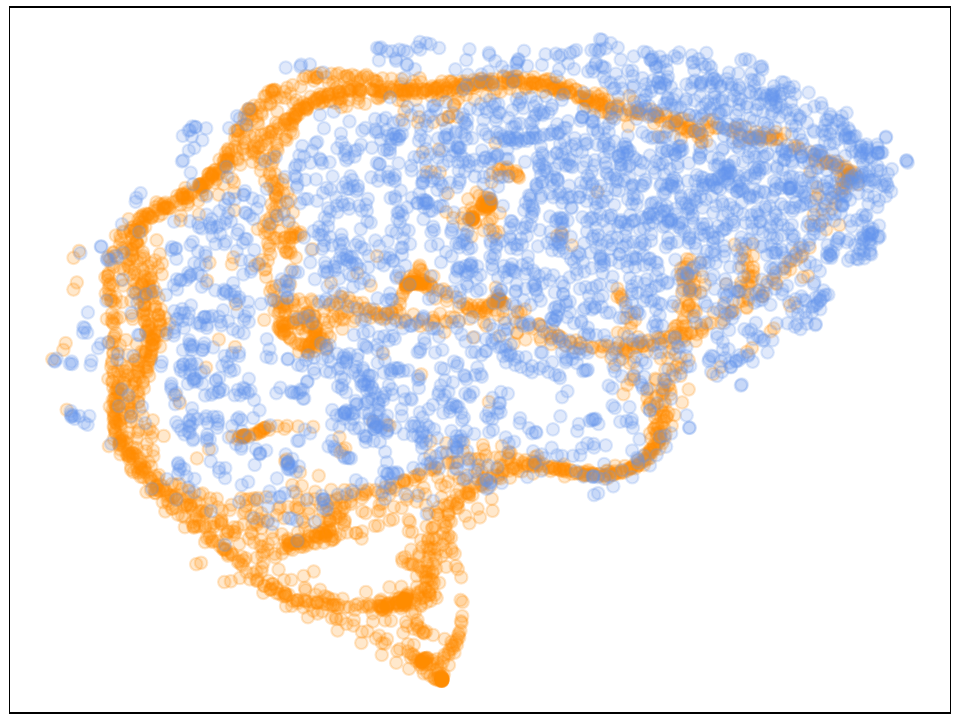}
		\end{minipage}
	}
	\subfigure{
		\begin{minipage}[t]{0.27\linewidth}
			\centering
			\includegraphics[width=1.12\linewidth]{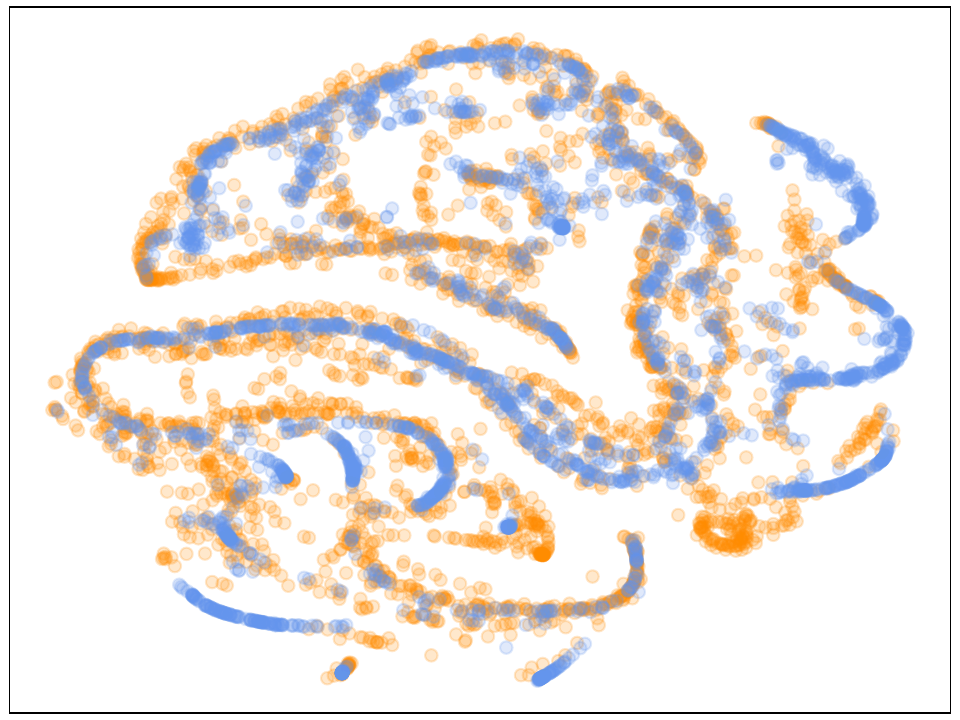}
		\end{minipage}
	}
	\subfigure{
		\begin{minipage}[t]{0.27\linewidth}
			\centering
			\includegraphics[width=1.12\linewidth]{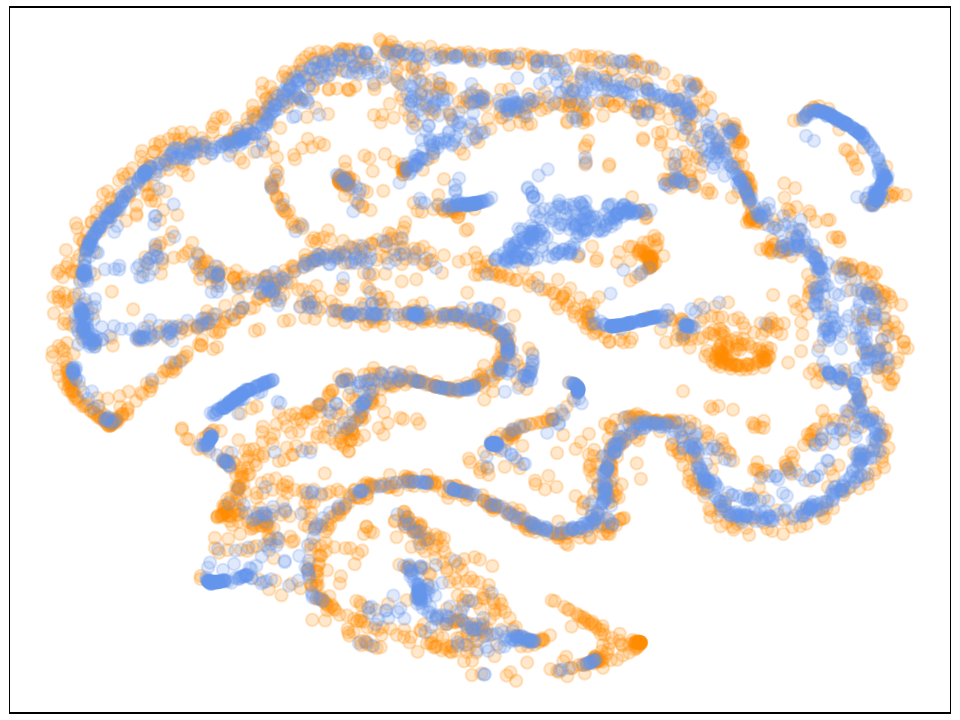}
		\end{minipage}
	}

	\vspace{-3mm}
	\setcounter{subfigure}{0}

    \subfigure{
        \rotatebox{90}{\scriptsize{~~~~~~~WGAN}}
		\begin{minipage}[t]{0.27\linewidth}
			\centering
			\includegraphics[width=1.12\linewidth]{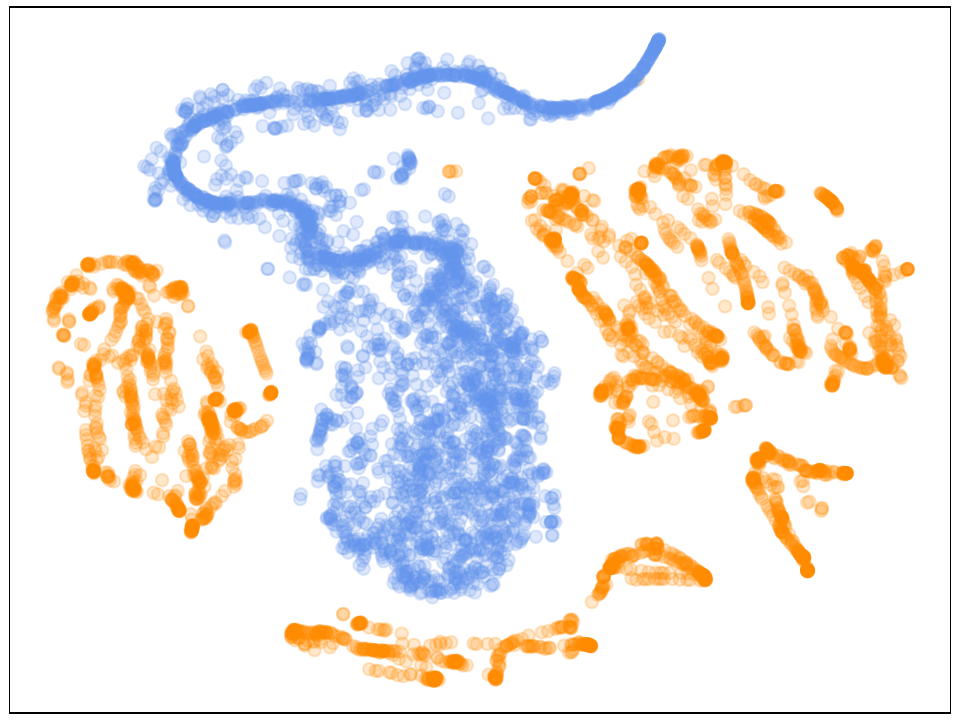}
		\end{minipage}
	}
	\subfigure{
		\begin{minipage}[t]{0.27\linewidth}
			\centering
			\includegraphics[width=1.12\linewidth]{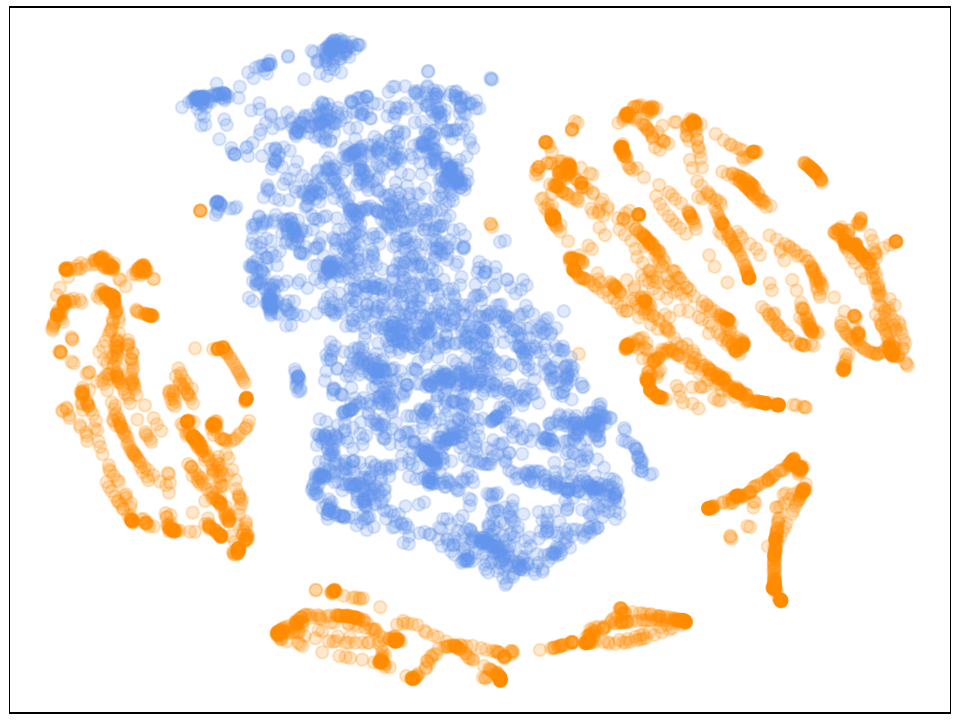}
		\end{minipage}
	}
	\subfigure{
		\begin{minipage}[t]{0.27\linewidth}
			\centering
			\includegraphics[width=1.12\linewidth]{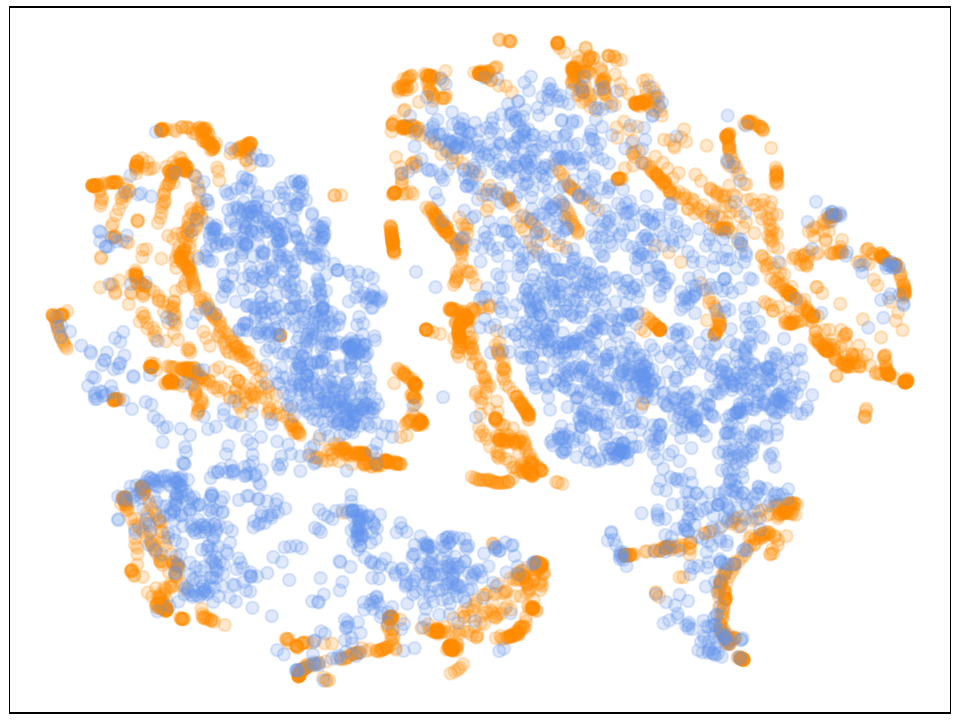}
		\end{minipage}
	}
 
        \vspace{-3mm}
	\setcounter{subfigure}{0}

 \subfigure[$2\%$]{
        \rotatebox{90}{\scriptsize{~~~~~~~~~VAE}}
		\begin{minipage}[t]{0.27\linewidth}
			\centering
			\includegraphics[width=1.12\linewidth]{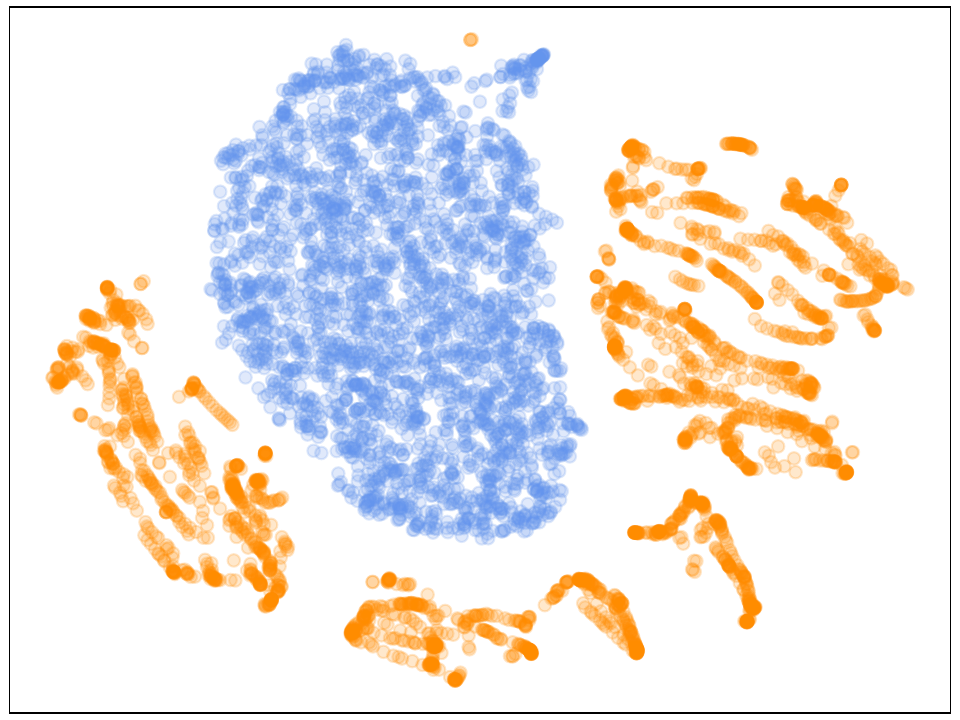}
		\end{minipage}
	}
	\subfigure[$50\%$]{
		\begin{minipage}[t]{0.27\linewidth}
			\centering
			\includegraphics[width=1.12\linewidth]{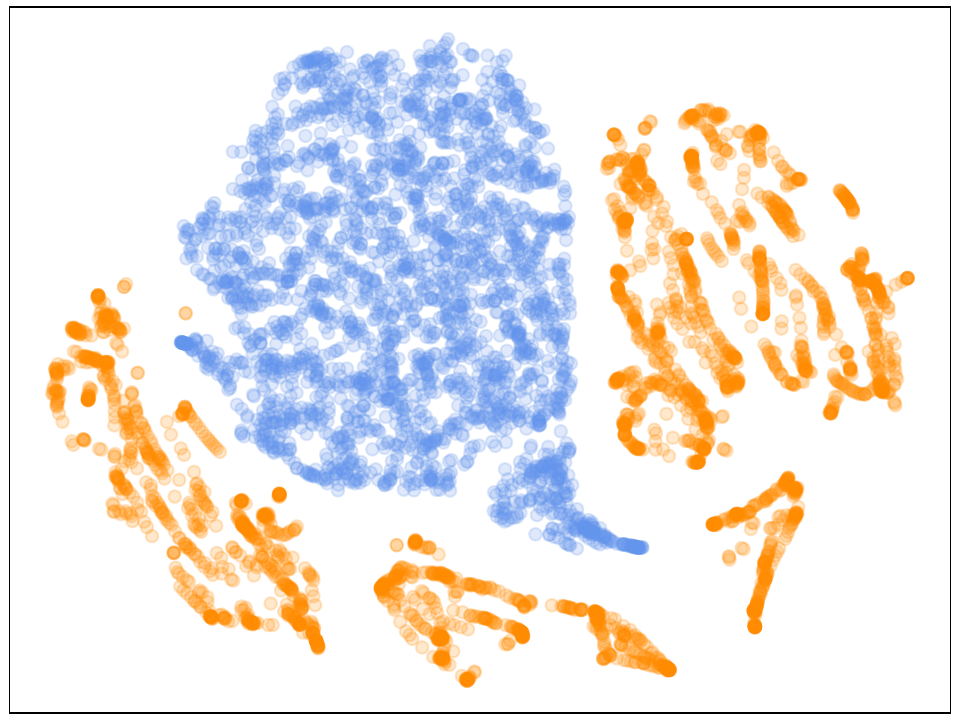}
		\end{minipage}
	}
	\subfigure[$100\%$]{
		\begin{minipage}[t]{0.27\linewidth}
			\centering
			\includegraphics[width=1.12\linewidth]{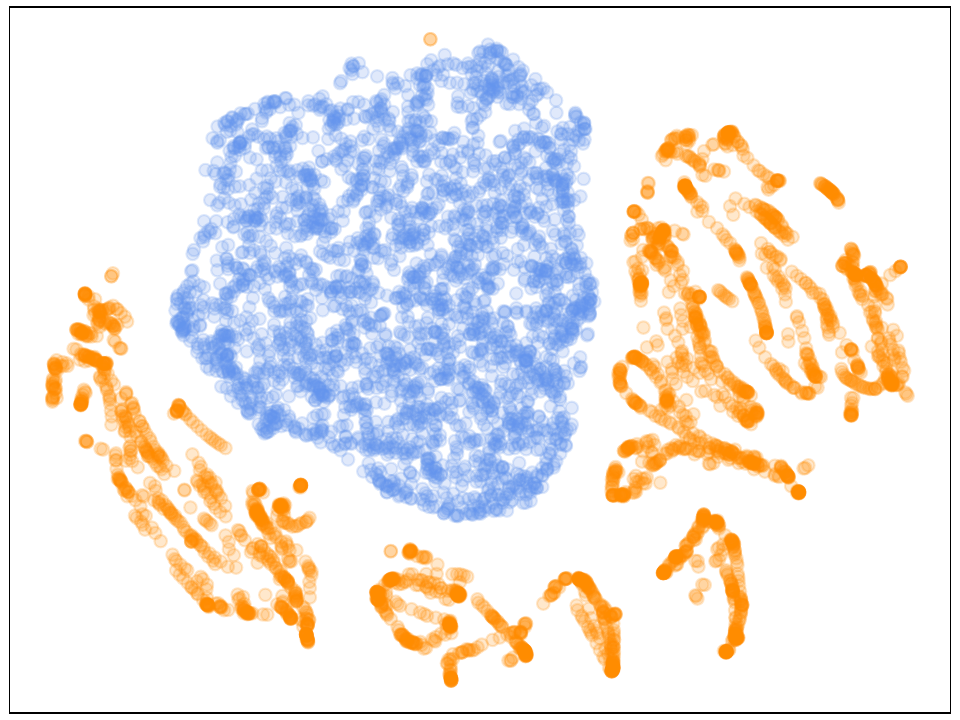}
		\end{minipage}
	}
\caption{t-SNE plots for Diffusion-TM (1st row), WGAN (2nd row), VAE (3rd row) in G\'{E}ANT dataset with different sampling rates ($2\%, 50\%, 100\%$) of training traffic traces.} 
\label{tsne_geant}
\end{figure}

We first report the quality of synthetic traffic data produced by the generative models with respect to the distributions over the original and generated data. We flatten the temporal dimension and use t-SNE plots to compress them into 2-dimensional space for visualization. Fig.~\ref{tsne_abilene} and Fig.~\ref{tsne_geant} present results with different sampling rates (i.e., the proportion of training data from the known entries of different datasets), where a greater overlap of blue (fake) and orange (real) dots shows a better distributional-similarity between the generated TMs and original TMs. There are several key observations: $\left({\rm i}\right)$ we observe that Diffusion-TM consistently matches the realistic distribution better than other benchmarks. Despite the powerful generative ability of diffusion models, all generative models except ours synthesize inferior data (covering a much smaller area across the original data) on real-world datasets, 
which indicates that previous GAN (or VAE) methods may not be able to model high-dimensional and complex network traffic distribution well. $\left({\rm ii}\right)$ One can also note that the performance of Diffusion-TM does not degrade significantly with the percentage of missing values, even in the extreme case of only $2\%$ observed data. The significant improvement of Diffusion-TM demonstrates our two-stage training algorithm can greatly ease the effect of missing values, making the learned distribution much more accurate. Overall, our proposed diffusion framework shows the best data diversity and robust quality, with the closest match to the original traffic data. Thus, the Diffusion-TM provides a promising approach to generating large-scale network measurement data in practical applications.

\vspace{-3mm}

\begin{table*}[htbp]
  \centering
  \caption{Completion Performance on Training Set of Abilene and G\'{E}ANT.}
  \begin{threeparttable}
    \resizebox{\textwidth}{!}{\begin{tabular}{c|c|cccccc|cccccc}
    \toprule
    \multicolumn{2}{c|}{Dataset} & \multicolumn{6}{c|}{Abilene}                  & \multicolumn{6}{c}{G\'{E}ANT} \\
    \midrule
    Metric & Model & 2\%   & 4\%   & 6\%   & 8\%   & 10\%  & 50\%  & 2\%   & 4\%   & 6\%   & 8\%   & 10\%  & 50\% \\
    \midrule
    \multirow{6}[2]{*}{NMAE} & Diffusion-TM & {\textbf{0.2712(32)}} & {\textbf{0.2446(25)}} & {\textbf{0.2329(22)}} & {\textbf{0.2261(15)}} & {\textbf{0.2111(12)}} & {\textbf{0.1623(7)}} & {\textbf{0.4703(56)}} & {\textbf{0.3152(42)}} & {\textbf{0.2514(33)}} & {\textbf{0.2360(26)}} & {\textbf{0.2185(15)}} & {\textbf{0.1627(9)}} \\
    % & Pre-AE &  &  &  &  &  &  &  &  &  &  &  &  \\
          & DATC  & 0.2954  & 0.2748  & 0.2631  & 0.2596  & 0.2541  & 0.2381  & 0.4939  & 0.4332  & 0.3651  & 0.3215  & 0.3037  & 0.2487  \\
          & NTC   & 0.3702  & 0.3362  & 0.3243  & 0.3115  & 0.3073  & 0.2497  & 0.7042  & 0.6796  & 0.6239  & 0.5781  & 0.4751  & 0.2871  \\
          & NTM   & 1.2214  & 0.8818  & 0.7633  & 0.6686  & 0.6178  & 0.3758  & 1.4547  & 1.3646  & 1.3266  & 1.2003  & 1.1987  & 0.5150  \\
          & NTF   & 0.7529  & 0.4147  & 0.3637  & 0.3486  & 0.3353  & 0.2639  & 0.9061  & 0.7783  & 0.7189  & 0.5958  & 0.5125  & 0.3208  \\
          & CoSTCo & 0.3857  & 0.3739  & 0.3381  & 0.3224  & 0.3145  & 0.2971  & 0.8588  & 0.7826  & 0.7291  & 0.7241  & 0.7042  & 0.6851  \\
    \midrule
    \multirow{6}[2]{*}{NRMSE} & Diffusion-TM & {\textbf{0.3115(37)}} & {\textbf{0.2901(28)}} & {\textbf{0.2777(30)}} & {\textbf{0.2611(21)}} & {\textbf{0.2467(17)}} & {\textbf{0.2034(11)}} & {0.5501(63)} & {\textbf{0.3383(44)}} & {\textbf{0.2795(35)}} & {\textbf{0.2457(20)}} & {\textbf{0.2250(14)}} & {\textbf{0.1643(6)}} \\
          & DATC  & 0.3453  & 0.3312  & 0.3226  & 0.3202  & 0.3178  & 0.3076  & \textbf{0.3818 } & 0.3415  & 0.2979  & 0.2547  & 0.2368  & 0.1742  \\
          & NTC   & 0.4388  & 0.4072  & 0.3951  & 0.3858  & 0.3766  & 0.3436  & 0.6412  & 0.6266  & 0.5888  & 0.5450  & 0.4310  & 0.2319  \\
          & NTM   & 0.9720  & 0.7573  & 0.7234  & 0.6613  & 0.6197  & 0.4473  & 0.9230  & 0.9057  & 0.8877  & 0.7854  & 0.7629  & 0.3976  \\
          & NTF   & 0.6856  & 0.4201  & 0.3882  & 0.3848  & 0.3723  & 0.3667  & 0.7110  & 0.6621  & 0.5838  & 0.4959  & 0.4128  & 0.2728  \\
          & CoSTCo & 0.4081  & 0.3929  & 0.3736  & 0.3666  & 0.3600  & 0.3579  & 0.7215  & 0.6579  & 0.6252  & 0.6037  & 0.5927  & 0.5827  \\
    \midrule
    % \multirow{6}[2]{*}{$10 \times {\rm MMD}_k^2$} & Diffusion-TM & \textbf{0.1833} & \textbf{0.1568} & \textbf{0.1492} & \textbf{0.0887} & \textbf{0.0595} & \textbf{0.0192} & \textbf{1.1684} & \textbf{0.3358} & \textbf{0.0750} & \textbf{0.0598} & \textbf{0.0580} & \textbf{0.0071} \\
    %       & DATC & 0.4465 & 0.3592 & 0.2750 & 0.2506 & 0.1226 & 0.0593 & 1.9785 & 0.9935 & 0.4607 & 0.1399 & 0.1042 & 0.0470 \\
    %       & NTC & 4.3793 & 2.3580 & 1.3319 & 0.8593 & 0.7116 & 0.1746 & 12.1562 & 4.6855 & 2.9075 & 1.2910 & 0.9411 & 0.2767 \\
    %       & NTM & 14.0916 & 10.1699 & 8.0668 & 4.7944 & 2.5515 & 0.4391 & 20.6559 & 10.6659 & 8.3274 & 7.6139 & 6.9558 & 1.9278 \\
    %       & NTF & 7.9598 & 4.5893 & 2.8991 & 1.8745 & 1.7717 & 0.3072 & 13.5843 & 6.7041 & 5.0165 & 1.9542 & 1.4547 & 0.5247 \\
    %       & CoSTCo & 4.2636 & 2.1414 & 1.4873 & 1.2282 & 0.9015 & 0.3332 & 11.2508 & 6.4661 & 4.2945 & 2.8334 & 2.0990 & 1.0012 \\
    \multirow{6}[2]{*}{MMD} & Diffusion-TM & {\textbf{0.0183(6)}} & {\textbf{0.0156(3)}} & {\textbf{0.0149(3)}} & {\textbf{0.0088(1)}} & {\textbf{0.0059(1)}} & {\textbf{0.0019(0)}} & {\textbf{0.1168(21)}} & {\textbf{0.0335(4)}} & {\textbf{0.0075(1)}} & {\textbf{0.0059(1)}} & {\textbf{0.0058(1)}} & {\textbf{0.0007(0)}} \\
          & DATC & {0.0446(13)} & {0.0359(8)} & {0.0275(10)} & {0.0250(9)} & {0.0122(3)} & {0.0049(1)} & {0.1978(27)} & {0.0993(18)} & {0.0460(9)} & {0.0139(4)} & {0.0104(4)} & {0.0047(1)} \\
          & NTC & {0.1565(30)} & {0.0426(11)} & {0.0297(8)} & {0.0184(4)} & {0.0120(6)} & {0.0055(1)} & {0.5598(35)} & {0.2010(23)} & {0.1458(25)} & {0.1265(12)} & {0.1259(22)} & {0.0074(3)} \\
          & NTM & {0.5883(87)} & {0.5249(68)} & {0.4887(59)} & {0.3806(42)} & {0.3256(37)} & {0.0983(9)} & {0.8679(98)} & {0.6743(76)} & {0.5619(64)} & {0.4594(55)} & {0.3951(34)} & {0.1271(11)} \\ 
          & NTF & {0.2959(36)} & {0.1899(11)} & {0.1571(10)} & {0.1089(8)} & {0.0512(6)} & {0.0151(2)} & {0.7584(57)} & {0.5704(39)} & {0.4016(30)} & {0.2554(25)} & {0.1854(27)} & {0.0212(6)} \\
          & CoSTCo & {0.1607(19)} & {0.0513(15)} & {0.0329(8)} & {0.0246(4)} & {0.0155(5)} & {0.0098(3)} & {0.6089(40)} & {0.3713(24)} & {0.3368(22)} & {0.2289(14)} & {0.1677(6)} & {0.0184(4)} \\
    \bottomrule
    \end{tabular}}
    \begin{tablenotes}
        \footnotesize
        \item {In the table, results (except for MMD) of all baselines are from~\cite{r42}. Additionally, here we use a concise error notation where the values in brackets\\ affect the least significant digits, e.g. 0.2712(32) signifies 0.2712 ± 0.032.}
    \end{tablenotes}
    \end{threeparttable}
  \label{TMC_train}
\end{table*}

\vspace{-3mm}

\begin{table*}[htbp]
  \centering
  \caption{Completion Performance on Testing Set of Abilene and G\'{E}ANT.}
    \resizebox{\textwidth}{!}{\begin{tabular}{c|c|cccccc|cccccc}
    \toprule
    \multicolumn{2}{c|}{Dataset} & \multicolumn{6}{c|}{Abilene}                  & \multicolumn{6}{c}{G\'{E}ANT} \\
    \midrule
    Metric & Model & 2\%   & 4\%   & 6\%   & 8\%   & 10\%  & 50\%  & 2\%   & 4\%   & 6\%   & 8\%   & 10\%  & 50\% \\
    \midrule
    \multirow{5}[2]{*}{NMAE} & Diffusion-TM & {0.3041(28)} & {0.2854(19)} & {0.2804(15)} & {0.2701(18)} & {0.2395(13)} & {0.2162(9)} & {0.4834(48)} & {0.4142(36)} & {0.3714(27)} & {0.3470(19)} & {0.2714(13)} & {0.2368(10)} \\
          & Diffusion-TM (10\%) & {0.2835(30)} & {0.2604(16)} & {0.2577(12)} & {0.2542(15)} & {0.2313(5)} & {0.2066(3)} & {0.4734(46)} & {0.3269(30)} & {0.3181(19)} & {0.3025(13)} & {0.2630(14)} & {0.2350(6)} \\
          & Diffusion-TM (50\%) & {0.2366(17)} & {0.2221(16)} & {0.2193(14)} & {0.2170(13)} & {0.1962(10)} & {0.1760(4)} & {0.4038(35)} & {0.3230(19)} & {0.3064(13)} & {0.2971(10)} & {0.2576(10)} & {0.2186(4)} \\
          & Diffusion-TM (100\%) & {\textbf{0.1981(16)}} & {\textbf{0.1871(21)}} & {\textbf{0.1853(9)}} & {\textbf{0.1802(7)}} & {\textbf{0.1715(11)}} & {\textbf{0.1577(3)}} & {\textbf{0.3492(27)}} & {\textbf{0.2831(14)}} & {\textbf{0.2820(13)}} & {\textbf{0.2660(8)}} & {\textbf{0.2155(5)}} & {\textbf{0.1829(3)}} \\
          & DATC & {0.4052(49)} & {0.3559(35)} & {0.3304(25)} & {0.3218(18)} & {0.2980(16)} & {0.2746(11)} & {0.5562(52)} & {0.4456(40)} & {0.4030(37)} & {0.3918(29)} & {0.3815(12)} & {0.2889(15)} \\
          % & NTC & 0.4752 & 0.4000 & 0.3559 & 0.3279 & 0.3275 & 0. & 0. & 0. & 0. & 0. & 0. & 0. \\
    \midrule
    \multirow{5}[2]{*}{NRMSE} & Diffusion-TM & {0.3463(35)} & {0.3192(26)} & {0.3123(19)} & {0.3048(22)} & {0.2792(14)} & {0.2501(7)} & {0.5401(52)} & {0.4686(49)} & {0.4239(34)} & {0.3893(23)} & {0.2923(17)} & {0.2416(15)} \\
          & Diffusion-TM (10\%) & {0.3167(31)} & {0.3017(21)} & {0.3055(18)} & {0.2941(16)} & {0.2708(10)} & {0.2422(2)} & {0.5418(54)} & {0.3476(35)} & {0.3393(23)} & {0.3158(10)} & {0.2631(14)} & {0.2368(7)} \\
          & Diffusion-TM (50\%) & {0.2708(22)} & {0.2610(19)} & {0.2602(15)} & {0.2597(19)} & {0.2269(16)} & {0.2022(4)} & {0.4275(38)} & {0.3263(24)} & {0.3182(16)} & {0.3056(12)} & {0.2411(9)} & {0.2083(3)} \\
          & Diffusion-TM (100\%) & {\textbf{0.2011(19)}} & {\textbf{0.2003(15)}} & {\textbf{0.1952(11)}} & {\textbf{0.1941(13)}} & {\textbf{0.1862(6)}} & {\textbf{0.1660(2)}} & {\textbf{0.3452(31)}} & {\textbf{0.2732(20)}} & {\textbf{0.2663(18)}} & {\textbf{0.2528(14)}} & {\textbf{0.2027(6)}} & {\textbf{0.1782(3)}} \\
          & DATC & {0.3972(42)} & {0.3882(38)} & {0.3614(28)} & {0.3536(17)} & {0.3314(14)} & {0.2958(14)} & {0.5050(63)} & {0.4686(48)} & {0.4574(41)} & {0.4486(33)} & {0.4507(24)} & {0.3258(17)} \\
          % & NTC & 0.4381 & 0.3748 & 0.3367 & 0.3284 & 0.3271 & 0. & 0. & 0. & 0. & 0. & 0. & 0. \\
    \midrule
    % \multirow{5}[2]{*}{$10 \times {\rm MMD}_k^2$} & Diffusion-TM & 0.6846 & 0.6185 & 0.5693 & 0.5198 & 0.3502 & 0.0817 & 3.3485 & 2.2185 & 2.0477 & 1.5419 & 0.4093 & 0.0872 \\
    %       & Diffusion-TM (10\%) & 0.7067 & 0.6101 & 0.5764 & 0.5281 & 0.2763 & 0.0679 & 2.4488 & 1.3069 & 1.1903 & 1.0145 & 0.3205 & 0.0814 \\
    %       & Diffusion-TM (50\%) & 0.4899 & 0.4317 & 0.4148 & 0.4064 & 0.1572 & 0.0352 & 2.2005 & 1.3460 & 1.2772 & 0.8406 & 0.3147 & 0.0668 \\
    %       & Diffusion-TM (100\%) & \textbf{0.1982} & \textbf{0.1836} & \textbf{0.1571} & \textbf{0.1314} & \textbf{0.1271} & \textbf{0.0277} & \textbf{1.4420} & \textbf{0.8087} & \textbf{0.7920} & \textbf{0.5377} & \textbf{0.2164} & \textbf{0.0475} \\
    %       & DATC & 1.2490 & 1.0872 & 0.7733 & 0.7391 & 0.5396 & 0.3503 & 5.5700 & 2.8525 & 2.5984 & 2.2544 & 1.8813 & 0.5764 \\
    \multirow{5}[2]{*}{MMD} & Diffusion-TM & {0.0684(12)} & {0.0618(14)} & {0.0569(9)} & {0.0519(11)} & {0.0350(5)} & {0.0081(2)} & {0.3348(33)} & {0.2218(25)} & {0.2047(21)} & {0.1541(16)} & {0.0409(10)} & {0.0087(1)} \\
          & Diffusion-TM (10\%) & {0.0706(29)} & {0.0610(16)} & {0.0576(12)} & {0.0528(10)} & {0.0276(4)} & {0.0067(1)} & {0.2448(35)} & {0.1306(18)} & {0.1190(13)} & {0.1014(14)} & {0.0320(11)} & {0.0081(2)} \\
          & Diffusion-TM (50\%) & {0.0489(10)} & {0.0431(8)} & {0.0414(11)} & {0.0406(14)} & {0.0157(5)} & {0.0035(0)} & {0.2200(23)} & {0.1346(14)} & {0.1277(18)} & {0.0840(15)} & {0.0314(7)} & {0.0066(1)} \\
          & Diffusion-TM (100\%) & {\textbf{0.0198(6)}} & {\textbf{0.0183(4)}} & {\textbf{0.0157(2)}} & {\textbf{0.0131(3)}} & {\textbf{0.0127(1)}} & {\textbf{0.0027(0)}} & {\textbf{0.1442(19)}} & {\textbf{0.0808(17)}} & {\textbf{0.0792(9)}} & {\textbf{0.0537(11)}} & {\textbf{0.0216(3)}} & {\textbf{0.0047(1)}} \\
          & DATC & {0.1249(36)} & {0.1087(23}) & {0.0773(21)} & {0.0739(15)} & {0.0539(17)} & {0.0350(5)} & {0.5570(49)} & {0.2852(38)} & {0.2598(25)} & {0.2254(16)} & {0.1881(13)} & {0.0576(9)} \\
    \bottomrule
    \end{tabular}}
  \label{TMC_test}
\end{table*}

\vspace{-3mm}

\subsection{Traffic Matrix Completion Performance Comparison}

In Table~\ref{TMC_train}, we present the experimental results of our Diffusion-TM and five neural network tensor completion algorithms (DATC, NTC, NTM, NTF, CoSTCo) on \textit{training} dataset, when varying the amount of known information range from 2 to 50 percent of the OD-flow entries. Except for NRMSE under $2\%$ sampling rate in G\'{E}ANT, we can see that the proposed Diffusion-TM yields the best performance on all the table items. Specifically, Diffusion-TM can reduce the average NMAE of the best baseline by $15\%$ in Abilene and $23\%$ in G\'{E}ANT. Even when the sampling rate is less than $5\%$ which is a very low ratio, Diffusion-TM is still effective thanks to the combination of pre-processing module and designed reconstruction loss on observed flow loads, which forces the diffusion model to optimize the distribution for unobserved flows during the training stage. Although four neural network (NN) based tensor completion algorithms exhibit some ability to handle a considerable amount of missing data, the overall imputation accuracy is still limited compared with our approach. Among them, the effect of NTC is generally better than other methods, because the algorithm designed for monitoring performs well to extract features in the network traffic while NTF, CoSTCo, and NTM are more suitable for recommender systems. On the contrary, the autoencoder-based method DATC performs marginally worse than ours when the sampling rate is low. However, the accuracy gap between Diffusion-TM and DATC widens for a small missing ratio  (e.g., $90\% \sim 50\%$), since the larger number of observed entities allows DMs to fit the distribution of training traffic better, so Diffusion-TM can achieve a completion error of 0.16 which significantly outperforms the best DATC with different datasets when the sampling ratio is $50\%$.

In addition, our Diffusion-TM achieves a much more excellent MMD score. In contrast, the peer NN-based algorithms perform poorly for all loss probabilities, as their performance largely depends on the position of the missing entries in the tensor (matrix) while unable to capture the data distribution. DATC may perform better than these low-rank methods owing to its adversarial nature in this case, but nevertheless, Diffusion-TM still shows the best performance over the widest range of missing ratios. The typically very large performance gap also suggests that our gradient-oriented completion algorithm can solve the recovery problem while preserving the learned distribution of powerful diffusion models to the greatest extent. 

Complex and dynamic network behavior can not guarantee that training data would not undergo distribution shift in the real world. Retraining on new traffic data may be a solution, but it cannot meet the needs of real-time filling. Thus we then perform additional infilling experiments on \textit{testing} dataset to validate the performance of the online TMC with an already trained model. Since NTC, NTF, NTM, and CoSTCo use the whole tensor data as input to complete it together, and have no concept of training set (or inference set), they can not run online to complete continuously arrived traffic matrix sequence (sliding window). We do not compare all baselines and eliminate them from the experiment. Moreover, in actual online executions, it is normal to observe some easily obtained link-load data during inference. As aforementioned, our algorithm can be easily extended to solve multi-objective problems. Thus we will also experience and show how the Diffusion-TM performs by combining the additional useful constraints. As shown in Table.~\ref{TMC_test}, we report results for newly collected data imputation on the test set, where Diffusion-TM ($p\%$) signifies Diffusion-TM with $p\%$ link loads measured. Due to the distribution drift and probable overfitting, both DATC and Diffusion-TM show inferior accuracy. The MMD score further supports the problem of distribution alignment. Nevertheless, Diffusion-TM can still improve the accuracy of DATC by $16\% \sim 25\%$ ($15\% \sim 29\%$) in Abilene (G\'{E}ANT). Moreover, if any link measurements of the network are provided, then Diffusion-TM’s performance improves dramatically. Finally, by combining 100 percent link loads and arbitrary known TM elements, Diffusion-TM gets the best of both networks and even outperforms itself on the training set in some cases.

These results demonstrate the superiority of our model in recovering missing data. To summarize, our Diffusion-TM has three key advantages: $\left({\rm i}\right)$ it does not require a complete training dataset which is generally expensive; $\left({\rm ii}\right)$ it can precisely produce unobserved flows that closely fit the prior distribution; and $\left({\rm iii}\right)$ it can also impose extra constraints in a plug-and-play way, preventing useful information waste.

\begin{figure}[htbp]
\centering
\subfigure[{Abilene}]{
\begin{minipage}[b]{.46\linewidth}
    \centering
    \includegraphics[width=1.05\linewidth]{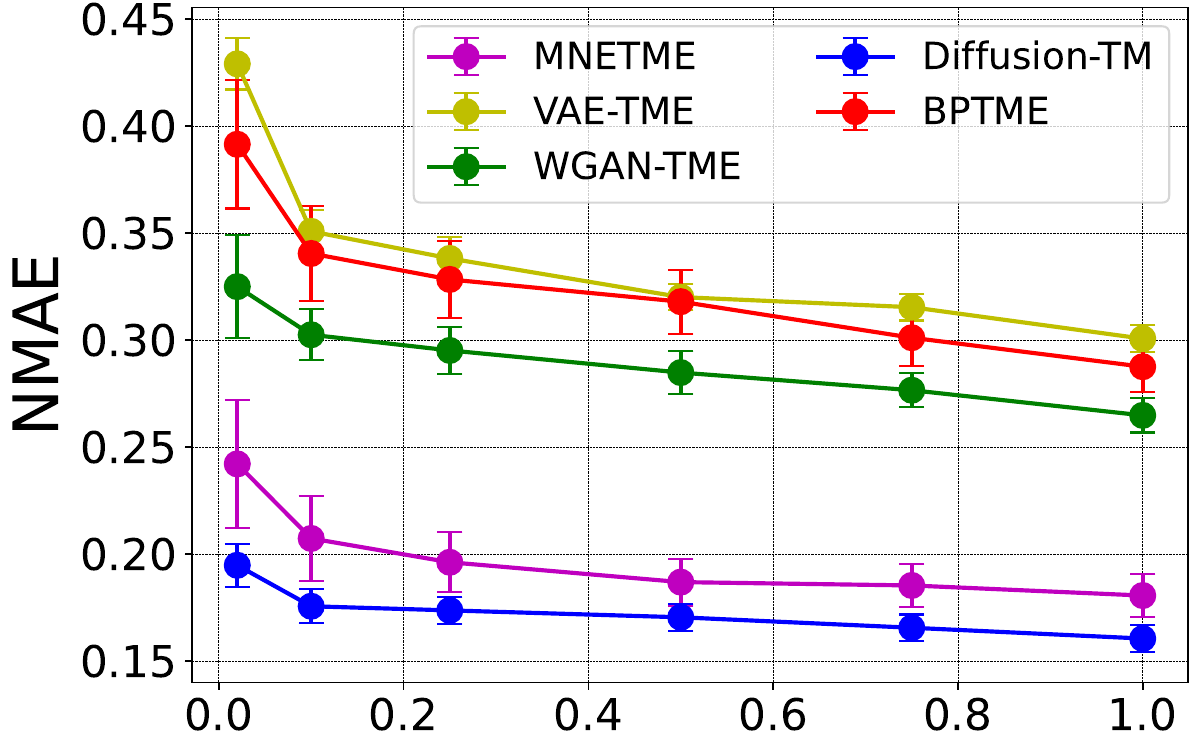}\\
    \includegraphics[width=1.05\linewidth]{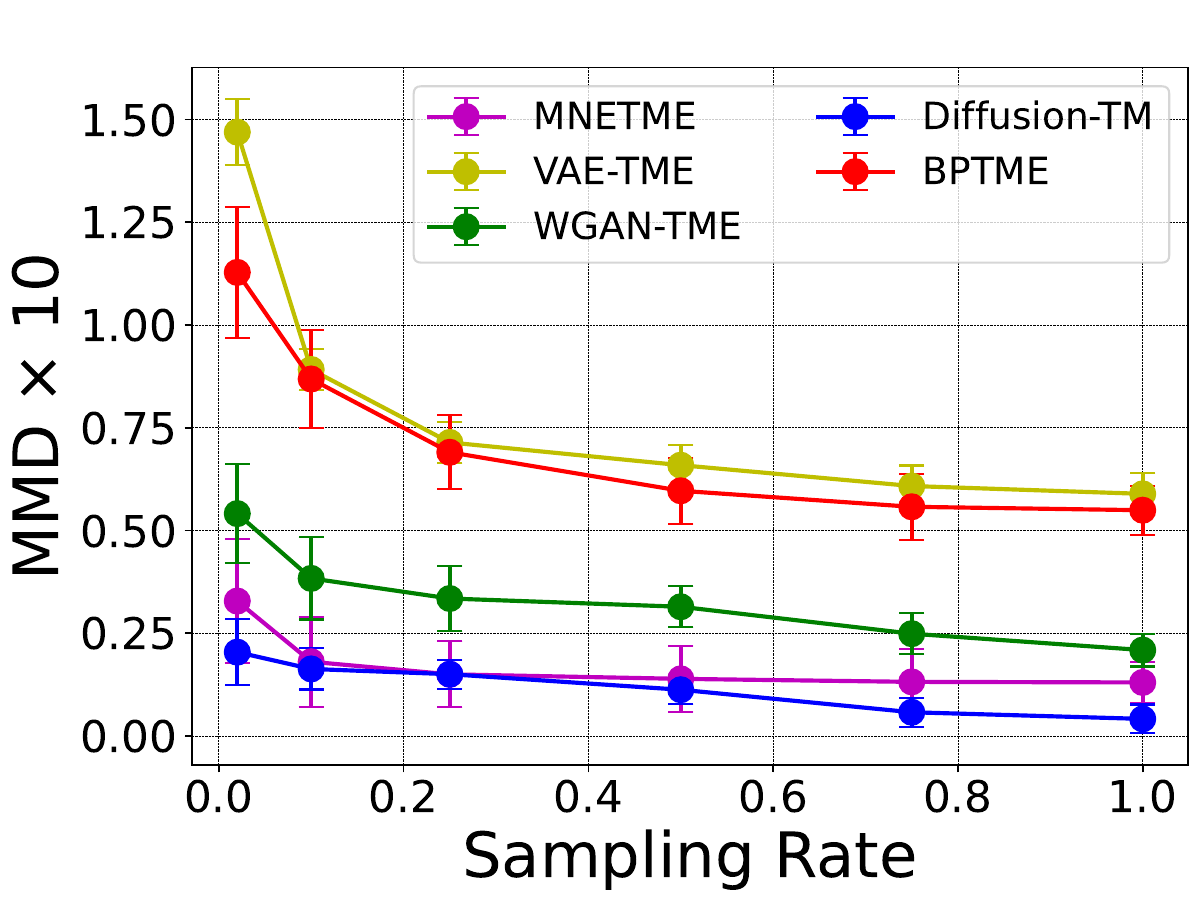}
\end{minipage}
}
\subfigure[{G\'{E}ANT}]{
\begin{minipage}[b]{.46\linewidth}
    \centering
    \includegraphics[width=1.05\linewidth]{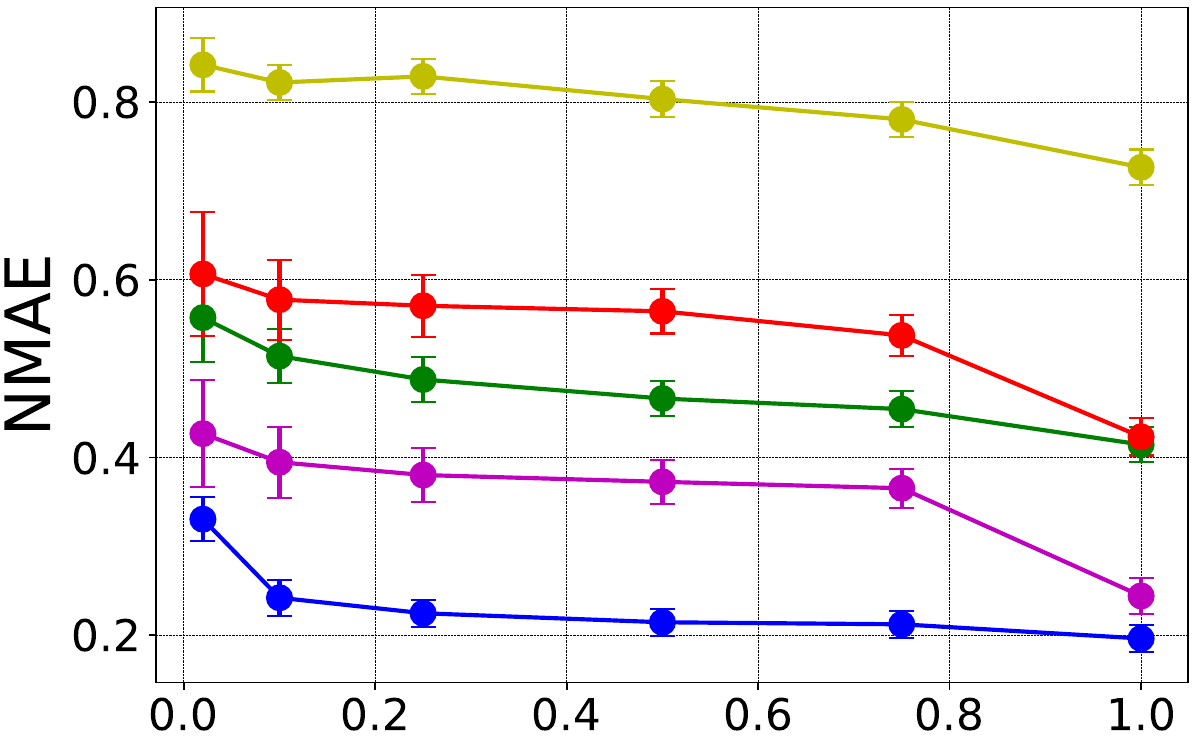}\\
    \includegraphics[width=1.05\linewidth]{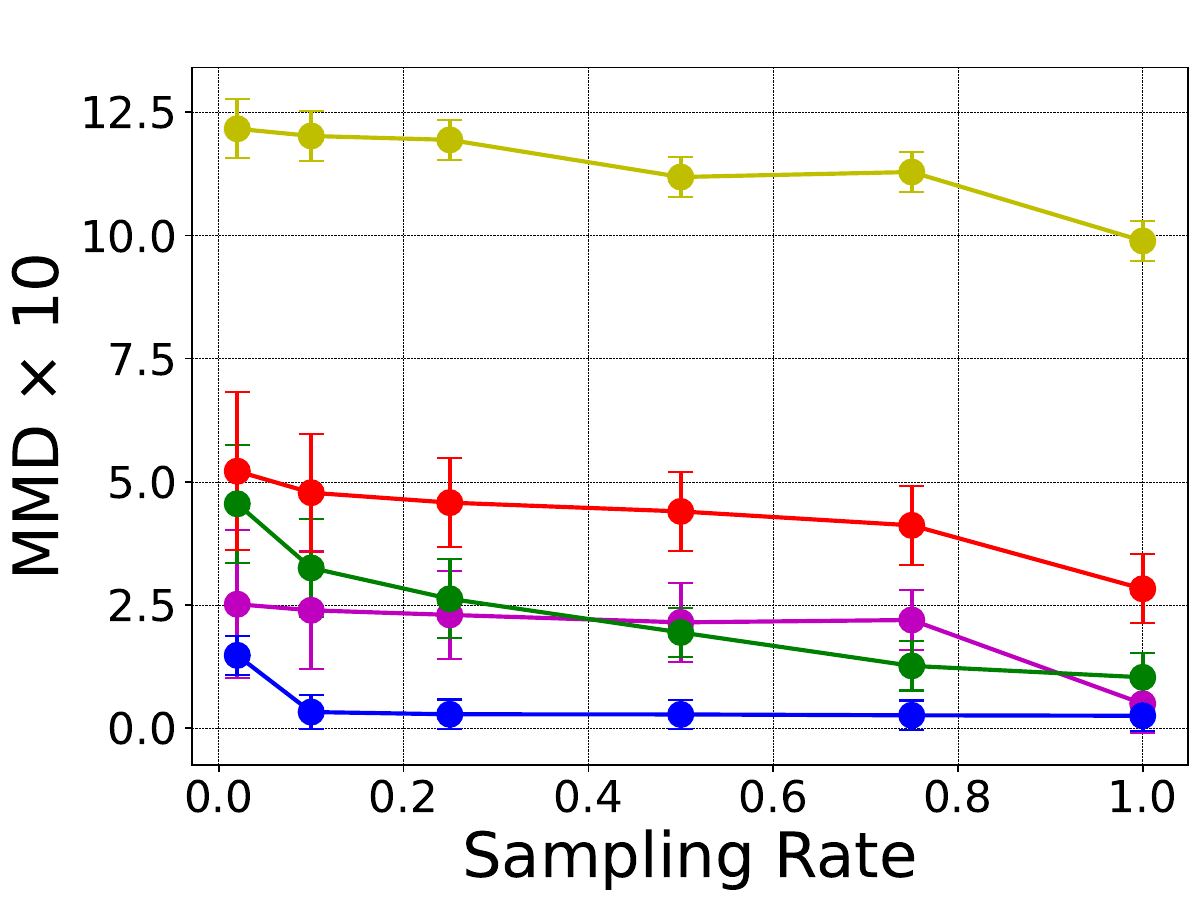}
\end{minipage}
}
\caption{{Network tomography performance for complete TM estimation under different sampling rates.}}
\label{TME_scalar}
\end{figure}

\vspace{-1mm}

\begin{figure*}[htbp]
\centering
\subfigure[Sampling Rate $ = 2\%$]{
\begin{minipage}[b]{.31\linewidth}
    \centering
    \includegraphics[width=1.\linewidth]{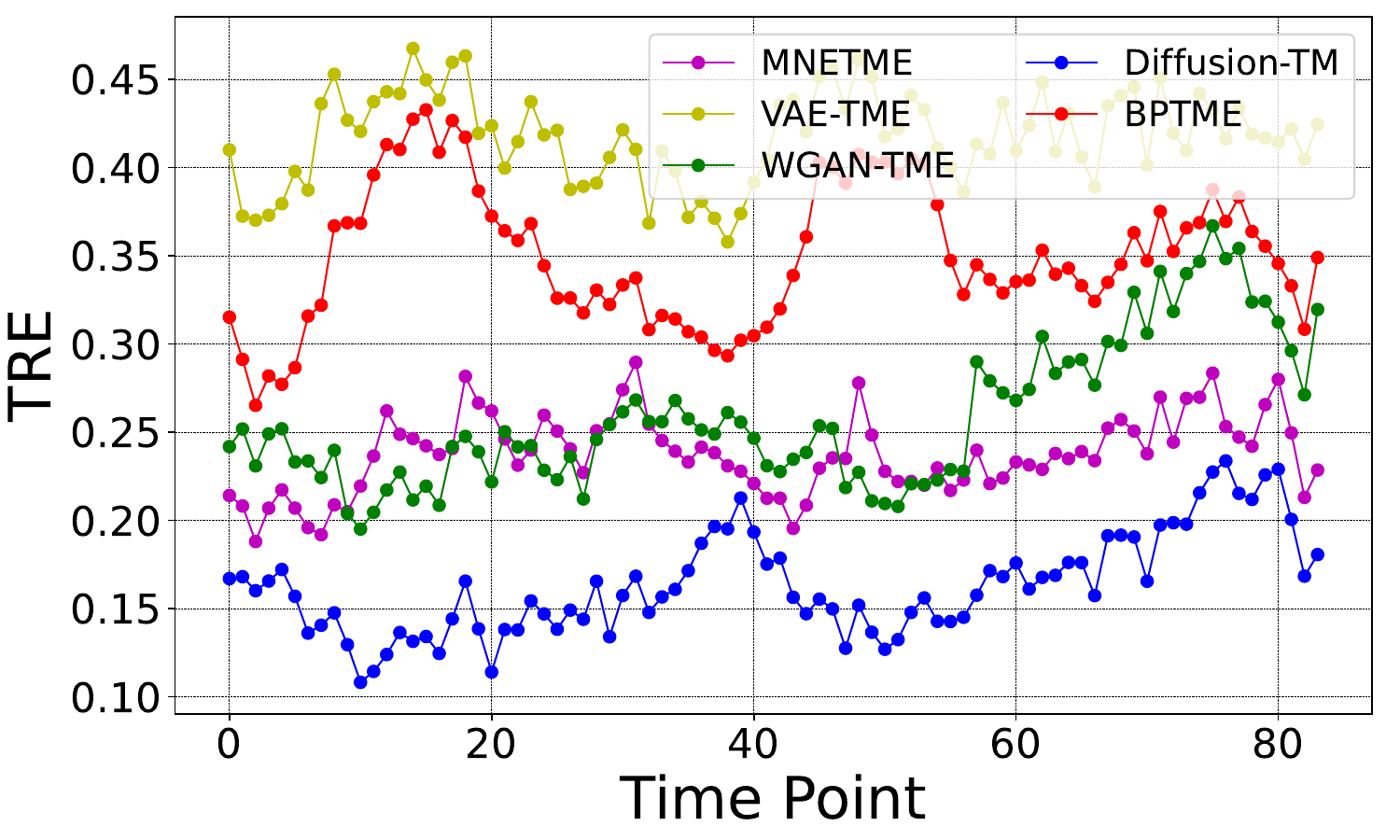}\\
    \includegraphics[width=1.\linewidth]{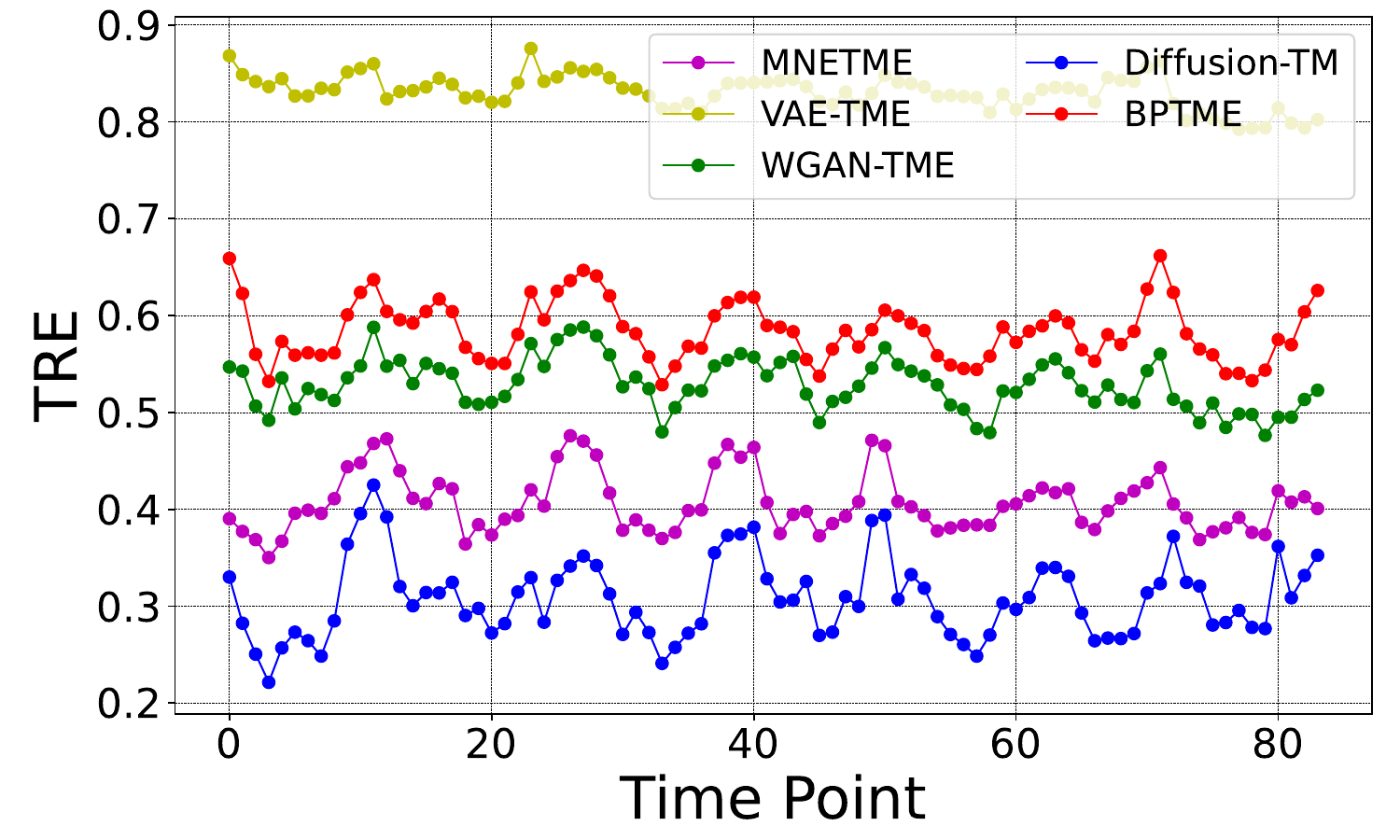}
\end{minipage}
}
\subfigure[Sampling Rate $ = 50\%$]{
\begin{minipage}[b]{.31\linewidth}
    \centering
    \includegraphics[width=1.\linewidth]{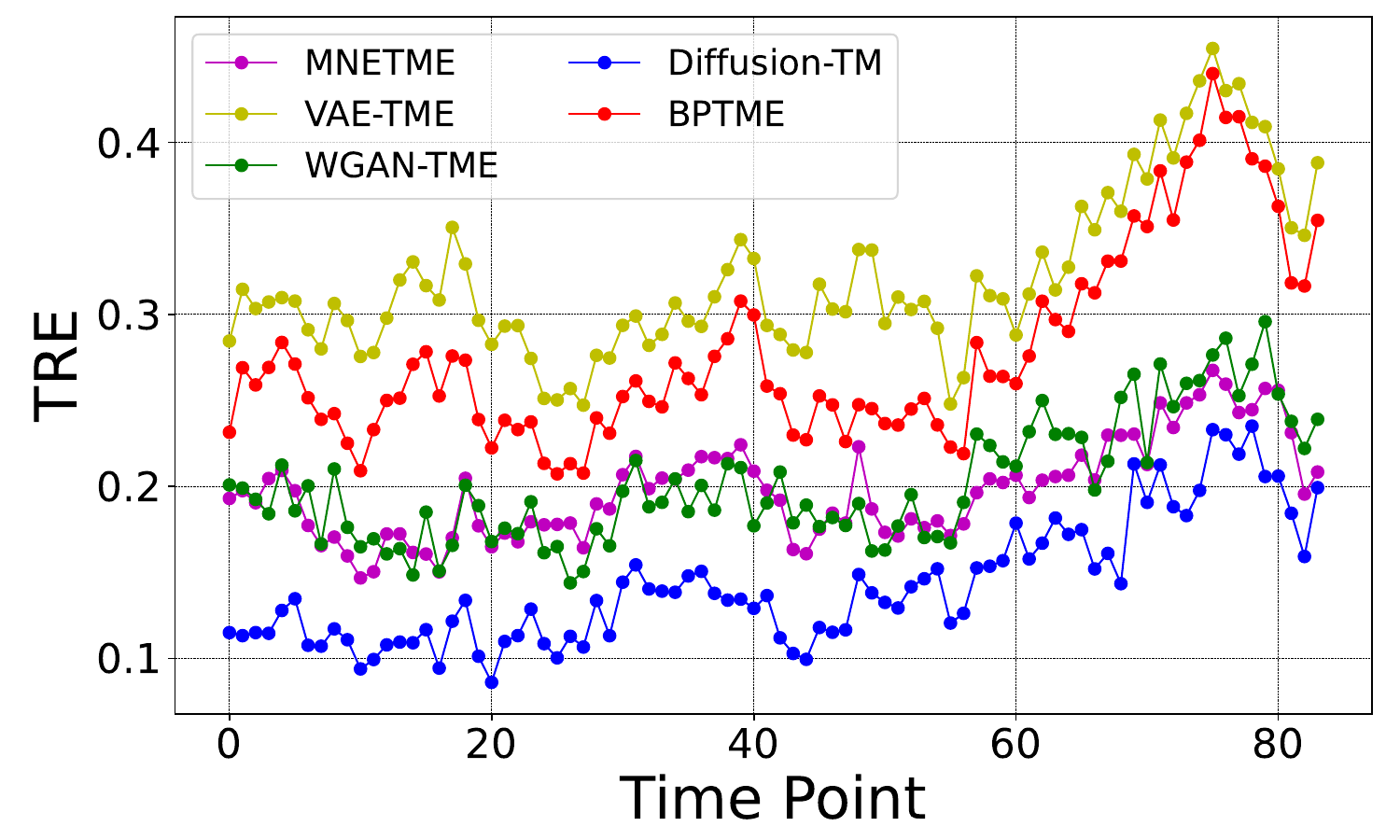}\\
    \includegraphics[width=1.\linewidth]{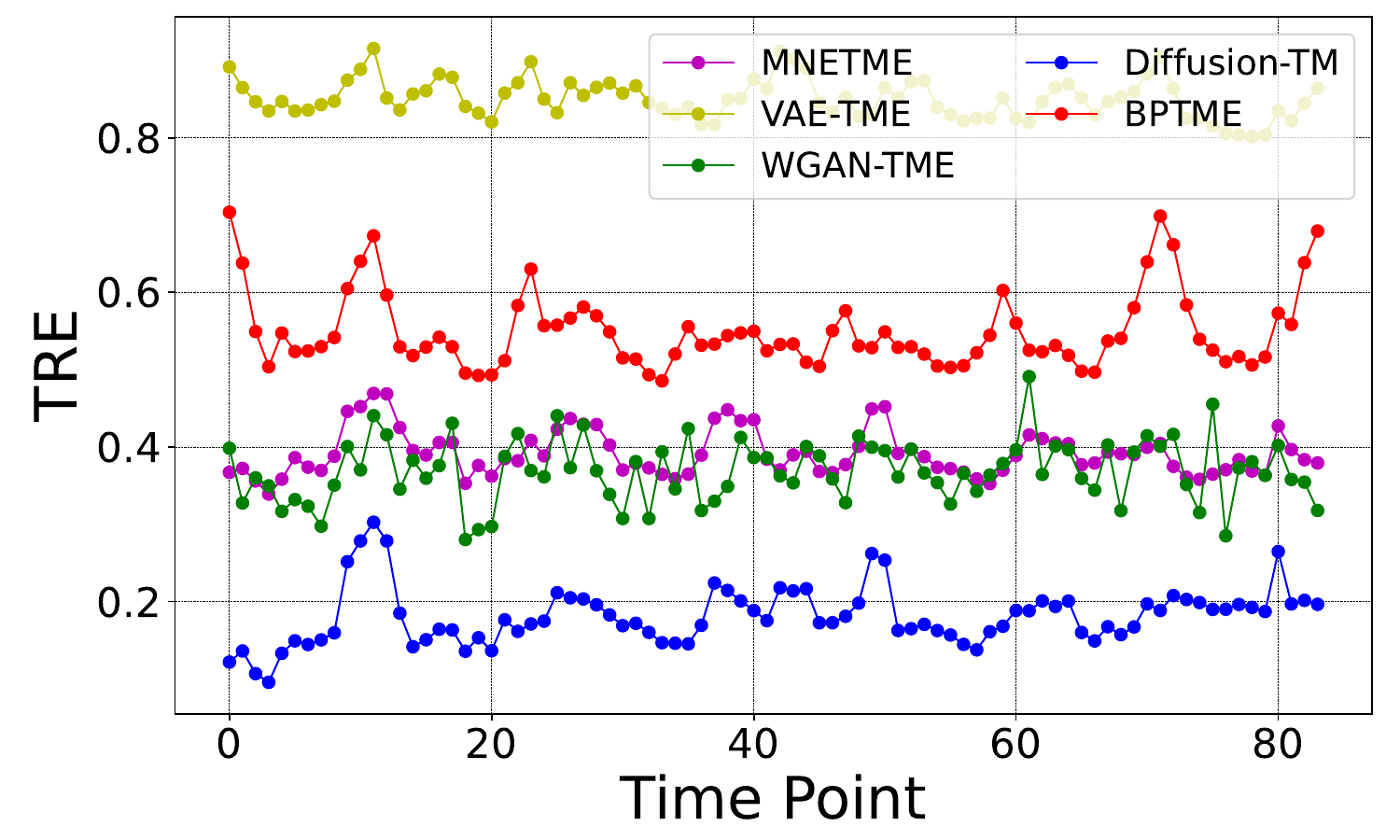}
\end{minipage}
}
\subfigure[Sampling Rate $ = 100\%$]{
\begin{minipage}[b]{.31\linewidth}
    \centering
    \includegraphics[width=1.\linewidth]{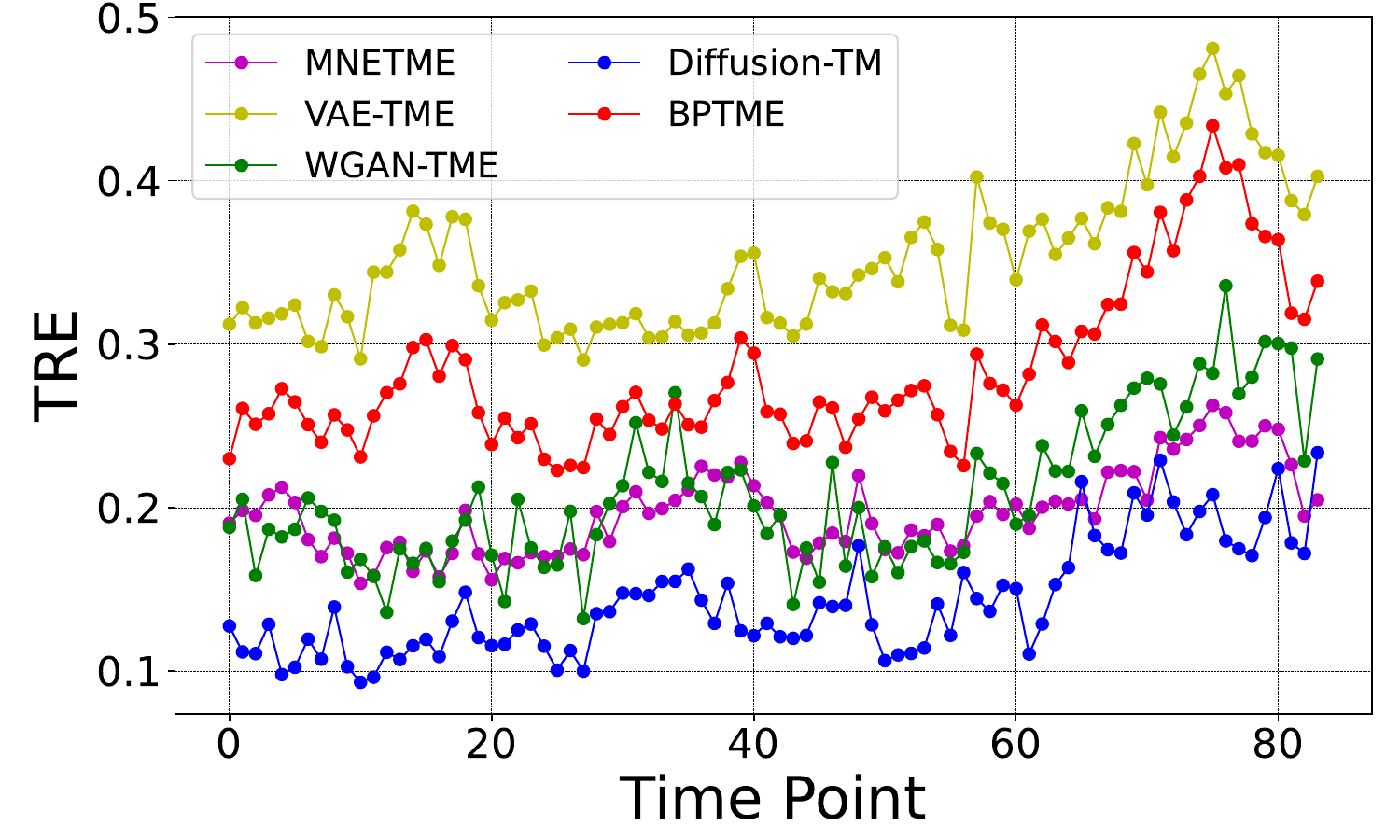}\\
    \includegraphics[width=1.\linewidth]{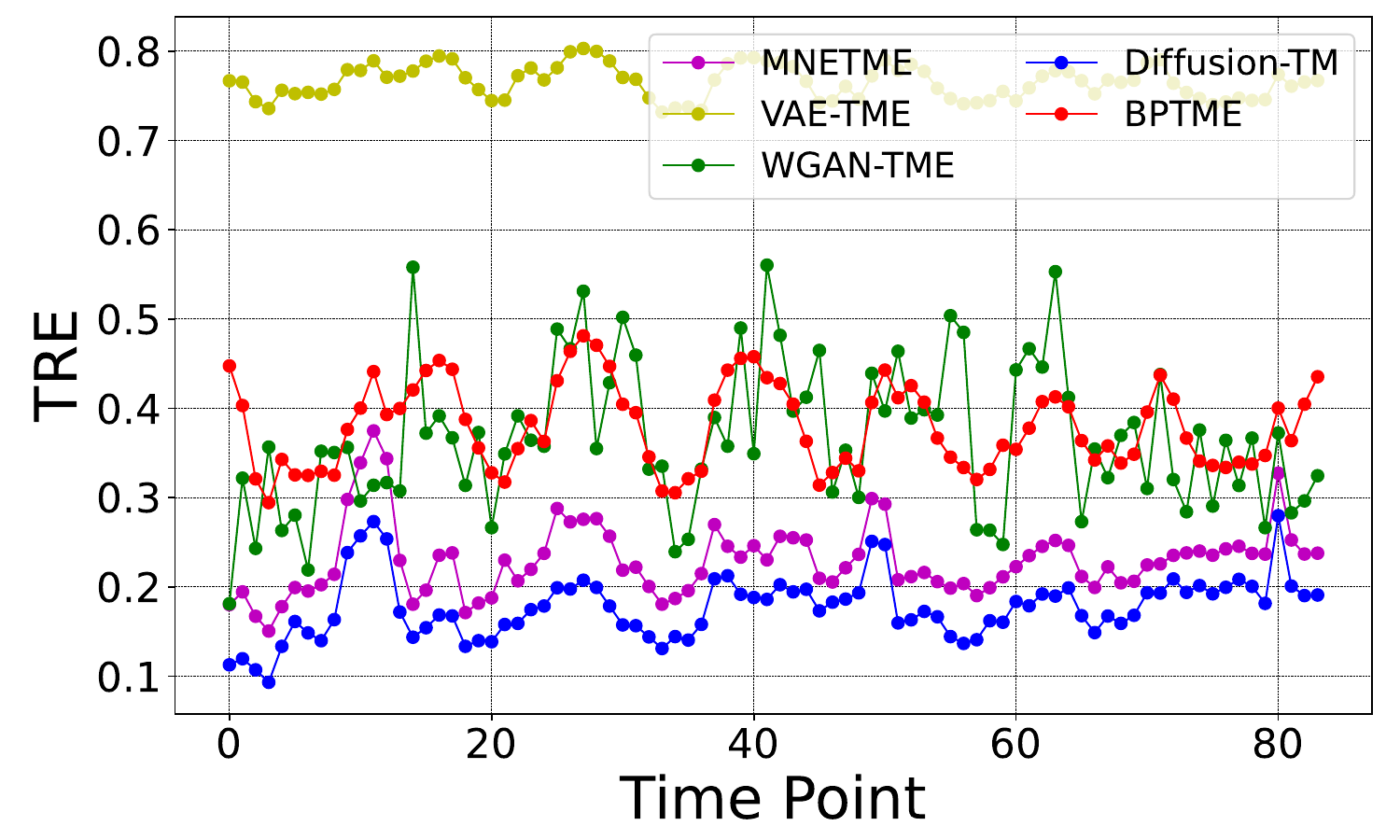}
\end{minipage}
}
% \caption{spatial relative errors (SREs) and temporal relative errors (TREs) on three sampling rates in Abilene dataset.}
\caption{{Temporal relative errors (TREs) on three sampling rates in Abilene (top) and GÉANT (bottom) dataset. For better visualization, the 672 instances were aggregated into 84 records in both datasets.}}
\label{TME_abliene_geant}
\end{figure*}

% \vspace{-1mm}

% \begin{figure*}[htbp]
% \centering
% \subfigure[Sampling Rate $ = 2\%$]{
% \begin{minipage}[b]{.31\linewidth}
%     \centering
%     \includegraphics[width=1.\linewidth]{Figures/TME_GEANT_0.02_TRE.pdf}
%     % \includegraphics[width=1.\linewidth]{Figures/TME_GEANT_0.02_TRE.pdf}\\
%     % \includegraphics[width=1.\linewidth]{Figures/TME_GEANT_0.02_SRE.pdf}
% \end{minipage}
% }
% \subfigure[Sampling Rate $ = 50\%$]{
% \begin{minipage}[b]{.31\linewidth}
%     \centering
%     \includegraphics[width=1.\linewidth]{Figures/TME_GEANT_0.5_TRE.pdf}
%     % \includegraphics[width=1.\linewidth]{Figures/TME_GEANT_0.5_TRE.pdf}\\
%     % \includegraphics[width=1.\linewidth]{Figures/TME_GEANT_0.5_SRE.pdf}
% \end{minipage}
% }
% \subfigure[Sampling Rate $ = 100\%$]{
% \begin{minipage}[b]{.31\linewidth}
%     \centering
%     \includegraphics[width=1.\linewidth]{Figures/TME_GEANT_1.0_TRE.pdf}
%     % \includegraphics[width=1.\linewidth]{Figures/TME_GEANT_1.0_TRE.pdf}\\
%     % \includegraphics[width=1.\linewidth]{Figures/TME_GEANT_1.0_SRE.pdf}
% \end{minipage}
% }
% % \caption{spatial relative errors (SREs) and temporal relative errors (TREs) on three sampling rates in GÉANT dataset.}
% \caption{Temporal relative errors (TREs) on three sampling rates in GÉANT dataset.}
% \label{TME_geant}
% \end{figure*}

\vspace{-1mm}

\subsection{Network Tomography Performance Comparison}

In this section, we consider the performance of Diffusion-TM with respect to the network tomography problem of inferring a TM from link-load measurements. Fig.~\ref{TME_scalar} presents the NMAE and MMD of all NT methods under different sampling rates. Note that we assume here we can measure all of the link loads on the networks. From the figures, the errors of all methods grow with the increasing of the unknown rate, indicating the missing values in the training set do affect the reliability of the estimations. However, we can see that Diffusion-TM consistently better approaches the ground truth and tracks the distribution compared with other methods. VAE-TME performs the worst as VAE cannot capture the statistics of the real-world traffic data, especially in a larger network such as G\'{E}ANT. It is uniformly shown that BPTME is better than VAE-TME, meanwhile, MNETME using the routing matrix's Moore–Penrose inverse and EM algorithm with a BPTME architecture provides a definite improvement that is closest to our Diffusion-TM. But note also that both methods belong to supervised learning that requires a fixed routing matrix for learning the inverse mapping directly, which means that the topology and routing of the network must be unchanged during training and testing, otherwise, they need to be retrained. So regardless of implicit distribution learning without theoretical support, the flexibility of these methods is greatly restricted. The last algorithm is WGAN-TME, which is similar to VAE-TME but uses the generator network of GAN to search for the optimal solution. It is apparent that the competition between the generator and discriminator results in the generator learning to produce a wide variety of more plausible outputs compared to VAEs. However, apart from the problem of training instability, the ability of GAN to learn the high-dimensional traffic distribution is still worse than that of Diffusion-TM, leading to significantly greater errors across all loss models. One can also observe that Diffusion-TM exhibits very stable estimation performance with $10\% \sim 100\%$ observed TM elements in the training set. In other words, if as few as $10\%$ of the TM elements are used to train our model, then Diffusion-TM achieves a performance similar to that requiring a sampling ratio of $90\%$ or more. The phenomenon again demonstrates the robustness of our proposed diffusion-based framework.

% \vspace{-3mm}

% Fig.~\ref{TME_abliene} and Fig.~\ref{TME_geant} plot TREs and SREs of all baselines under three sampling rates, where the SRE were sorted by flows' averaged true volumes in an ascending order to reveal the relationships between errors and OD flow volume. 
Fig.~\ref{TME_abliene_geant} plots the TREs of all baselines under three sampling rates. 
In both two datasets, the curves of all methods are gradually showing periodicity as the measurement time goes beyond. Overall, our method improves the TRE of the best baseline by $10\%$ in Abilene dataset,
and $27\%$ in G\'{E}ANT, averaging over all unknown ratios. Interestingly, the frequency and amplitude of WGAN-TME's curve suffer from a rapid change when the sampling rate is high, especially in high-dimensional G\'{E}ANT. That is because, on the one hand, it does not take into account temporal characteristics (but it may also lead to more unstable training). On the other hand, the synthetic data of WGANs fail to capture the diversity of the complex traffic data, which means any traffic estimated on that basis would also fail to diversify with increasing of the observed elements. 
% From the figures, we can see that all methods have a lower SREs in estimating large flows but OD-pairs with small volumes are usually difficult to reconstruct. We can also find the SRE of Diffusion-TM is less than other methods for even small flows in general. 
That indicates our method can achieve high accuracy for all flows without concerns on specific monitoring for small traffic measurements.

% \subsection{Robustness Study}
% .

% \subsubsection{Noisy traffic-level measurements}
% .

% \subsubsection{Noisy link-level measurements}
% .

% \subsubsection{Missing link-level measurements}
% .

\subsection{Computational Times}

{The theoretical complexity of different methods is compared in Table.~\hyperref[Table_R]{\Rmnum{3}}, where $L$ denotes the sequence length, $R$ denotes the number of optimization iterations, $s$ denotes the sampling steps, $k$ denotes the interval for updating generator network in GANs, $T_\theta$ denotes the propagation time of the underlying network, $T_{em}$ denotes the computation time of the EM algorithm, $E$ denotes the number of training epochs, $B$ denotes the batch size, $p$ denotes the sampling rate, $D_1$ and $D_2$ denotes the number of samples in training and testing set, respectively. The testing complexity of Diffusion-TM is proportional w.r.t $2sT_\theta + T_{em}$, however, the $\frac{1}{L}$ term and fewer reverse steps can considerably reduce its practical complexity. We also evaluate the complexity w.r.t baseline models. The inference of the direct-mapping methods (MNETME, BPTME, DATC) is fast in general, but the efficiency of generative method is constrained by $R$ which is often very large. For tensor factorization methods, ratio $p$ has significant effect as the training time is positively correlated with it during inference.} 

% \vspace{-3mm}

\begin{table}[htbp]
  \centering
  {\caption{Computation complexity w.r.t. benchmarks}}
    \resizebox{0.5\textwidth}{!}{\begin{tabular}{c|c|c}
    \toprule
    {Methods} & {Training Complexity} & {Inference Complexity} \\
    \midrule
    {MNETME} & {$\mathcal{O}\left( {E{D_1}{T_\theta }}/{B} \right)$} & {$\mathcal{O}\left( {{D_2}\left({T_\theta + T_{em}}\right)}/{B}\right)$} \\
           {BPTME} & {$\mathcal{O}\left( {E{D_1}{T_\theta }}/{B} \right)$} & {$\mathcal{O}\left( {{D_2}{T_\theta }}/{B} \right)$} \\
    \midrule
           {WGAN-TME} & {$\mathcal{O}\left( {\left(1+k\right)E{D_1}{T_\theta}}/{B} \right)$} & {$\mathcal{O}\left( {R{D_2}{T_\theta }}/{B} \right)$} \\
           {VAE-TME} & {$\mathcal{O}\left( {E{D_1}{T_\theta }}/{B} \right)$} & {$\mathcal{O}\left( {R{D_2}{T_\theta }}/{B} \right)$} \\
    \midrule
   {Tensor} & \multirow{2}*{{\textbackslash{}}} & {$\mathcal{O}( {pE{D_2}{T_\theta }}/{B} \ \ \ \ \ \ \ \ $} \\
           {Factorization} & & {$ \ \ \ \ \ + {\left(1-p\right){D_2}{T_\theta }}/{B})$} \\
    \midrule
           {DATC}  & {$\mathcal{O}\left( {\left(1+k\right)E{D_1}{T_\theta}}/{B} \right)$} & {$\mathcal{O}\left( {{D_2}{T_\theta }}/{B} \right)$} \\
    \midrule
           \rowcolor[rgb]{ .906,  .902,  .902} {Diffusion-TM} & {$\mathcal{O}\left( {E{D_1}{T_\theta }}/{B} \right)$} & {$\mathcal{O}\left( {{D_2}(2s{T_\theta} + {T_{em}})}/{LB} \right)$} \\
    \bottomrule
    \end{tabular}}
  \label{Table_R}
\end{table}

We then measure the actual computation times of Diffusion-TM and its competitors on both traffic matrix completion and network tomography tasks. For a fair comparison of these tasks, we collected 3000 samples ($10\%$ entities observed) as a training set, and then tested the inference times on them. As shown in Table.~\ref{Table_time}, we list all detailed computation times. Here the total diffusion step is set as 300, thus there is still space to reduce the running time of Diffusion-TM. Regarding NT solutions, it can be seen that Diffusion-TM outperforms other generative-model-based TME algorithms (WGAN-TME, VAE-TME) in terms of sampling time because they leverage extra thousands of iterations to adjust each estimation to be consistent with both the learned distribution and the NT constraints. The full-supervised TME methods (MNETME, BPTME) have the fastest computation times, although they also require intact label and routing information during training. For TMC times, tensor factorization methods usually take the longest recovering time as expected, at the cost of no off-line training requirements. DATC achieves the shortest inference time among its competitors, but similar to WGAN-TME, the algorithm suffers from longer training time than Diffusion-TM due to its adversarial nature. Overall, our Diffusion-TM is capable of providing more accurate estimates within a reasonable time, indicating our approach can be applied in practical scenarios.

\begin{table}[htbp]
  \centering
  \arrayrulecolor{black}
  \caption{Average Training and Inference Time}
    \resizebox{0.5\textwidth}{!}{\begin{tabular}{c|c|c|c|c}
    \toprule
    \multicolumn{5}{c}{\textbf{Solving Network Tomography Problem}} \\
    \midrule
    \multirow{2}[4]{*}{Methods} & \multicolumn{2}{c|}{Abilene} & \multicolumn{2}{c}{G\'{E}ANT} \\
\cmidrule{2-5}          & \multicolumn{1}{p{3.94em}|}{Training \newline{}Time (s)} & \multicolumn{1}{p{4.19em}|}{Inference \newline{}Time (s)} & \multicolumn{1}{p{3.94em}|}{Training \newline{}Time (s)} & \multicolumn{1}{p{4.19em}}{Inference \newline{}Time (s)} \\
    \midrule
    MNETME &   84.22    &   5.50    &   105.01    & 5.61 \\
    \midrule
    BPTME &    72.98   &   0.03    &    93.47   &  0.04 \\
    \midrule
    WGAN-TME &   1331.65    &   120.54    &    1614.18   & 123.72 \\
    \midrule
    VAE-TME &   109.70    &   156.25    &   130.48    &  162.30 \\
    \midrule
    \rowcolor[rgb]{ .906,  .902,  .902} Diffusion-TM &   542.08    &   11.45    &    550.64   & 12.29 \\
    \midrule
    \multicolumn{5}{c}{\textbf{Solving Traffic Matrix Completion Problem}} \\
    \midrule
    \multirow{2}[4]{*}{Methods} & \multicolumn{2}{c|}{Abilene} & \multicolumn{2}{c}{G\'{E}ANT} \\
\cmidrule{2-5}          & \multicolumn{1}{p{3.94em}|}{Training \newline{}Time (s)} & \multicolumn{1}{p{4.19em}|}{Inference \newline{}Time (s)} & \multicolumn{1}{p{3.94em}|}{Training \newline{}Time (s)} & \multicolumn{1}{p{4.19em}}{Inference \newline{}Time (s)} \\
    \midrule
    NTC   & \textbackslash{} &    339.03   & \textbackslash{} & 6290.32 \\
    \midrule
    NTM   & \textbackslash{} &    101.86   & \textbackslash{} & 1683.34 \\
    \midrule
    NTF   & \textbackslash{} &   8903.63    & \textbackslash{} & 17045.31 \\
    \midrule
    CoSTCo & \textbackslash{} &    41.90   & \textbackslash{} & 202.56 \\
    \midrule
    DATC  &   1602.92   &   1.33    &   2914.15  & 1.35 \\
    \midrule
    \rowcolor[rgb]{ .906,  .902,  .902} Diffusion-TM &   542.08    & 10.26      &    550.64   & 10.73 \\
    \bottomrule
    \end{tabular}}
  \label{Table_time}
\end{table}

\subsection{Ablation and Sensitivity Analysis}
The section mainly focuses on evaluating the impact of the crucial choices. In what follows, we will first test the following diffusion-related hyperparameters: a) trade-off scale parameter; b) diffusion and sampling steps. Then we conduct an ablation study including different algorithm components. Where not otherwise stated, a fixed scenario with $25\%$ link loads observed in the target set will be implemented thorough our analysis.

\subsubsection{Impact of scaling coefficients}
\begin{figure}[htbp]
\centering
\subfigure[Abilene]{
\includegraphics[width=0.8\linewidth]{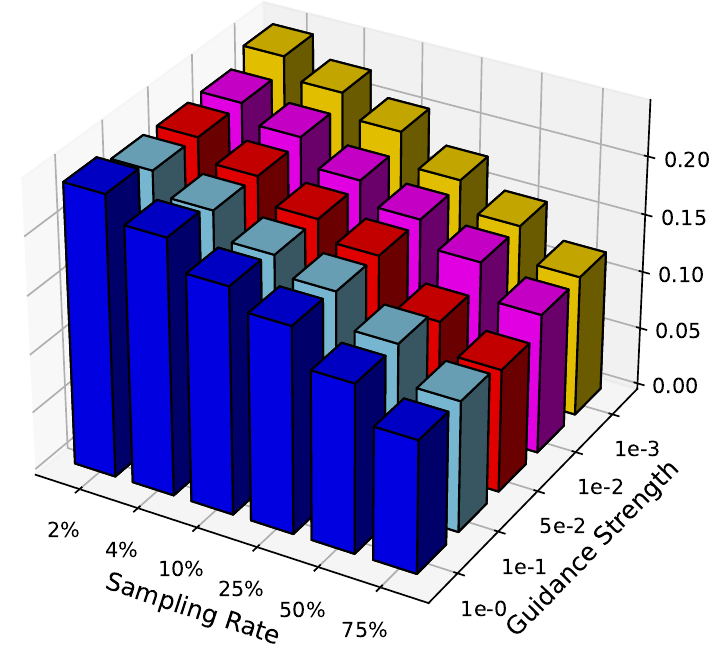}
}
\\
% \vspace{-0.3cm}
\subfigure[G\'{E}ANT]{
\includegraphics[width=0.8\linewidth]{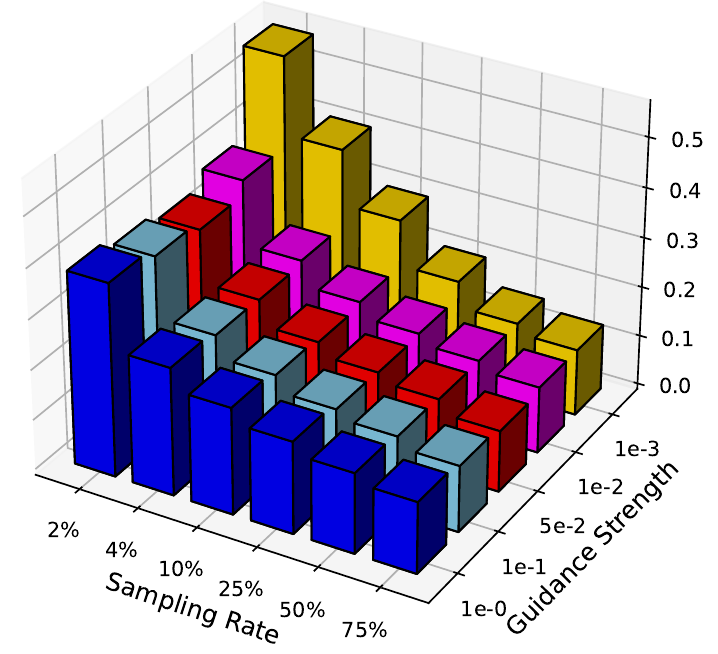}
}
\caption{{The performance (NMAE) of Diffusion-TM with different guidance strength $\rho$ under different known rates.}}
\label{Ablation_strength}
\end{figure}

Fig.~\ref{Ablation_strength} draws the recovery performance of our Diffusion-TM with different fixed scaling coefficient $\rho$, which is an important hyper-parameter that directly affects the consistency of sampling results. From the figure, we can find that the parameter is not very sensitive in Abilene. However, the choice of $\rho$ tends to have a significant influence on TME accuracy in G\'{E}ANT, especially when a smaller portion of the data is collected. Also note that there are a number of cases where the optimal $\rho$ is around 0.05. Therefore, we use this value in our experiments.

% \vspace{-0.5cm}

\subsubsection{Ablational Study}
\begin{figure*}[htbp]
\centering
\subfigure[Abilene]{
\includegraphics[width=0.48\linewidth]{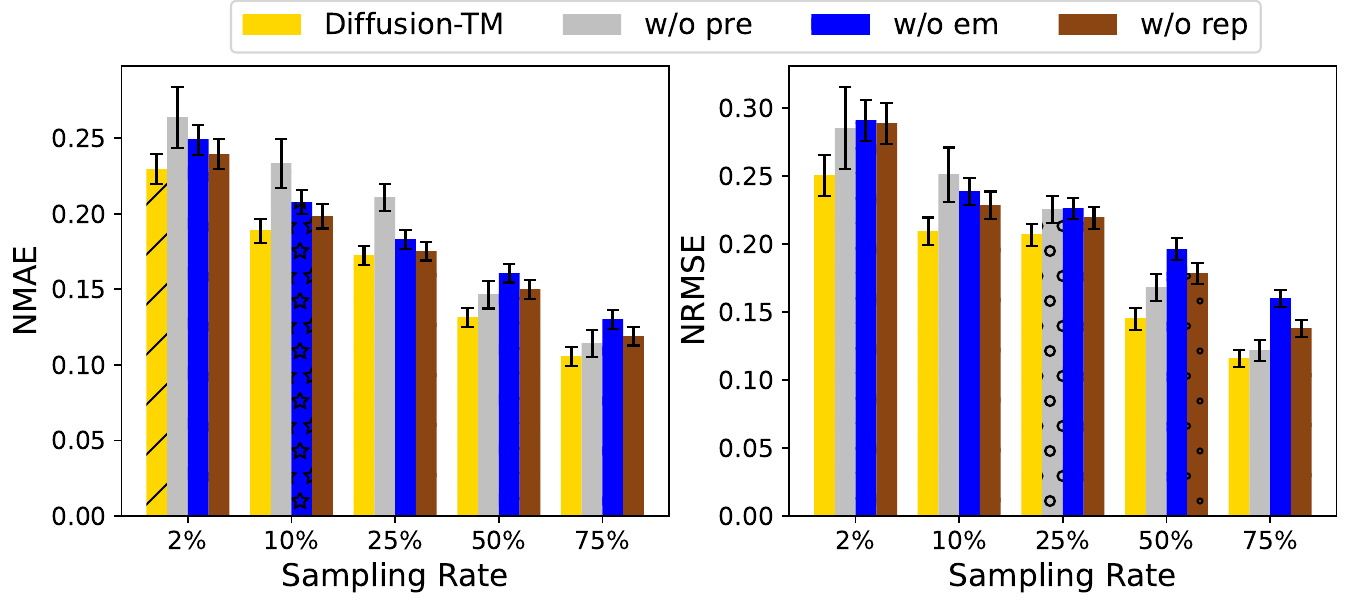}
}
\subfigure[G\'{E}ANT]{
\includegraphics[width=0.48\linewidth]{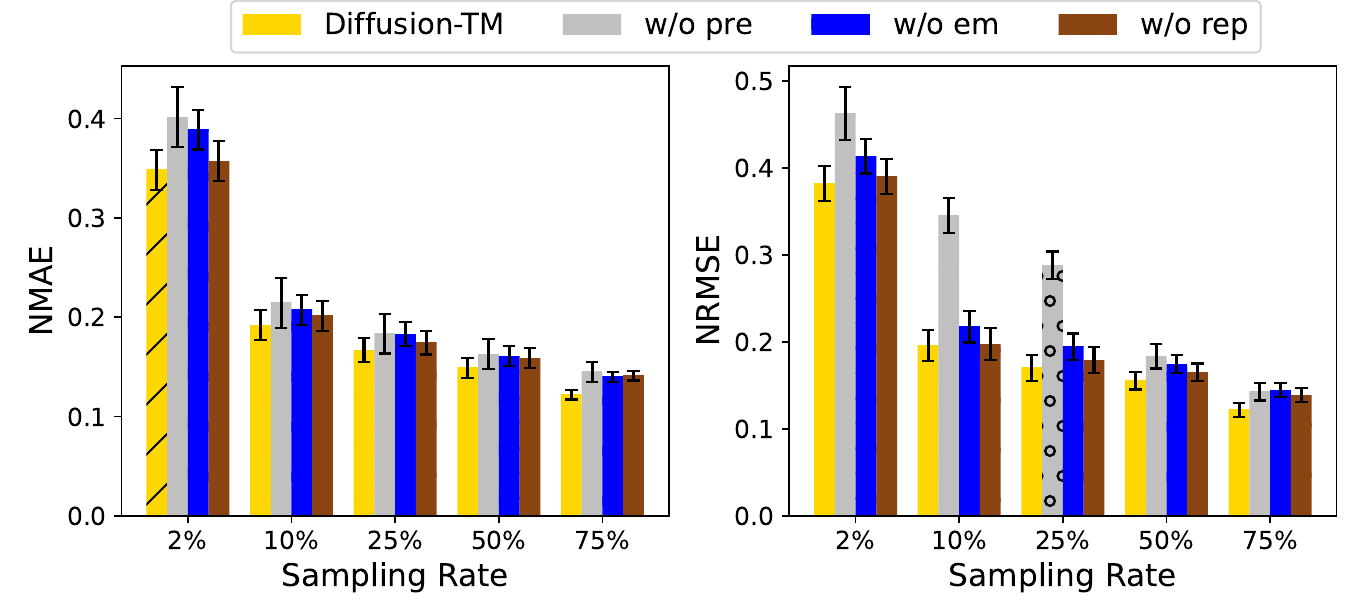}
}
\caption{{Results of Diffusion-TM and its three variants.}}
\label{Ablation}
\end{figure*}

We start by analyzing the effect of each part in our approach. Here, three variants of the Diffusion-TM were investigated: $\left( \rm i \right)$ \textbf{w/o pre}. We replace the pre-processing module using Eqn.~\ref{average_pre_processing} to train the diffusion model; $\left( \rm ii \right)$ \textbf{w/o em}. We remove the EM algorithm after each sampling through solving NT equations; $\left( \rm iii \right)$ \textbf{w/o rep}. We cancel additional correction steps to further clarify the efficacy of replace-based guidance. 
% $\left( \rm 4 \right)$ \textbf{w/o att}. we replace all the transformer layers by using a convolutional U-Net~\cite{r71}, which is common for diffusion models in image synthesis~\cite{r14}. 
Their ablation results are shown in Fig.~\ref{Ablation}. First, we see our model works across different known ratios and generates better quality for real-world network TM estimation in general, indicating the effectiveness of the combination of its components. It can also be seen our missing-data-aware strategy does boost the robust performance of Diffusion-TM. When there is a large number of network nodes (i.e. G\'{E}ANT), the dimension dependency is more prosperous, and the pre-processing module performs much better, especially under a sampling rate $< 50\%$. Moreover, we notice a clear positive impact of the expectation maximization, even with only $25\%$ of link measurements known. Diffusion-TM outperforms the counterpart without replace-based correction, which verifies the additional guidance is beneficial for recovering real-world traces.

\subsubsection{Impact of diffusion steps}

\begin{figure}[htbp]
\centering
\subfigure[Abilene]{
\includegraphics[width=0.8\linewidth]{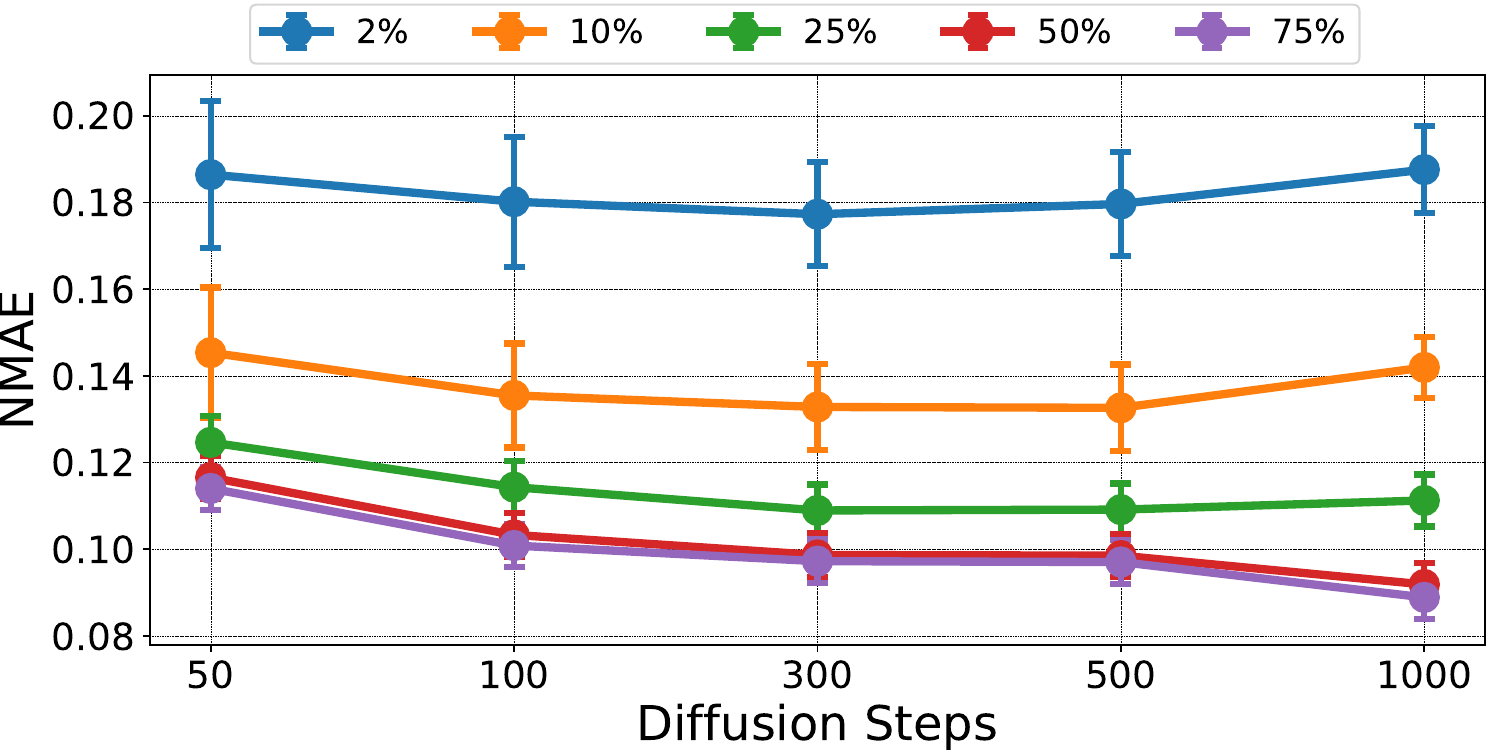}
}
\\
\vspace{-0.2cm}
\subfigure[G\'{E}ANT]{
\includegraphics[width=0.8\linewidth]{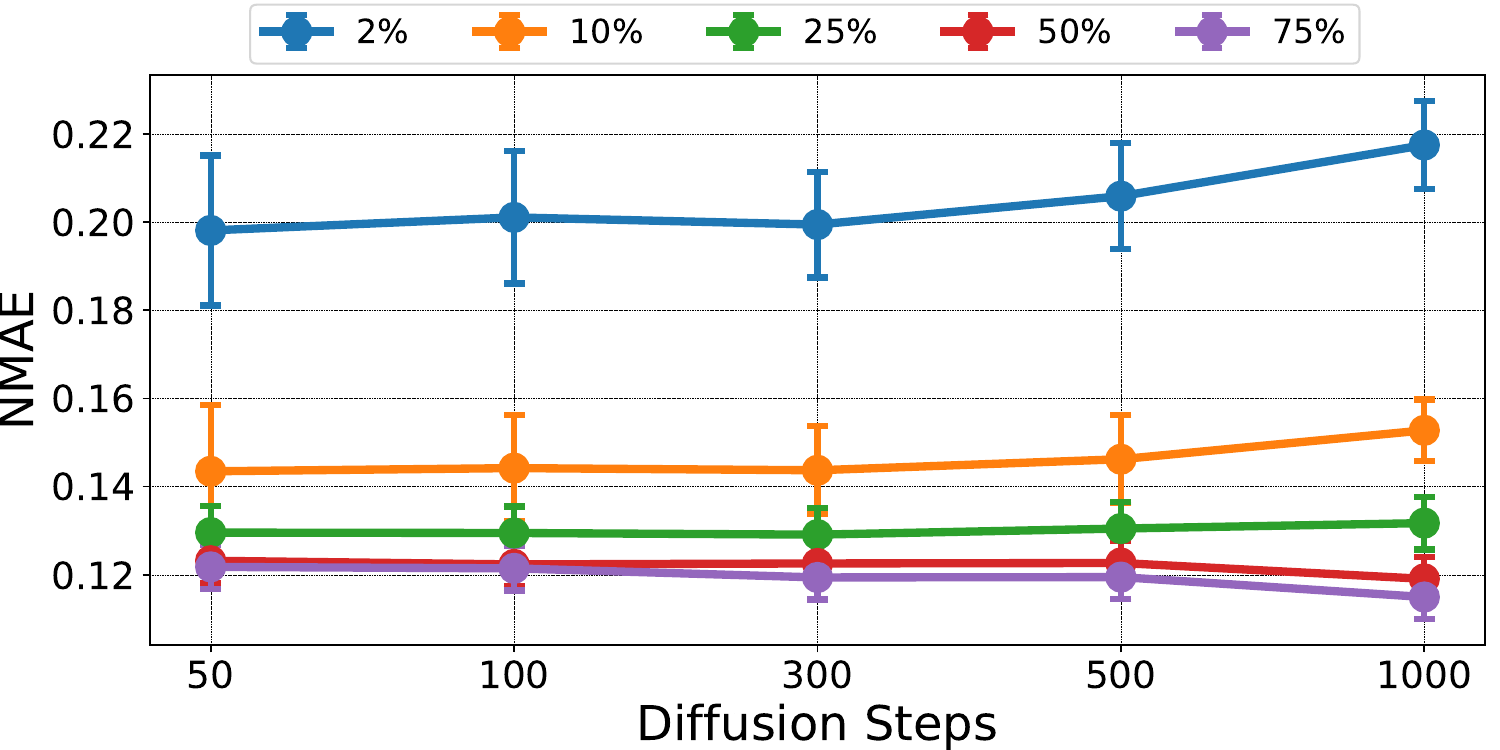}
}
\caption{{Diffusion-TM for different diffusion steps.}}
\label{Ablation_step}
\end{figure}

Here we evaluate the impact of a different number of sampling steps in the diffusion models. To do that, we first train a network with forward steps randomly selected in $\left\{1, \cdots, 1000\right\}$. Then when it comes to the inference period, we vary the diffusion steps in [50, 100, 300, 500, 1000] using DDIM shown in Eqn.~\ref{ddim}. Fig.~\ref{Ablation_step} reports the quantitative results versus number of diffusion steps. Overall, The recovery performance of different diffusion steps is close and as the known ratio increases, the diffusion model with more steps brings a better result. Also in this figure, one can observe that increasing the diffusion steps does not improve or even exacerbate the results when the sampling rate is low, due to potential overfitting. Therefore, to adapt to fast estimation in practice, in all our experiments we considered the most economical but effective setup of step number = 300.

% \subsubsection{Impact of network hyper-parameters}
% .

% \subsubsection{Impact of noise variance}
% .

\vspace{-3mm}

\section{Conclusion}\label{s8}

We presented in this paper Diffusion-TM, a novel diffusion framework to traffic matrix (TM) analysis in computer networks. Diffusion-TM bridges the gap between denoising diffusion models and traditional TM-related problems. By refining the generative process with available measurements and sampling from the space of plausible TMs, our diffusion-based approach achieves outstanding performance over state-of-the-art models on various tasks while avoiding expensive problem-specific training. We prove the feasibility of our method theoretically, then optimize the sampling procedure using the EM algorithm and replace-based guidance. Additionally, we proposed a two-stage training scheme to adapt Diffusion-TM to practical scenarios with a large number of missing values in the training set. Finally, we conducted extensive experiments on two real-world traffic datasets. Our results demonstrate the superiority of our Diffusion-TM on producing qualified TMs in different scenarios with incomplete training instances, offering an attractive alternative to the mainstream TM analysis methods.

Unlike existing TM analysis solutions, Diffusion-TM is versatile in that it can be flexibly implemented for multiple tasks (TM synthesis, tomography, and completion) at the same time by only collecting the traffic data of a subset of OD-flows within a short period for offline training. We thus believe that the method provides profound insights into a broad range of traffic matrix related applications. However, it is also known that the key limitation of our method is the high computational cost of the iterative denoising process. So exploring faster solvers that provide a trade-off between performance and computational overhead leaves considerable work to do in the future.

\section*{Acknowledgment}
This work was supported by Open Foundation of State key Laboratory of Networking and Switching Technology (Beijing University of Posts and Telecommunications) (SKLNST-2024-1-02).

% \begin{IEEEbiography}[{\includegraphics[width=1in,height=1.25in,clip,keepaspectratio]{fig1.png}}]{Xinyu Yuan}
% In this paragraph you can place your educational, professional background and research and other interests.\end{IEEEbiography}

% \begin{IEEEbiography}[{\includegraphics[width=1in,height=1.25in,clip,keepaspectratio]{fig1.png}}]{Yan Qiao}
% In this paragraph you can place your educational, professional background and research and other interests.\end{IEEEbiography}

% \clearpage

\vspace{-3mm}

\appendix

\textbf{Proof for Lemma 1:} We first compute the derivative of the marginal distribution $p\left( \boldsymbol{x}\right)$ with respect to $\boldsymbol{x}$ which could be expressed as
\begin{equation}
    \begin{aligned}
        &{\nabla _{\boldsymbol{x}}} p\left( \boldsymbol{x} \right) = {\nabla _{\boldsymbol{x}}}\int {p\left( {\boldsymbol{x}|\boldsymbol{\eta} } \right)g\left( \boldsymbol{\eta}  \right)d\boldsymbol{\eta} } \\
        &= {\nabla _{\boldsymbol{x}}}\int {{p_0}\left( \boldsymbol{x} \right)\exp \left( {{\boldsymbol{\eta} ^T}F\left( \boldsymbol{x} \right) - \psi \left( \boldsymbol{\eta} \right)} \right) g\left( \boldsymbol{\eta}  \right)d\boldsymbol{\eta} } \\
        &= {{\left( {{\nabla _{\boldsymbol{x}}}F\left( \boldsymbol{x} \right)} \right)}^T} \int {\boldsymbol{\eta} {p_0}\left( \boldsymbol{x} \right)\exp \left( {{\boldsymbol{\eta} ^T}F\left( \boldsymbol{x} \right) - \psi \left( \boldsymbol{\eta} \right)} \right)g\left( \boldsymbol{\eta} \right)d\boldsymbol{\eta} } \\
        & \ \ \ + {\nabla _{\boldsymbol{x}}}{p_0}\left( \boldsymbol{x} \right)\int {\exp \left( {{\boldsymbol{\eta} ^T}F\left( \boldsymbol{x} \right) - \psi \left( \boldsymbol{\eta}  \right)} \right)g\left( \boldsymbol{\eta} \right)d\boldsymbol{\eta} } \\
        &= {\left( {\nabla _x}F{\left( \boldsymbol{x} \right)}\right)}^T \int {\boldsymbol{\eta} p\left( {\boldsymbol{x},\boldsymbol{\eta} } \right)} d\boldsymbol{\eta} + \frac{{{\nabla _{\boldsymbol{x}}}{p_0}\left( \boldsymbol{x} \right)}}{{{p_0}\left( \boldsymbol{x} \right)}}\int {p\left( {\boldsymbol{x}|\boldsymbol{\eta} } \right)g\left( \boldsymbol{\eta}  \right)d\boldsymbol{\eta}} \\
        &= {\left( {\nabla _x}F{\left( \boldsymbol{x} \right)}\right)}^T \int {\boldsymbol{\eta} p\left( {\boldsymbol{x},\boldsymbol{\eta} } \right)} d\boldsymbol{\eta} + \frac{{{\nabla _{\boldsymbol{x}}}{p_0}\left( \boldsymbol{x} \right)}}{{{p_0}\left( \boldsymbol{x} \right)}}p\left( {\boldsymbol{x}} \right).
    \end{aligned}
\end{equation}
As a consequence, 
\begin{equation}
    {\left( {{\nabla _{\boldsymbol{x}}}F\left( \boldsymbol{x} \right)} \right)^T}\int {\boldsymbol{\eta} p\left( {\boldsymbol{\eta} |\boldsymbol{x}} \right)d\boldsymbol{\eta} } = \frac{{{\nabla _{\boldsymbol{x}}}p\left( \boldsymbol{x} \right)}}{{p\left( \boldsymbol{x} \right)}} - \frac{{{\nabla _{\boldsymbol{x}}}{p_0}\left( \boldsymbol{x} \right)}}{{{p_0}\left( \boldsymbol{x} \right)}}.
\end{equation}
Then, we have
\begin{equation}
    {\left( {{\nabla _{\boldsymbol{x}}}F\left( \boldsymbol{x} \right)} \right)^T}\boldsymbol{\hat \eta}  = {\nabla _{\boldsymbol{x}}}\log p\left( \boldsymbol{x} \right) - {\nabla _{\boldsymbol{x}}}\log {p_0}\left( \boldsymbol{x} \right),
\end{equation}
which concludes the proof. $\hfill \Box$

\textbf{Proof for Proposition 1:} If we consider a diffusion model in which the forward step can be modeled as Eqn.~\ref{forward}, the corresponding formula is then given by $\boldsymbol{x}|\boldsymbol{\eta} \sim {\cal N}\left( {\sqrt {{{\bar \alpha }_t}} \boldsymbol{\eta},\left( {1 - {{\bar \alpha }_t}} \right) \mathbf{I}}\right)$.
Therefore, using Eqn.~\ref{TF_conclusion}, we have
\begin{equation}
    \mathbb{E}\left( \boldsymbol{x_0}|\boldsymbol{x_t} \right) = \frac{1}{{\sqrt {{{\bar \alpha }_t}} }} \boldsymbol{x_t} + \frac{1 - {{\bar \alpha }_t}}{{\sqrt {{{\bar \alpha }_t}} }} {\nabla _{\boldsymbol{x_t}}}\log p_t\left( \boldsymbol{x_t} \right).
\end{equation}
This concludes the proof. $\hfill \Box$

\textbf{Lemma 2}: \textit{Let vector function ${\cal Q}$ is the orthogonal projection onto a subspace ${\cal M} \subset \mathbb{R}^n$. Then $\boldsymbol{J}_{\cal Q}$, the Jacobian matrix of ${\cal Q}$, is symmetric, i.e., $\boldsymbol{J}_{\cal Q}=\boldsymbol{J}_{\cal Q}^T$.}

\textit{Proof.} We start by defining ${T_s}\left( {{\cal M}, {\cal Q}\left( \boldsymbol{x}\right)} \right)$ as the tangent space at a point ${\cal Q}\left( \boldsymbol{x}\right)$ on ${\cal M}$. Let $\boldsymbol{u_1} \in {T_s}\left( {{\cal M}, {\cal Q}\left( \boldsymbol{x}\right)} \right)$, $\boldsymbol{u_2} \perp {T_s}\left( {{\cal M}, {\cal Q}\left( \boldsymbol{x}\right)} \right)$, and $\boldsymbol{u} = \boldsymbol{u_1} + \boldsymbol{u_2}$. Then given a constant $k$,
\begin{equation}
    {\cal Q}\left( \boldsymbol{x} + k\boldsymbol{u}\right) = {\cal Q}\left( \boldsymbol{x}\right) + k\boldsymbol{u_1},
\end{equation}
which comes from the only tangent vector $\boldsymbol{u_1}$ influence the orthogonal projection onto ${\cal M}$. And by differentiating the equation with respect to $k$, we can write ${\boldsymbol{J}}_{\cal Q}{\boldsymbol{u}} = \boldsymbol{u_1}$. Now considering another vector $\boldsymbol{v} = \boldsymbol{v_1} + \boldsymbol{v_2}$ with the same settings: $\boldsymbol{v_1} \in {T_s}\left( {{\cal M}, {\cal Q}\left( \boldsymbol{x}\right)} \right)$, $\boldsymbol{v_2} \perp {T_s}\left( {{\cal M}, {\cal Q}\left( \boldsymbol{x}\right)} \right)$. We have
\begin{equation}
    \begin{aligned}
        {{\boldsymbol{u}}^T}{{\boldsymbol{J}}_{\cal Q}}\boldsymbol{v} &= {\left( {\boldsymbol{u_1} + \boldsymbol{u_2}} \right)^T}\boldsymbol{v_1} = \boldsymbol{u_1}^T\boldsymbol{v_1} \\
         &= \left( {\boldsymbol{J}}_{\cal Q}{\boldsymbol{u}}\right)^T \boldsymbol{v_1} = {{\boldsymbol{u}}^T}{{\boldsymbol{J}}_{\cal Q}^T}\boldsymbol{v}.
        % &= \boldsymbol{v_1}^T\boldsymbol{u_1} = {\left( {\boldsymbol{v_1} + \boldsymbol{v_2}} \right)^T}\boldsymbol{u_1} \\
        % &= {{\boldsymbol{v}}^T}{{\boldsymbol{J}}_{\cal Q}}\boldsymbol{u}.
    \end{aligned}
\end{equation}
Thus, ${\boldsymbol{J}}_{\cal Q} = {\boldsymbol{J}}_{\cal Q}^T$ which concludes the proof. $\hfill \Box$

\textbf{Proof for Theorem 1:} First by combining with the condition that ${\cal M}$ has linear structure, we have ${\cal D}_t\left( \boldsymbol{x_t}\right) \in {\cal M}$ as ${\cal D}_t\left( \boldsymbol{x_t}\right) = \mathbb{E}\left( \boldsymbol{x_0}|\boldsymbol{x_t} \right) = \int {\boldsymbol{x_0}p\left( {\boldsymbol{x_0}|\boldsymbol{x_t}} \right)} d\boldsymbol{x_0}$ is the weighted average of points on the traffic data manifold.

Note that $p\left( \boldsymbol{x_0}| \boldsymbol{x_t}\right)$ is not only a Gaussian, but also a radial function $r\left( \boldsymbol{x_0} \right) = \hat r\left( {\left\| {\boldsymbol{x_0} - c} \right\|} \right)$ with center $c=\mathbb{E}\left( \boldsymbol{x_0}|\boldsymbol{x_t} \right)$. Thus intuitively, ${\cal D}_t\left( \boldsymbol{x_t}\right)$ should be the nearest point on ${\cal M}$ to $\boldsymbol{x_t}$ on the noisy data manifold ${\cal M}_t$. Then since the distance is usually the closest, we can say that ${\cal D}_t$ is locally an orthogonal projection onto $\cal M$, and $\boldsymbol{J}_{{\cal D}_t}$ is the orthogonal projection onto ${T_s}\left( {{\cal M}, {\cal D}_t\left( \boldsymbol{x_t}\right)} \right)$. Considering the constrained gradient term, we have
\begin{equation}
    \begin{aligned}
        &{\rho _t}{\nabla _{\boldsymbol{x_t}}}\left\| {\boldsymbol{y} - {\cal A}\left( {\boldsymbol{{\hat x}_0}} \right)} \right\|_2^2 = -2{\rho _t}{\boldsymbol{J}_{\boldsymbol{H{D_t}}}^T}\left( {y - \boldsymbol{H}{\boldsymbol{\hat x}_0}} \right) \\ 
        &= -2{\rho _t}{\boldsymbol{J}_{\boldsymbol{D_t}}^T}\boldsymbol{M}^T\left( {\boldsymbol{y} - \boldsymbol{H}{\boldsymbol{\hat x}_0}} \right)
    \end{aligned}
\end{equation}
Therefore, ${\rho _t}{\nabla _{\boldsymbol{x_t}}}\left\| {\boldsymbol{y} - {\cal A}\left( {\boldsymbol{{\hat x}_0}} \right)} \right\|_2^2 = \boldsymbol{J_{{D_t}}}s \in {T_s}\left( {{\cal M}, {\cal D}_t\left( \boldsymbol{x_t}\right)} \right)$ where $s = -2{\rho _t}\boldsymbol{M^T}\left( {\boldsymbol{y} - \boldsymbol{M}{\boldsymbol{\hat x}_0}} \right)$, which comes from the result of Lemma 2. As the gradient is a vector on ${T_s}\left( {{\cal M}, {\cal D}_t\left( \boldsymbol{x_t}\right)} \right)$, we finally conclude that constraint term would guide the diffusion model to lie on the data manifold ${\cal M}$, which may lead to more accurate inference. $\hfill \Box$

\bibliographystyle{IEEEtran}
\bibliography{main}

\end{document}